\documentclass[sigconf]{acmart}
\AtBeginDocument{%
  }

\setcopyright{acmlicensed}
\copyrightyear{2026}
\acmYear{2026}

\setcopyright{none}
\setcctype{by}
\acmConference[KDD '26]{Proceedings of the 32nd ACM SIGKDD Conference on Knowledge Discovery and Data Mining V.2}{August 09--13, 2026}{Jeju Island, Republic of Korea}
\acmBooktitle{Proceedings of the 32nd ACM SIGKDD Conference on Knowledge Discovery and Data Mining V.2 (KDD '26), August 09--13, 2026, Jeju Island, Republic of Korea}
\acmDOI{10.1145/3770855.3817528}
\acmISBN{979-8-4007-2259-2/2026/08}

\usepackage{tabularx}
\usepackage{booktabs} 
\usepackage{array}    
\usepackage{colortbl}  
\usepackage{amsmath}   
\usepackage{bm}         
\usepackage{multirow}
\newcolumntype{C}{>{\centering\arraybackslash}X}
\newcolumntype{L}[1]{>{\raggedright\arraybackslash}p{#1}}
\usepackage{booktabs}
\usepackage{array}
\usepackage{tabularx}
\usepackage{tcolorbox}    
\usepackage{listings}   
\usepackage{enumitem}   
\tcbuselibrary{breakable} 
\usepackage{listings}
\usepackage{bbm}
\usepackage{cleveref}
 \usepackage[normalem]{ulem}
\useunder{\uline}{\ul}{}
\newcommand{\greencheck}{{\color{green!70!black}$\boldsymbol{\checkmark}$}}
\newcommand{\redcross}{{\color{red!70!black}$\boldsymbol{\times}$}}

\definecolor{ldr-light-primary}{RGB}{234, 242, 255}
\definecolor{ldr-light-secondary}{RGB}{220, 236, 255}
\definecolor{ldr-light-accent}{RGB}{189, 215, 238}
\definecolor{ldr-light-border}{RGB}{173, 216, 230}

\definecolor{ldr-dark-primary}{RGB}{70, 130, 180}  
\definecolor{ldr-dark-secondary}{RGB}{100, 149, 237}
\definecolor{ldr-dark-accent}{RGB}{30, 144, 255}   
\definecolor{ldr-dark-border}{RGB}{65, 105, 225}  

\definecolor{ec-light-primary}{RGB}{255, 240, 245}  
\definecolor{ec-light-secondary}{RGB}{255, 228, 235} 
\definecolor{ec-light-accent}{RGB}{255, 209, 220} 
\definecolor{ec-light-border}{RGB}{255, 192, 203}

\definecolor{ec-dark-primary}{RGB}{219, 112, 147}     
\definecolor{ec-dark-secondary}{RGB}{199, 21, 133}    
\definecolor{ec-dark-accent}{RGB}{255, 105, 180}     
\definecolor{ec-dark-border}{RGB}{219, 112, 147}     

\definecolor{lt-light-primary}{RGB}{240, 255, 240}    
\definecolor{lt-light-secondary}{RGB}{220, 255, 220}
\definecolor{lt-light-accent}{RGB}{189, 236, 182}  
\definecolor{lt-light-border}{RGB}{144, 238, 144}

\definecolor{lt-dark-primary}{RGB}{60, 179, 113}   
\definecolor{lt-dark-secondary}{RGB}{50, 205, 50}   
\definecolor{lt-dark-accent}{RGB}{34, 139, 34}      
\definecolor{lt-dark-border}{RGB}{0, 128, 0}      

\definecolor{validate-light}{RGB}{255, 251, 230} 
\definecolor{validate-dark}{RGB}{218, 165, 32}

\definecolor{score-light}{RGB}{250, 245, 255}
\definecolor{score-dark}{RGB}{106, 90, 205}

\definecolor{gray-light}{RGB}{245, 245, 245}  
\definecolor{gray-medium}{RGB}{211, 211, 211}   
\definecolor{gray-dark}{RGB}{128, 128, 128}    

\definecolor{eval-light}{RGB}{255, 245, 230}  
\definecolor{eval-dark}{RGB}{236, 158, 84}   
\definecolor{ourrow}{RGB}{230, 240, 255}
\begin{document}

\title{Fin-RATE: A Real-world Financial Analytics and Tracking Evaluation Benchmark for LLMs on SEC Filings}

\settopmatter{authorsperrow=4} 
\author{Yidong Jiang}
\authornote{This work was done during the author's time at Yale University.}
\email{2253899@tongji.edu.cn}
\orcid{0009-0000-4033-8709}
\affiliation{%
 \institution{Tongji University}
 \city{Shanghai}
 \country{China}
}

\author{Junrong Chen}
\email{juc036@ucsd.edu}
\orcid{0009-0009-2529-3114}
\affiliation{%
  \institution{University of California, San Diego}
  \city{San Diego}
  \state{CA}
  \country{USA}
}

\author{Eftychia Makri}
\email{eftychia.makri@yale.edu}
\orcid{0009-0003-0469-4812}
\affiliation{%
  \institution{Yale University}
  \city{New Haven}
  \state{CT}
  \country{USA}
}

\author{Jialin Chen}
\email{jialin.chen@yale.edu}
\orcid{0009-0007-0909-4620}
\affiliation{%
  \institution{Yale University}
  \city{New Haven}
  \state{CT}
  \country{USA}
}

\author{Peiwen Li}
\email{peiwen.li@yale.edu}
\orcid{0009-0009-8318-2420}
\affiliation{%
  \institution{Yale University}
  \city{New Haven}
  \state{CT}
  \country{USA}
}

\author{Ali Maatouk}
\email{ali.maatouk@yale.edu}
\orcid{0000-0002-3436-7068}
\affiliation{%
  \institution{Yale University}
  \city{New Haven}
  \state{CT}
  \country{USA}
}

\author{Leandros Tassiulas}
\email{leandros.tassiulas@yale.edu}
\orcid{0000-0003-0932-774X}
\affiliation{%
  \institution{Yale University}
  \city{New Haven}
  \state{CT}
  \country{USA}
}

\author{Eliot Brenner}
\email{eliotpbrenner@gmail.com}
\orcid{0000-0002-9176-0595}
\affiliation{%
  \institution{Goldman Sachs}
  \city{New York}
  \state{NY}
  \country{USA}
}

\author{Bing Xiang}
\email{bxiang05@gmail.com}
\orcid{0009-0006-4028-4935}
\affiliation{%
  \institution{Goldman Sachs}
  \city{New York}
  \state{NY}
  \country{USA}
}

\author{Rex Ying}
\authornote{Corresponding author.}
\email{rex.ying@yale.edu}
\orcid{0000-0002-5856-5229}
\affiliation{%
  \institution{Yale University}
  \city{New Haven}
  \state{CT}
  \country{USA}
}

\renewcommand{\shortauthors}{Yidong Jiang et al.}

\begin{abstract}
With the increasing deployment of Large Language Models (LLMs) in the finance domain, LLMs are increasingly expected to parse complex regulatory disclosures. However, existing benchmarks often focus on isolated details, failing to reflect the complexity of professional analysis that requires synthesizing information across multiple documents, reporting periods, and corporate entities. Furthermore, these benchmarks do not disentangle whether errors arise from retrieval failures, generation inaccuracies, domain-specific reasoning mistakes, or misinterpretation of the query or context, making it difficult to precisely diagnose performance bottlenecks. To bridge these gaps, we introduce Fin-RATE, a benchmark built on U.S. Securities and Exchange Commission (SEC) filings and mirroring financial analyst workflows through three pathways: detail-oriented reasoning within individual disclosures, cross-entity comparison under shared topics, and longitudinal tracking of the same firm across reporting periods. We benchmark 17 leading LLMs, spanning open-source, closed-source, and finance-specialized models, under both ground-truth context and retrieval-augmented settings. Results show substantial performance degradation, with accuracy dropping by 18.60\% and 14.35\% as tasks shift from single-document reasoning to longitudinal and cross-entity analysis. This degradation is associated with increased comparison hallucinations, temporal and entity mismatches, and is further reflected in declines in reasoning quality and factual consistency—limitations that existing benchmarks have yet to formally categorize or quantify.
\end{abstract}



\maketitle
{%
  \renewcommand{\thefootnote}{}
  \footnotetext{GitHub: \url{https://github.com/jyd777/Fin-RATE}}
  \footnotetext{HuggingFace: \url{https://huggingface.co/datasets/GGLabYale/Fin-RATE}}
}
\section{Introduction}
In recent years, the application of general-purpose large language models (LLMs) in the financial domain has rapidly expanded, with an increasing number of financial institutions exploring their use in high-value tasks such as financial report analysis, risk monitoring, and investment research assistance. According to the McKinsey 2025 report \cite{mckinsey2025_finance_ai}, several finance teams have already leveraged LLMs to improve the accuracy of budgeting and the timeliness of financial reporting. LLMs are viewed as promising analytical assistants, expected to undertake parts of professional analysis workflows, particularly in parsing structurally complex and semantically dense regulatory disclosures such as SEC filings. However, effective understanding of such documents goes far beyond fact-level extraction or isolated question answering. Instead, it involves cross-document, cross-temporal, and cross-entity information integration and structured reasoning, imposing significantly higher demands on a model’s consistency, depth of reasoning, information aggregation, and robustness in expression.

\begin{table*}[htbp]
\centering
\small
\setlength{\tabcolsep}{2.95pt}
\caption{The criteria columns of the table include: (1) cross-company comparison, (2) cross-year tracking, (3) multi-document reasoning, (4) analysis-oriented question formulation, (5) fine-grained evaluation beyond answer-level correctness, and (6) data source coverage. Fin-RATE incorporates a wide range of SEC filings beyond 10-K, 10-Q, and 8-K (see Appendix~\ref{sec:dataset_detail} for details).}
\begin{tabular}{c c c c c c c}
\toprule
\textbf{Benchmark} & \textbf{Cross-Comp.} & \textbf{Cross-Year} & \textbf{Multi-Doc.} & \textbf{Analytic} & \textbf{Fine-grained Eval.}&\textbf{Data Source Details} \\ 
\midrule
\textbf{FinQA} \cite{chen2021finqa}& \redcross & \redcross & \redcross & \redcross & \redcross & annual report (source not disclosed) \\ 
\textbf{Tab-CQA} \cite{liu-etal-2023-tab} & \redcross & \redcross & \redcross & \redcross & \redcross & annual report (source not disclosed)\\ 
\textbf{ConvFinQA} \cite{chen-etal-2022-convfinqa}& \redcross & \redcross & \redcross & \redcross & \redcross & annual report (source not disclosed) \\ 
\textbf{Financial Touchstone} \cite{10.1145/3768292.3770417}& \redcross & \redcross & \redcross & \greencheck & \redcross & annual report (source not disclosed)\\ 
\textbf{PACIFIC} \cite{deng-etal-2022-pacific}& \redcross & \redcross & \redcross & \redcross & \redcross & SEC filings (10-K) \\ 
\textbf{TAT-QA} \cite{zhu-etal-2021-tat}& \redcross & \redcross & \redcross & \redcross & \redcross & SEC filings (10-K) \\ 
\textbf{FinDER} \cite{10.1145/3768292.3770361}& \redcross & \redcross & \redcross & \greencheck & \redcross & SEC filings (10-K) \\ 
\textbf{DocFinQA} \cite{reddy-etal-2024-docfinqa}& \redcross & \redcross & \redcross & \redcross & \redcross & SEC filings (10-K) \\ 
\textbf{SECQUE} \cite{benyoash-etal-2025-secque}& \greencheck & \redcross & \redcross & \greencheck & \redcross & SEC filings (10-K, 10-Q) \\ 
\textbf{FinanceBench} \cite{islam2023financebench}& \redcross & \redcross & \redcross & \greencheck & \redcross & SEC filings (10-K, 10-Q, 8-K) \\ 
\textbf{SEC-QA} \cite{lai-etal-2025-sec}& \greencheck & \greencheck & \greencheck & \redcross & \redcross & SEC filings (10-K, 10-Q, 8-K) \\ 

\midrule
\rowcolor{ourrow}
\textbf{Fin-RATE (Ours)} & \greencheck & \greencheck & \greencheck & \greencheck & \greencheck& \textbf{SEC filings} (10-K, 10-Q, 8-K, 6-K, etc.) \\ 
\bottomrule
\end{tabular}
\label{tab:benchmarks}
\end{table*}

While prior studies have shown that LLMs exhibit promising capabilities in financial knowledge question answering (QA) and structured numerical reasoning \cite{guo2025fineval, tang2025financereasoning}, multiple benchmarks grounded in real-world financial filings \cite{islam2023financebench,reddy-etal-2024-docfinqa,10.1145/3768292.3770361,lai-etal-2025-sec,zhang2025faith,ji2025phantom} have highlighted persistent limitations in several core dimensions: (1) difficulty in integrating dispersed information from long documents and establishing reasoning paths across paragraphs and time periods; (2) challenges in maintaining factual consistency between answers and the original disclosures, especially under complex semantic contexts; and (3) absence of explanatory reasoning to justify comparisons, interpret trends, or verify claims based on financial disclosures.

Although these limitations have been partially identified in prior benchmarks, existing evaluations still suffer from systematic shortcomings in both task design and evaluation methodology.
\textbf{(1) Most benchmarks oversimplify SEC filings as flat, isolated text spans,} focusing on information extraction, numerical computation over tabular data, and QA restricted to a single section and reporting year \cite{reddy-etal-2024-docfinqa,chen2021finqa,islam2023financebench,10.1145/3768292.3770417}. This narrow view ignores competitive dynamics between peer firms, the evolution of a single company across years, and the mutually reinforcing relationship between sections. As a result, the depth and breadth of the constructed questions fall far short of real-world analyst workflows, failing to effectively evaluate LLMs for practical financial analysis.
\textbf{(2) Evaluation protocols remain coarse-grained,} focusing primarily on final answer correctness. Even benchmarks that incorporate LLM-as-Judge modules \cite{10.1145/3768292.3770361,benyoash-etal-2025-secque} remain confined to generic axes like faithfulness. They provide little visibility into how and why models degrade on complex financial analysis, struggling to quantify capability loss or reveal concrete failure modes.

\begin{figure}[t]
    \centering
    \includegraphics[width=\linewidth]{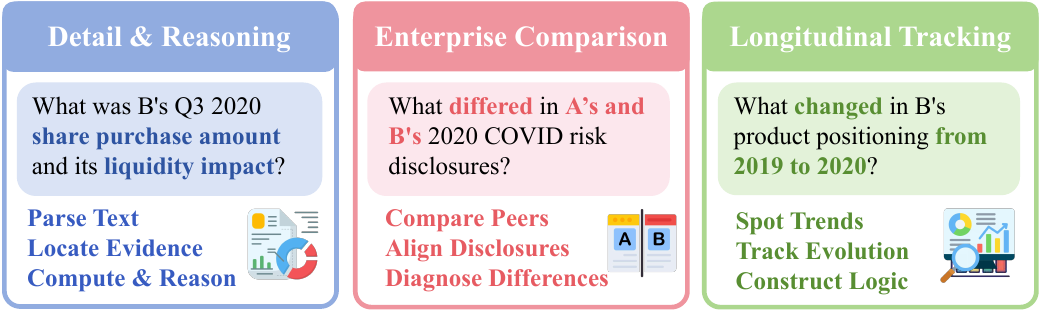}
\caption{An overview of Fin-RATE with core tasks and corresponding evaluated capabilities.}
    \label{fig:dataset_overview}
\end{figure}

To close these gaps, we propose \textbf{Fin-RATE}, a benchmark designed to model multi-dimensional alignment across corporate entities, temporal sequences, and heterogeneous disclosure structures as shown in Figure~\ref{fig:dataset_overview}. \textbf{Detail \& Reasoning QA (DR-QA)} assesses if a model can extract and integrate related information within a financial disclosure section to interpret specific financial impacts or perform calculations.
\textbf{Enterprise Comparison QA (EC-QA)} evaluates the model's ability to align disclosures from different companies and identify salient differences across structured metrics and narrative content, despite variations in reporting style.
\textbf{Longitudinal Tracking QA (LT-QA)} examines whether the model can detect trends and shifts across filings from the same company over time, including structural or semantic changes under evolving disclosure frames. 
Furthermore, to systematically diagnose model failures, we construct a fine-grained error taxonomy derived from extensive manual examination of model outputs on SEC filing tasks. Based on recurring failure patterns, we define thirteen core types across four categories: retrieval errors, generation inconsistencies, financial reasoning failures, and comprehension issues. We further observe persistent entity- and year-level retrieval mismatches in EC-QA and LT-QA. We summarize our key contributions below:

\begin{itemize}
\item We introduce \textbf{Fin-RATE}, the first benchmark that models realistic financial analysis workflow, integrating information across documents, reporting periods, and corporate entities.
\item We incorporate a domain-grounded Likert scoring scheme together with a 13-type error taxonomy grounded to quantify model degradation and identify bottlenecks.
\item We assess 17 LLMs under ground-truth context and retrieval-augmented settings, observing distinct capability profiles. Finance-specific models perform well in numeric reasoning and domain terminology but break down under complex reasoning demands.
\item Our error taxonomy reveals task-specific flaws: EC-QA is prone to hallucinated comparisons and entity confusion, and LT-QA to temporal mismatches and trend distortion.
\end{itemize}

\begin{figure*}
    \centering
    \includegraphics[width=0.9\textwidth]{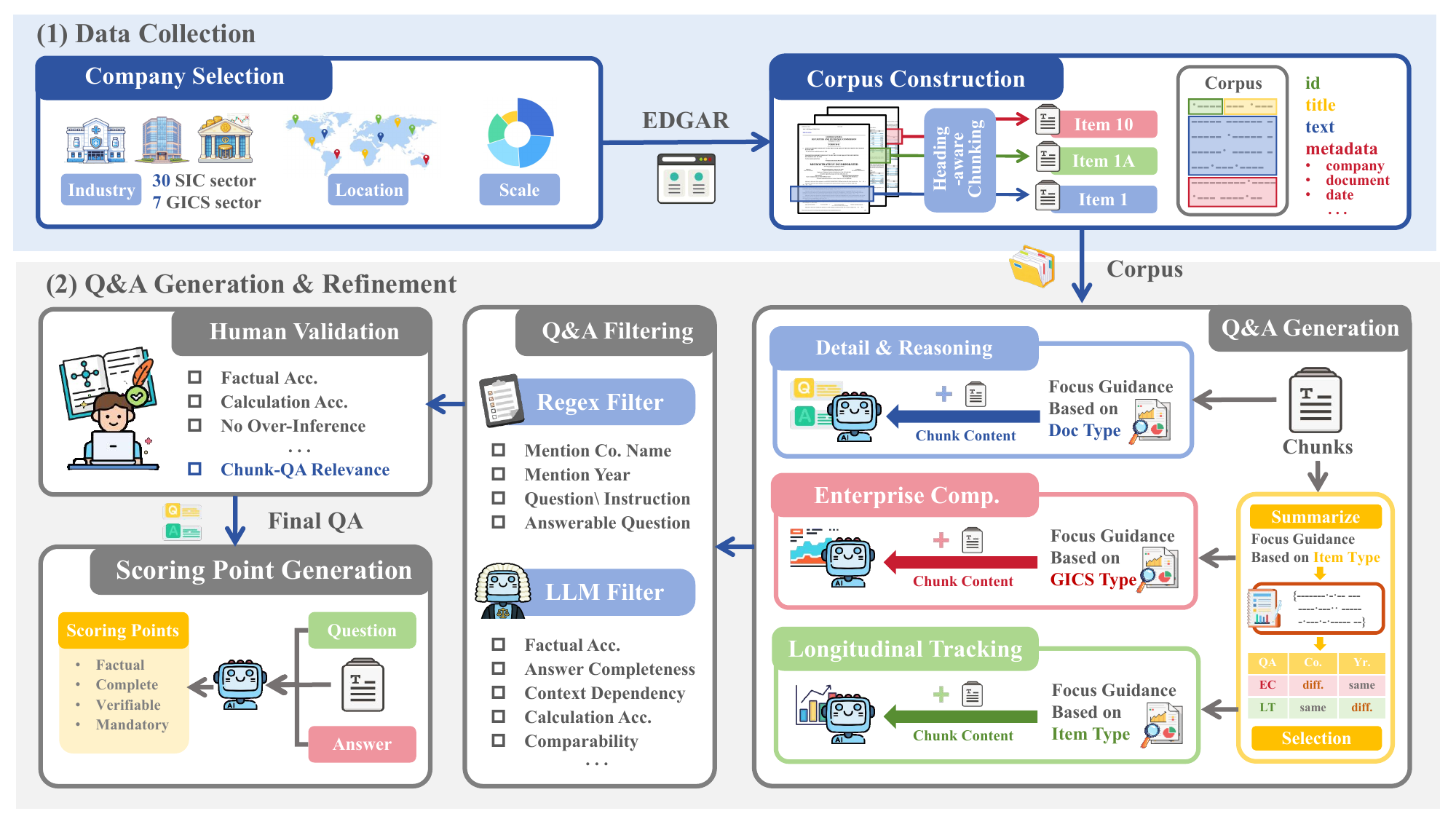}
    \caption{Overview of the Fin-RATE dataset construction framework. Stage (1) builds a heterogeneous corpus from SEC filings via systematic collection, parsing, and chunking. Stage (2) generates and refines QA pairs through task-specific prompting, multi-step filtering (regex and LLM-based), human validation, and scoring point extraction.}
    \label{fig:framework}
\end{figure*}

\section{Related Work}
Early financial QA benchmarks, such as FinQA~\cite{chen2021finqa}, TAT-QA~\cite{zhu-etal-2021-tat}, and PACIFIC~\cite{deng-etal-2022-pacific}, rely on curated tables and text excerpts, treating financial information as isolated knowledge units. Such atomization reduces analysis to symbolic computation, shielding models from the semantic noise and narrative complexity inherent in full SEC filings.
With the advancement of long-context language models, subsequent benchmarks, including FinanceBench~\cite{islam2023financebench}, FinDER~\cite{10.1145/3768292.3770361}, and DocFinQA~\cite{reddy-etal-2024-docfinqa} began to incorporate full-length SEC disclosure documents. These works focus on evaluating information retrieval and factual extraction under long-document settings.

Despite incorporating richer contexts, these benchmarks still fundamentally model filings as flattened databases, where core tasks are limited to point-wise fact lookup or single-period detail verification. As a result, they do not test whether models can reason about logical consistency, structural alignment, or semantic evolution across disclosures.
To extend beyond single-document retrieval, SEC-QA~\cite{lai-etal-2025-sec} introduces cross-company queries through a template-based generation pipeline. While enabling limited cross-entity comparison, the resulting questions are highly rigid in structure and struggle to capture the heterogeneous and narrative-driven nature of disclosures across different companies. Although SECQUE~\cite{benyoash-etal-2025-secque} further incorporates expert-authored analytical questions spanning tasks such as trend analysis, its comparative settings focus on financial metrics and remain restricted to single reporting periods, leaving non-financial dimensions like risk factors largely unexplored. 
Beyond task construction, existing benchmarks also suffer from a pronounced black-box issue in their evaluation mechanisms. As illustrated in Table~\ref{tab:benchmarks}, prior work relies solely on answer-level accuracy or coarse-grained labels such as Correct, Partial, or Incorrect. Such shallow evaluation protocols obscure the true failure modes of language models in complex financial reasoning scenarios, making it difficult to distinguish whether errors stem from long-document perception degradation, breakdowns in multi-step financial logic, or misalignment across temporal and structural contexts.

\section{Fin-RATE Dataset Construction}

\subsection{Data Source and Statistics}

As illustrated in Figure~\ref{fig:framework}, Fin-RATE is built entirely from official filings retrieved via the U.S.\ SEC EDGAR database\footnote{\url{https://www.sec.gov/edgar/search-and-access}}, ensuring authoritative, traceable, and consistent data sourcing. We select a representative set of filing types that together cover the key dimensions required for realistic financial analysis, including periodic reports (10-K, 10-Q), event-driven disclosures (8-K), governance materials (DEF~14A), foreign issuer updates (6-K), equity incentive plans (S-8), securities offerings (424B series), and shareholder communications (SC series). 
To support cross-year and cross-company reasoning, we filter companies by:
(1) ranking candidates by the number of available 10-K filings and retaining those with sufficient multi-year coverage; and
(2) sampling across GICS sectors to ensure broad industry and reporting-style diversity.
The collection spans sectors ranging from relatively stable industries such as utilities and consumer goods to dynamic or newly regulated areas such as cryptocurrency services and biopharmaceutical R\&D. 

All the filings are converted into Markdown, retaining section headings, tables, and lists. We segment each filing with SEC-standard item boundaries into coherent analytical units to support reasoning and retrieval at multiple levels of granularity. The final corpus comprises 15,311 document chunks derived from 2,472 filings, covering 43 companies across 36 industries from 2020 to 2025, with each chunk containing an average of 2,649 words. Periodic and event-driven reports like 10-K, 8-K, and 10-Q forms constitute the majority of the collection, with their distributions and regulatory functions detailed in Appendix~\ref{sec:dataset_detail}.
Based on this corpus, we construct three QA tasks: Detail Reasoning (DR-QA), Enterprise Comparison (EC-QA), and Longitudinal Tracking (LT-QA). Each task contains 2,500 instances. DR-QA questions are grounded in a single chunk, while EC-QA and LT-QA require aggregating evidence from an average of 5.6 and 3.0 chunks, respectively.

\subsection{Task Definition and Construction}

\subsubsection{Detail \& Reasoning QA (DR-QA)}
DR-QA examines a model’s capability for fine-grained financial reasoning within a single, self-contained document chunk. It focuses on local-context comprehension, precise detail extraction, and analytical interpretation of both structured and unstructured financial content. Questions involve numerical computation, semantic understanding, and professional judgment based on dense textual disclosures, without reliance on external retrieval. 
A core challenge lies in precision. The model must ground its reasoning strictly in textual evidence, avoid unsupported assumptions, and correctly interpret nuanced yet materially significant expressions, including technical accounting terms, nested tabular data, and qualitative risk descriptions. 

To ensure the analytical depth of each generated QA pair, we design adaptive prompt templates guided by the official section definitions of various SEC filing types. Prompts for 10-K filings emphasize annual ratio analysis, 10-Q filings highlight quarterly trend dynamics, and 8-K filings direct attention to the quantitative impact of material events. This ensures that each question focuses on the most analytically salient components of the disclosure, effectively reducing the generation of low-value questions and establishing a high-fidelity benchmark for local financial comprehension.

\subsubsection{Enterprise Comparison QA (EC-QA)}
EC-QA aims to evaluate a model’s ability to perform cross-company comparative reasoning under consistent disclosure contexts. The task involves aligning multi-entity information and identifying salient differences across shared analytical dimensions, covering both structured financial metrics and unstructured narrative disclosures.
A core challenge lies in accurate entity alignment and multi-source comparison. The model must distinguish between disclosures from different companies, identify comparable dimensions, and avoid misalignment caused by variations in disclosure styles or contextual framing.

To ensure interpretability and regulatory consistency, EC-QA is framed within the same Global Industry Classification Standard (GICS) sector category and limited to companies filing 10-K reports in the same fiscal year, ensuring consistency in disclosure frameworks and data granularity. During text selection, we pinpoint semantically aligned sections using structured headings from SEC filings, prioritizing high-comparability modules such as ``Item 1A: Risk Factors'' and ``Item 7: Management Discussion and Analysis.'' Besides, cross-sectional combinations are introduced to support multi-dimensional comparative questions across business operations, risk exposures, governance structures, and financial performance. To mitigate the risks of information dilution and hallucination caused by long input contexts, EC-QA construction process adopts a two-stage generation strategy. It first generates structured summaries for each selected company section and then constructs comparison QA pairs from these aligned summaries. In this way, EC-QA balances analytical depth with input complexity control.

\subsubsection{Longitudinal Tracking QA (LT-QA)}
LT-QA aims to evaluate a model’s ability to perform longitudinal integration and trend analysis over multi-year financial disclosures for a single company. It emphasizes structured understanding and temporal continuity across reporting periods, requiring models to identify trends, year-over-year changes, and strategic or risk-related shifts across multiple thematic sections.
The core challenge lies in tracking semantic consistency over time despite variations in terminology, measurement units, or narrative framing. Models must synthesize information across years to derive generalized conclusions, rather than relying on isolated, single-period evidence, which requires strong temporal sensitivity and multi-level semantic abstraction. 

LT-QA is constructed based on long-form disclosures, primarily annual 10-K reports and DEF 14A proxy statements, and adopts a two-stage, summary-first generation pipeline similar to that used in EC-QA. First, structured summaries are generated for each company–year–section unit, capturing key aspects such as strategic orientation. Subsequently, QA instances are constructed based on these summaries, with each question explicitly linking the same section across multiple years and reflecting temporal comparisons and evolutionary reasoning.
\subsection{Rigorous Quality Control}
Fin-RATE includes three quality control stages designed to ensure the reliability, factual grounding, and financial relevance of all QAs.
In the automated quality control phase, we adopt a dual-model generation-verification mechanism. First, the generation model (DeepSeek-V3.2) produces multiple candidate QAs under either chunk-level or document-level contexts, based on different prompting strategies. Then, two structurally distinct and high-performing LLMs (DeepSeek-V3.2 and Qwen3-235B) independently assess each candidate QA, focusing on whether the answer is sufficiently supported by the context, whether it has numerical or logical errors, and whether it relies on unsupported inference. Based on task-specific validation rules and a standardized rubric, each QA is labeled as Correct, Partial, Incorrect, or Failure. Any QA flagged as erroneous by either model is discarded, and only those rated as correct proceed to the human review stage.
In the final manual review phase, each QA that passed automated filtering is confirmed one by one to ensure strict adherence to the original disclosure without subjective interpretation. For each QA type, this process is repeated iteratively until 2,500 valid QAs are retained. 
For QAs involving multiple chunks or cross-document contexts, we examine whether each referenced chunk contributes substantively to the reasoning chain. If any referenced chunk is irrelevant, the QA is rewritten or discarded. For numerical questions, we recalculate all values to avoid potential errors. During this process, we observed recurring error patterns, which motivated the error type taxonomy used in our following evaluation.

\section{Experiments}
\begin{table*}[htbp]
\centering
\small
\setlength{\tabcolsep}{2pt}
\caption{Each model with Ground Truth Context is evaluated on three task types (DR-QA, EC-QA, LT-QA) using four label distributions (Corr / Partial / InCorr / Fail) and overall accuracy (Acc. (\%)= Corr/Total).\textbf{Bold} indicates the highest score across all models; \underline{Underlined} indicates the second-highest overall. Colored values are the highest score within each category.}
\label{tab:GT}
\begin{tabular}{@{}l|ccccc|ccccc|ccccc@{}}
\toprule
\multicolumn{1}{c|}{}                        & \multicolumn{5}{c|}{DR-QA}                                                                                                     & \multicolumn{5}{c|}{EC-QA}                                                      & \multicolumn{5}{c}{LT-QA}                                                                                                      \\ \cmidrule(l){2-16} 
\multicolumn{1}{c|}{\multirow{-2}{*}{Model}} & Corr & Partial & Incorr & Failure & Acc. (\%)                                  & Corr & Partial& Incorr & Failure & Acc. (\%)                                  & Corr & Partial & Incorr & Failure & Acc. (\%)                                  \\ \midrule
\multicolumn{16}{c}{\cellcolor[HTML]{FFF3CA}Closed-Source}                                                                                                                                                                                                                                                                                                                                                                                      \\\midrule
GPT-4.1                                     & 964     & 1156                                                   & 379       & 1       & 38.56                                 & 722     & 949                                                    & 812       & 17      & {\ul 28.88}                           & 807     & 1091                                                   & 584       & 18      & {\ul 32.28}                           \\
GPT-4.1-websearch                           & 939     & 1174                                                   & 385       & 2       & 37.56                                 & 683     & 988                                                    & 811       & 18      & 27.32                                 & 763     & 1111                                                   & 600       & 26      & 30.52                                 \\
GPT-5-websearch                             & 1074    & 1036                                                   & 384       & 6       & {\color[HTML]{CD9934} 42.96}          & 1091    & 683                                                    & 716       & 10      & {\color[HTML]{CD9934} \textbf{43.64}} & 1088    & 791                                                    & 592       & 29      & {\color[HTML]{CD9934} \textbf{43.52}} \\\midrule
\multicolumn{16}{c}{\cellcolor[HTML]{D9E1F4}Open-Source General}                                                                                                                                                                                                                                                                                                                                                                               \\\midrule
MIMO-V2-Flash                               & 587     & 1381                                                   & 520       & 12      & 23.48                                 & 339     & 1114                                                   & 1009      & 38      & {\color[HTML]{3166ff} 13.56}          & 428     & 1129                                                   & 923       & 20      & {\color[HTML]{3166ff} 17.12}          \\
DeepSeek-V3                                 & 453     & 1634                                                   & 412       & 1       & 18.12                                 & 77      & 536                                                    & 370       & 1517    & 3.08                                  & 206     & 1016                                                   & 314       & 964     & 8.24                                  \\
DeepSeek-V3.2                               & 505     & 1574                                                   & 420       & 1       & 20.20                                 & 318     & 1050                                                   & 932       & 200     & 12.72                                 & 401     & 1189                                                   & 683       & 227     & 16.04                                 \\
DeepSeek-R1                                 & 578     & 1469                                                   & 420       & 33      & 23.12                                 & 197     & 966                                                    & 925       & 412     & 7.88                                  & 390     & 1162                                                   & 604       & 344     & 15.60                                 \\
GPT-OSS-20B                                 & 1088    & 913                                                    & 498       & 1       & {\color[HTML]{3166ff} {\ul 43.52}}    & 60      & 283                                                    & 1506      & 651     & 2.40                                  & 254     & 130                                                    & 1914      & 202     & 10.16                                 \\
Llama-3.3-70B-Instruct                      & 873     & 1063                                                   & 563       & 1       & 34.92                                 & 107     & 1092                                                   & 1174      & 127     & 4.28                                  & 277     & 1250                                                   & 799       & 174     & 11.08                                 \\
Qwen3-8B                                    & 320     & 873                                                    & 1057      & 250     & 12.80                                 & 8       & 226                                                    & 920       & 1346    & 0.32                                  & 83      & 684                                                    & 1101      & 632     & 3.32                                  \\
Qwen3-14B                                   & 660     & 881                                                    & 843       & 116     & 26.40                                 & 42      & 440                                                    & 657       & 1361    & 1.68                                  & 142     & 636                                                    & 552       & 1170    & 5.68                                  \\
Qwen3-30B-A3B-Insturct-2507                 & 792     & 1105                                                   & 602       & 1       & 31.68                                 & 199     & 1030                                                   & 1262      & 9       & 7.96                                  & 331     & 1189                                                   & 957       & 23      & 13.24                                 \\
Qwen3-235B                             & 998     & 1060                                                   & 441       & 1       & 39.92                                 & 338     & 909                                                    & 791       & 462     & 13.52                                 & 493     & 983                                                    & 782       & 242     & 19.72                                 \\\midrule
\multicolumn{16}{c}{\cellcolor[HTML]{E3F2D9}Open-Source Finance}                                                                                                                                                                                                                                                                                                                                                                               \\\midrule
Fin-R1                                      & 1437    & 618                                                    & 444       & 1       & {\color[HTML]{036400} \textbf{57.48}} & 83      & 555                                                    & 1498      & 364     & {\color[HTML]{036400} 3.32}           & 243     & 680                                                    & 1559      & 18      & {\color[HTML]{036400} 9.72}           \\
Fino1-14B                                   & 767     & 989                                                    & 652       & 92      & 30.68                                 & 41      & 761                                                    & 904       & 794     & 1.64                                  & 177     & 1036                                                   & 706       & 581     & 7.08                                  \\
FinanceConnect-13B                          & 178     & 1206                                                   & 1094      & 22      & 7.12                                  & 1       & 559                                                    & 654       & 1286    & 0.04                                  & 20      & 909                                                    & 790       & 781     & 0.80                                  \\
TouchstoneGPT-7B-Instruct                   & 9       & 556                                                    & 1864      & 71      & 0.36                                  & 2       & 401                                                    & 846       & 1251    & 0.08                                  & 20      & 63                                                     & 1256      & 1161    & 0.80                                  \\ \midrule
\multicolumn{16}{c}{\cellcolor[rgb]{0.894,0.894,0.894} Human Expert}\\
\midrule
Human Expert  & 2310 & 172 & 18 & 0 & 92.40 & 2067 & 370 & 63 & 0 & 82.68 & 2147 & 313 & 40 & 0 & 85.88 \\ \bottomrule
\end{tabular}
\end{table*}
We evaluate 17 LLMs on DR-QA, EC-QA, and LT-QA.
Under the gold-context setting, we combine Likert-scale scores
with systematic error taxonomy to analyze task-level challenges
in EC-QA and LT-QA, as well as category-specific behaviors across
closed-source, general open-source, and finance-tuned models.
Under the RAG setting, we further investigate how retrieval
quality affects end-to-end performance using standard IR metrics
and retrieval-stage error analysis.
Based on these insights, we design a hierarchical retrieval
framework to improve evidence coverage and ranking robustness.
\subsection{Setup}
\subsubsection{Models}

To comprehensively evaluate financial analytics capabilities, we select 17 LLMs categorized into three groups:

\begin{itemize}
\item \textbf{Closed-source LLMs:} To evaluate top-tier financial reasoning LLMs, we evaluate GPT-5 with Web Search \cite{singh2025openai}, GPT-4.1 \cite{achiam2023gpt}, and GPT-4.1 with Web Search.
\item \textbf{Open-Source LLMs:} For reproducibility, we test DeepSeek-V3.2/V3 \cite{liu2024deepseek}, DeepSeek-R1 \cite{guo2025deepseek}, the Qwen3 family \cite{yang2025qwen3} (8B, 14B, 30B, 235B), Llama-3.3-70B \cite{dubey2024llama}, GPT-OSS-20B \cite{agarwal2025gpt}, and the MoE-based MiMo-V2-Flash \cite{xiao2026mimo}.
\item \textbf{Finance-Specific LLMs:} To assess the effect of domain adaptation, we evaluate TouchstoneGPT-7B \cite{wu-etal-2025-golden}, Fin-R1 \cite{liu2026finr1largelanguagemodel}, Fino1-14B \cite{qian2025fino1}, and FinanceConnect-13B \cite{ceadar_2023}.
\end{itemize}
\subsubsection{Retrieval Settings}
To simulate realistic document access and knowledge grounding in professional financial analysis, we evaluate models under a retrieval-augmented generation (RAG) setting:

\begin{itemize}

\item \textbf{Lexical Retrieval:} A sparse BM25-based retrieval method \cite{robertson2009probabilistic} focusing on exact term matching, essential for formal financial terminology and specific entity identifiers.

\item \textbf{Dense Vector Retrieval:} A semantic approach using the most downloaded general embedding model (all-MiniLM-L6-v2 \cite{wang2020minilm}) and domain-specific embedding model (finance-embeddings-investopedia \cite{FinLangInvestopediaEmbedding}) on HuggingFace to evaluate financial domain adaptation's impact.

\item \textbf{Hybrid Retrieval:} An equal-weighted fusion of BM25 and finance-embeddings-investopedia semantic vectors within Elasticsearch \cite{elasticsearch2018elasticsearch}, thereby balancing keyword precision with semantic depth.

\item \textbf{Reranking Stage:} To further refine the candidate set, we employ the bge-reranker-v2-m3 \cite{li2023making, chen2024m3} to re-scores the retrieved documents for high contextual relevance

\end{itemize}

\subsection{Evaluation Protocol}
\begin{figure*}
    \centering
    \includegraphics[width=\textwidth]{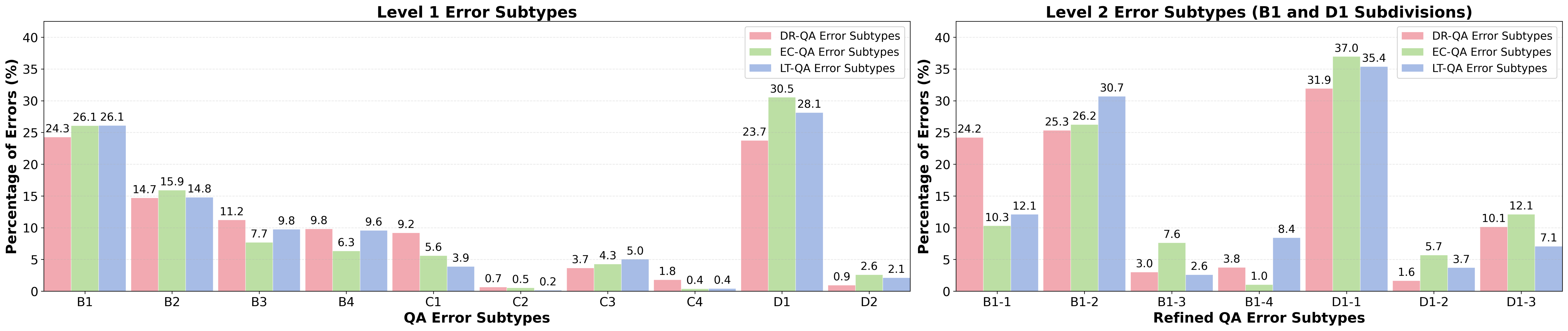}
    \caption{Distribution of QA error subtypes across three task. 
Left: Level 1 error types, covering broad categories B1–B4, C1–C4, and D1–D2.  
Right: Level 2 refinements of B1 (Hallucination) and D1 (Query Misunderstanding).}
    \label{fig:error_st}
\end{figure*}
\subsubsection{Generation Evaluation Metrics}
We evaluate model-generated answers using a unified LLM-as-Judge framework. Three LLMs with distinct 
architectures—GPT-5 \cite{singh2025openai}, DeepSeek-V3.2 \cite{liu2024deepseek}, and Qwen3-235B \cite{yang2025qwen3}—serve as independent judges. Each judge assesses the response strictly based on the provided reference answer or context, following standardized instructions that prohibit the use of external knowledge or speculative inference. 
To ensure reliable judgment, we derive weights through an alignment analysis (Appendix~\ref{consistency_verification}) and fuse their outputs using weighted averaging for correctness labels and Likert scores, and a union rule for
error-type attribution (Appendix~\ref{judge_fusion_strategy}). 
The judge ensemble provides three assessments:

\textbf{Correctness Rate} is the proportion of CORRECT responses among all answers, based on four mutually exclusive labels provided by LLM Judges: \textbf{CORRECT:} fully accurate and complete with no factual errors; \textbf{PARTIAL:} partially correct, missing key information but factually sound; \textbf{INCORRECT:} containing factual errors or contradictions; and \textbf{FAILURE:} off-topic, irrelevant, or empty.

\textbf{Fine-grained Likert Scores} assess each answer on five dimensions—information coverage, reasoning quality, factual consistency, clarity, and analytical depth, with each rated on a 1–5 Likert scale.

\textbf{Error Type Attribution} captures the failure cause for all non-CORRECT responses, enabling diagnostic analysis. Definitions of error categories are provided in Appendix~\ref{error_cases}.

\subsubsection{Retrieval Component Evaluation}
We evaluate the retriever's ability to surface relevant gold evidence with three metrics:

\textbf{Recall@K} measures the proportion of queries where any gold evidence appears in the top-$K$ results:
\[
\text{Recall@K} = \frac{1}{N} \sum_{i=1}^{N} \mathbb{1}(\text{rank}_i^{\text{gold}} \leq K),
\]
where $N$ is the number of queries.

\textbf{Precision@K} computes the proportion of gold evidences among top-$K$ candidates:
\[
\text{Precision@K} = \frac{1}{K \cdot N} \sum_{i=1}^{N} \sum_{j=1}^{K} \mathbb{1}(c_{ij} \in \mathcal{G}_i),
\]
with $c_{ij}$ denoting the $j$-th candidate and $\mathcal{G}_i$ the gold set for query $i$.

\textbf{MRR} (Mean Reciprocal Rank) evaluates how early the first correct evidence appears:
\[
\text{MRR} = \frac{1}{N} \sum_{i=1}^{N} \frac{1}{\text{rank}_i^{\text{first}}},
\]
where $\text{rank}_i^{\text{first}}$ is the position of the first correct retrieval.

\subsection{Main Results}
\subsubsection{Task-Level Challenge Landscape}
Table~\ref{tab:GT} and Figure~\ref{fig:error_st} demonstrate distinct challenges for EC-QA and LT-QA.

\paragraph{Human expert baseline.}
To calibrate task difficulty, we establish a human expert baseline on the full Fin-RATE dataset. As shown in Table~\ref{tab:GT}, human experts achieve 92.40\%, 82.68\%, and 85.88\% accuracy on DR-QA, EC-QA, and LT-QA, respectively. These results indicate that Fin-RATE is solvable by domain experts rather than being dominated by ambiguous or unanswerable questions. However, even the best-performing LLMs still lag behind human experts by 34.92, 39.04, and 42.36 percentage points on DR-QA, EC-QA, and LT-QA, respectively. The expert--model gap, especially on EC-QA and LT-QA, indicates that 
current LLMs still struggle with evidence-grounded cross-entity comparison and longitudinal 
tracking, even when all relevant context is provided.

\paragraph{EC-QA: Challenges in multi-entity integration and comparison.}
Most generative error types increase significantly after task transition. Comparative stance hallucination
(B1-3) and entity misidentification (D1-2) increase by 4,834 and 3,964 cases, and entity-attribute hallucination (B1-2) reaches 22,885 cases.
These trends indicate that models struggle to maintain consistent comparative logic, often extending partial evidence into unsupported conclusions or confusing the entities being contrasted.
These errors lead to inconsistent comparative reasoning and fabricated judgments, reflected in declines in clarity (-1.26) and factual
consistency (-1.07).
\paragraph{LT-QA: Challenges in temporal alignment and trend reasoning.}

\begin{table*}[htbp]
\centering
\small
\setlength{\tabcolsep}{2pt}
\caption{
Retrieval performance across DR-QA, EC-QA, and LT-QA. \textbf{Bold} indicates the highest score across all models; \underline{Underlined} indicates the second-highest overall.
Abbreviation key: R@K = Recall; P@K = Precision@K; MRR = Mean Reciprocal Rank; BM25+R = BM25 retrieval with reranking;  
VM+R = vector retrieval using MiniLM-L6-v2 embeddings with reranking;   
VF+R = vector retrieval using finance-embeddings-investopedia embeddings with reranking;   
VF = vector retrieval using finance-embeddings-investopedia embeddings;  
Hybrid+R = hybrid Retrieval. 
}
\label{tab:R}
\begin{tabular}{@{}l|ccc|ccc|c|ccc|ccc|c|ccc|ccc|c@{}}
\toprule
\multirow{2}{*}{\textbf{Retriever}}
& \multicolumn{7}{c|}{\textbf{DR-QA}}
& \multicolumn{7}{c|}{\textbf{EC-QA}}
& \multicolumn{7}{c}{\textbf{LT-QA}} \\
\cmidrule(lr){2-8} \cmidrule(lr){9-15} \cmidrule(lr){16-22}
& R@1 & R@5 & R@10 & P@1 & P@5 & P@10 & MRR
& R@1 & R@5 & R@10 & P@1 & P@5 & P@10 & MRR
& R@1 & R@5 & R@10 & P@1 & P@5 & P@10 & MRR \\\midrule
BM25                                            & 17.08                      & 33.36                      & \underline{41.16}                 & 17.08                      & 6.67                       & \underline{4.12}                  & 24.06                      & 0.73                       & 2.66                       & 4.46                        & 3.08                       & 2.30                        & 1.92                        & 6.42                       & \textbf{11.05}             & \underline{33.71}                & \underline{43.86}                 & \textbf{29.76}             & \textbf{18.55}             & \textbf{12.12}              & \textbf{40.43}             \\
BM25 + R                                        & \underline{24.44}                & \textbf{39.84}             & \textbf{44.84}              & \underline{24.44}                & \textbf{7.97}              & \textbf{4.48}               & \textbf{30.82}             & 1.63                       & 4.35                       & 5.94                        & 7.12                       & 3.84                       & 2.61                        & 11.31                      & \underline{10.85}                & \textbf{34.87}             & \textbf{44.25}              & \underline{26.92}                & \underline{17.76}                & \underline{11.81}                 & \underline{38.49}                \\
Hybrid                                          & 5.12                       & 12.48                      & 17.24                       & 5.12                       & 2.50                        & 1.72                        & 8.30                        & 0.58                       & 2.28                       & 3.83                        & 2.68                       & 2.00                          & 1.73                        & 5.56                       & 5.94                       & 23.08                      & 32.53                       & 15.96                      & 12.38                      & 8.84                        & 26.04                      \\
Hybrid + R                                      & 13.76                      & 20.32                      & 22.04                       & 13.76                      & 4.06                       & 2.20                         & 16.45                      & 1.61                       & 3.86                       & 5.42                        & 7.40                        & 3.59                       & 2.45                        & 11.18                      & 10.11                      & 30.73                      & 36.74                       & 25.28                      & 15.72                      & 9.81                        & 34.79                      \\
VF                                              & 7.48                       & 20.16                      & 27.44                       & 7.48                       & 4.03                       & 2.74                        & 12.8                       & 1.40                        & 4.57                       & \underline{7.81}                  & 5.08                       & 3.94                       & \underline{3.43}                  & 11.16                      & 3.29                       & 14.00                         & 20.86                       & 8.08                       & 6.73                       & 5.04                        & 14.33                      \\
VF + R                                          & 22.12                      & 32.76                      & 35.72                       & 22.12                      & 6.55                       & 3.57                        & 26.6                       & \textbf{2.4}               & \textbf{6.73}              & \textbf{9.73}               & \textbf{9.88}              & \textbf{5.81}              & \textbf{4.27}               & \textbf{17.01}             & 7.65                       & 22.85                      & 27.39                       & 18.24                      & 10.85                      & 6.62                        & 25.17                      \\
VM                                              & 11.00                         & 23.96                      & 30.52                       & 11.00                         & 4.79                       & 3.05                        & 16.33                      & 0.71                       & 2.68                       & 4.67                        & 3.56                       & 2.67                       & 2.34                        & 7.86                       & 3.10                        & 13.92                      & 20.47                       & 7.36                       & 6.70                        & 5.02                        & 13.66                      \\
VM + R                                          & \textbf{24.48}             & \underline{34.80}                 & 37.04                       & \textbf{24.48}             & \underline{6.96}                 & 3.70                         & \underline{28.84}                & \underline{1.87}                 & \underline{4.77}                 & 6.22                        & \underline{9.40}                  & \underline{4.82}                 & 3.18                        & \underline{14.63}                & 7.82                       & 21.73                      & 25.34                       & 18.84                      & 10.39                      & 6.20                         & 25.03                      \\ \bottomrule
\end{tabular}
\end{table*}
LT-QA's time mismatch (C3, 5.03\%) and trend hallucination (B1-4, 8.43\%) are
the highest among all tasks. Intent misunderstanding
(D1-1) further accounts for 35.38\% of LT-QA errors, showing that models
often treat trend-oriented questions as requests for yearly summaries. Besides, information coverage and reasoning scores
decline by 0.88 and 1.03, respectively, compared with DR-QA. 
Models lack mechanisms for temporal abstraction and cross-year
integration. They processes yearly information as independent units rather
than linking them into a continuous temporal sequence. Without a
representation that captures how indicators evolve over time, the model
cannot infer stable trends or causal relations and instead relies on
isolated factual recall, which fragments its temporal reasoning.
As a result, generated answers tend to include fabricated temporal claims, exhibit reduced clarity (-0.99), and become verbose without providing actionable trend insights.
\begin{figure}
    \centering
    \includegraphics[width=\linewidth]{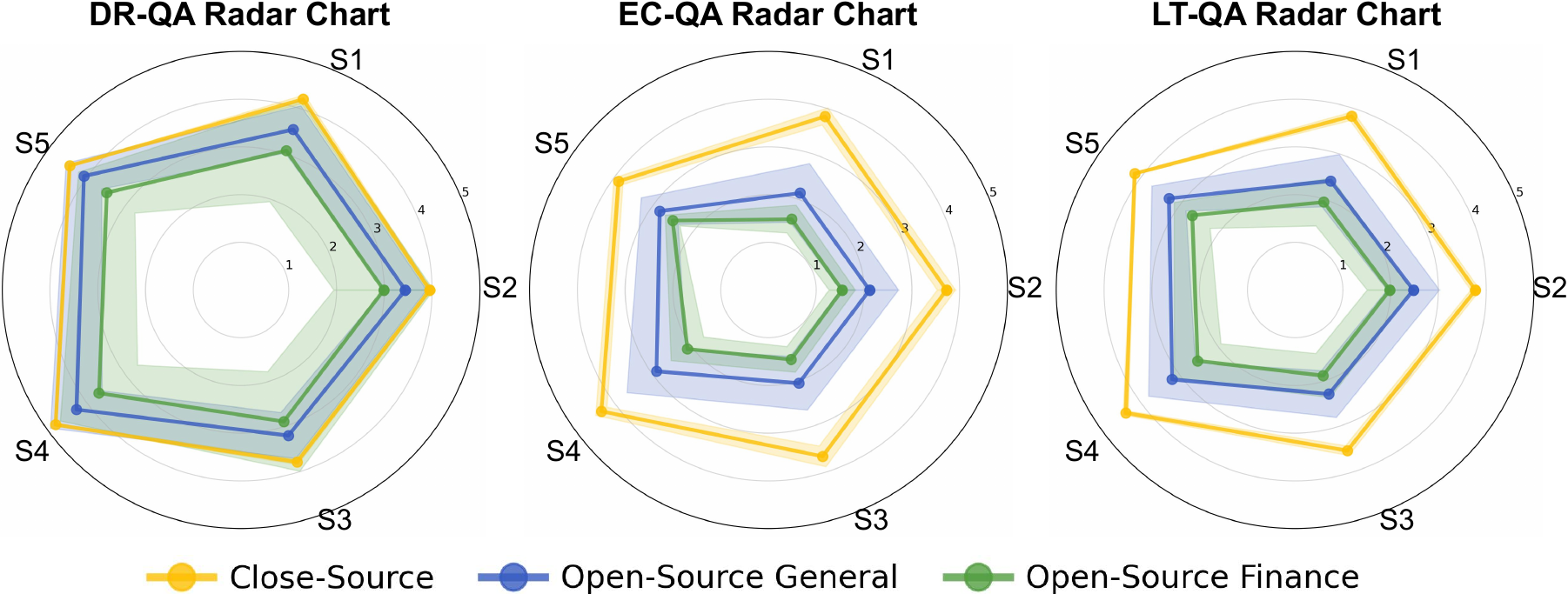}
    \caption{Radar charts showing model categories' performance across DR-QA, EC-QA, and LT-QA, with solid lines for mean and shaded areas for standard deviation. S1 = Reasoning Chain, S2 = Information Coverage, S3 = Analytical Depth, S4 = Clarity of Expression, S5 = Factual Consistency.} 
    \label{fig:radar}
\end{figure}

\subsubsection{Model category-Specific Behavior}
All three model categories degrade on cross-document tasks with distinct severities, error patterns, and failure mechanisms, as shown in Figure~\ref{fig:radar} and Table~\ref{tab:GT}.

\paragraph{Closed-source models: Stable performance with reasoning inconsistencies.}
Closed-source models exhibit the highest robustness across all tasks.
Their Correct rate decreases only moderately from 
39.69\% in DR-QA to 33.28\% in EC-QA, with Failure consistently below 1\%. Radar plots show a nearly intact circular profile, with cross-dimensional 
degradation limited to 3.28\%–12.49\%. Such robustness likely stems from large parameter counts and extended training chains, along with engineering advances such as position encoding that support superior long-context parsing and minimize information loss. Despite low hallucination rates, closed-source models frequently misattribute reasoning steps. Hallucination (B1) errors decrease 
substantially from DR-QA to EC-QA, while Contradicts Evidence (B2) rises to 1,980 cases, revealing a core bottleneck in maintaining self-consistent reasoning, despite correct retrieval.

\paragraph{General open-source models: Uniform degradation under cross-document reasoning.} Although achieving 27.42\% Correct in DR-QA, their Failure rate spikes from 1.63\% to 24.49\% in EC-QA.
EC-QA triggers simultaneous increases in multiple error types, with surges in Query Misunderstanding (D1), Hallucination (B1), and Contradicts Evidence (B2), along with a $6.8\times$ rise in Context Window Abuse (D2).
This pattern arises because broad-domain pretraining and smaller capacity lead these models to rely on shallow lexical cues and short-range attention, preventing stable tracking of entities, temporal references, or section relations.
With errors spreading across dimensions, the radar plots show uniform collapses across all evaluation axes as semantic drift and information loss cause early misalignments to compound along the reasoning chain.

\paragraph{Finance-tuned open-source models: Strong domain grounding but complete collapse on cross-entity tasks.}

\begin{table*}
\centering
\small
\setlength{\tabcolsep}{6pt}
\caption{
Each model with the top-5 retrieved chunks is evaluated on three task types (DR-QA, EC-QA, LT-QA). Acc., CI-L, and CI-U are reported in percentages, with 95\% bootstrap confidence intervals based on 1,000 resamples.
}
\label{tab:rag}
\begin{tabular}{cl|ccc|ccc|ccc}
\toprule
\multirow{2}{*}{Model}       & \multirow{2}{*}{Retrieve} & \multicolumn{3}{c|}{DR-QA}                                            & \multicolumn{3}{c|}{EC-QA}   & \multicolumn{3}{c}{LT-QA}   \\\cmidrule(lr){3-11}
                             & \multicolumn{1}{c|}{}                          & Acc.          & CI-L                     & CI-U                      & Acc.          & CI-L & CI-U & Acc.          & CI-L & CI-U \\\midrule
\multirow{5}{*}{GPT-5}       & BM25+R                                        & 21.60          & 18.00                       & 25.40                      & 6.00             & 4.00    & 8.00    & 12.40          & 9.40  & 15.40 \\
                             & Hybrid+R                                      & 8.20           & 5.80                      & 10.80                      & \underline{9.40}     & 7.00    & 12.00   & \textbf{18.80} & 15.40 & 22.40 \\
                             & VF                                            & \underline{25.00}      & 21.20                     & 28.80                      & 9.00             & 6.60  & 11.60 & 11.60          & 8.80  & 14.40 \\
                             & VF+R                                          & \textbf{26.20} & 22.40                     & 30.00                        & \textbf{10.80} & 8.20  & 13.60 & \underline{14.00}      & 11.00   & 17.20 \\
                             & VM+R                                          & 22.20          & 18.60                     & 25.80                      & 8.40           & 6.00    & 10.8 & 11.40          & 8.60  & 14.20 \\\midrule
\multirow{5}{*}{DeepSeek-V3} & BM25+R                                        & 3.20           & 1.80                      & 4.80                       & 0.00             & 0.00    & 0.00    & 0.40           & 0.00    & 1.00    \\
                             & Hybrid+R                                      & 1.40           & 0.40                      & 2.40                       & 0.00             & 0.00    & 0.00    & 0.00             & 0.00    & 0.00    \\
                             & VF                                            & 7.60           & 5.40                      & 10.00                        & 1.00             & 0.20  & 2.00    & 1.20           & 0.40  & 2.20  \\
                             & VF+R                                          & 8.20           & 5.80                      & 10.60                      & 0.80           & 0.20  & 1.60  & 0.80           & 0.20  & 1.60  \\
                             & VM+R                                          & 8.20           & 5.80                      & 10.60                      & 0.60           & 0.00    & 1.40  & 1.60           & 0.60  & 2.80  \\\midrule
\multirow{5}{*}{Qwen3-235B}  & BM25+R                                        & 14.40          & 11.40                     & 17.60                      & 0.40           & 0.00    & 1.00    & 3.20           & 1.80  & 4.80  \\
                             & Hybrid+R                                      & 11.00            & 8.40                      & 13.80                      & 1.60           & 0.60  & 2.80  & 5.20           & 3.40  & 7.20  \\
                             & VF                                            & 18.00            & 14.80                     & 21.40                      & 2.80           & 1.40  & 4.20  & 5.00             & 3.20  & 7.00    \\
                             & VF+R                                          & 23.80          & 20.20                     & 27.40                      & 3.60           & 2.00    & 5.40  & 5.60           & 3.60  & 7.80  \\
                             & VM+R                                          & 21.60          & 18.00                       & 25.40                      & 3.40           & 2.00    & 5.2  & 4.40           & 2.60 & 6.20  \\\midrule
\multirow{5}{*}{Fin-R1}      & BM25+R                                        & 4.60           & 2.80 & 6.60  & 1.20           & 0.40  & 2.20  & 3.00             & 1.60  & 4.60  \\
                             & Hybrid+R                                      & 2.20           & 1.00 & 3.60  & 0.80           & 0.20  & 1.60  & 2.40           & 1.20  & 3.80  \\
                             & VF                                            & 5.40           & 3.40 & 7.60  & 1.20           & 0.40  & 2.20  & 1.80           & 0.80  & 3.00    \\
                             & VF+R                                          & 7.40           & 5.20 & 9.80  & 1.00             & 0.20  & 2.00    & 2.20           & 1.00    & 3.60  \\
                             & VM+R                                          & 10.80         & 8.20 & 13.60 & 0.80           & 0.20  & 1.60  & 1.20           & 0.40  & 2.20 \\\bottomrule
\end{tabular}
\end{table*}
Correct rates drop from 23.91\% in DR-QA to just 1.27\% in EC-QA, with both Failure and Incorrect sharply rising. Radar plots reveal a lopsided profile, where nearly 50\% of metrics degrade, and Reasoning Chain drops by 49.16\%. 
In DR-QA, Query Misunderstanding (D1) errors are only 804.5, comparable to top closed-source models. In EC-QA and LT-QA, they also produce fewer numerical errors (C1, C2, C4) than general open-source models. However, Hallucination (B1) rates remain consistently higher than in other model categories.
Current finance-tuning pipelines rely on single-document supervision focused on lexical cues, terminology, and numerical patterns. While this enhances domain grounding, it also reinforces template-based guessing: models infer missing relations or numbers from familiar financial patterns, causing frequent hallucinations dominated by Entity Attribute Hallucination (B1-2).

\subsubsection{Retrieval Behavior Across QA Tasks}
\label{sec:rag_analysis}

As shown in Table~\ref{tab:R}, retrieval performance varies by task, and the low overlap between sparse and dense retrievers leads hybrid retrieval to amplify noise.

\paragraph{Task-dependent retrieval preferences.}
Different QA scenarios show distinct retrieval preferences. DR-QA and LT-QA depend heavily on localized, explicitly anchored information, while EC-QA requires broader semantic correspondence across 
corporate narratives.
BM25, which relies on lexical matching, achieves 41.2\% recall 
in DR-QA and 94.3\% accuracy in locating year-aligned evidence in LT-QA, outperforming VF at 
27.4\% and 74.7\%. However, in EC-QA, VF retrieves multi-company 
evidence in 23.8\% of cases, more than doubling BM25’s 10.4\%.
These patterns reflect fundamental differences in how the retrievers operate.
BM25 prioritizes distinctive tokens such as years and financial terms, making it effective when explicit anchors exist. VF, in contrast, encodes semantic similarity and therefore better captures heterogeneous cross-company narratives, but loses precision on fine-grained temporal cues.

\paragraph{Failure mode of hybrid retrieval.} Hybrid retrieval does not achieve complementary gains, underperforming both BM25 and VF.
Hybrid degradation is caused by the extremely low overlap between BM25 and VF candidate sets. The Jaccard similarity of their top-ranked passages is only 0.0412, 0.0113, and 0.0603 in DR-QA, EC-QA, and LT-QA. With almost no shared candidates, linear fusion combines mismatched ranking signals, disrupting each retriever’s strengths and ultimately weakening retrieval quality.

\subsubsection{End-to-End RAG Failure Attribution}
\begin{figure}[htbp]
    \centering
    \includegraphics[width=0.9\linewidth]{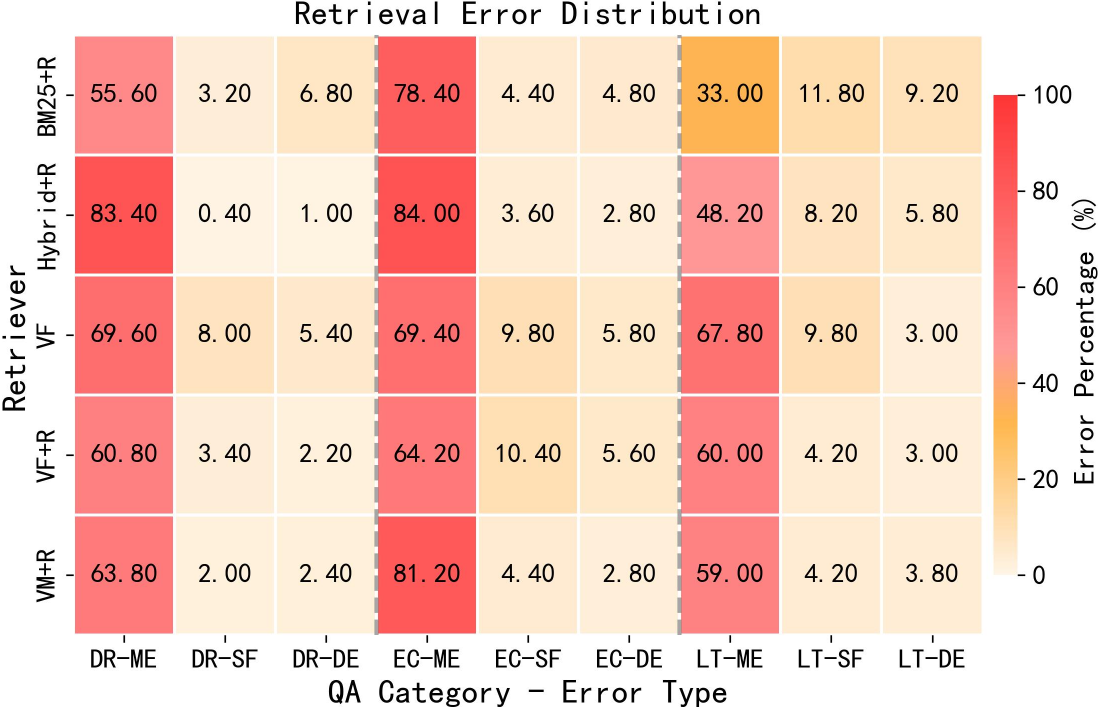}
    \caption{Retrieval error distribution. ME = Missing Evidence, SF = Sorting Failure, DE = Distractor Evidence.} 
    \label{fig:r-error}
\end{figure}
To evaluate end-to-end performance under retrieved context, we randomly select 500 examples for each QA type to evaluate model performance under retrieved context. For EC-QA and LT-QA, we restrict our evaluation to the subset of questions that contain at most five gold chunks, which serve as the ground-truth context.

Across all QA tasks, end-to-end degradation under RAG is driven primarily by retrieval rather than generation. As illustrated in Table~\ref{tab:rag}, end-to-end accuracy under retrieved context remains below 27\% across all tasks. In contrast, Table~\ref{tab:GT} shows that with gold context, the same models reach 57.48\% on DR-QA and 43–44\% on EC-QA and LT-QA, demonstrating that adequate reasoning is achievable once relevant evidence is available. Moreover, all retrievers perform weakly on EC-QA, with the best $R@10 \le 9.73\%$, leaving LLMs without the evidence needed to answer the questions. Thus, the dominant bottleneck in the RAG pipeline is the retriever’s failure to surface essential evidence, not deficiencies in generation.

To further analyze task-specific failure patterns under RAG, we categorize retrieval-stage errors into three types: Missing Evidence (no gold chunk retrieved), Sorting Failure 
(gold retrieved but ranked below top-3), and Distractor Evidence (gold appears within the top-3 but accounts for no more than 20\% of the top-5 context). 
EC-QA exhibits the most severe breakdowns. Figure~\ref{fig:r-error} shows that Missing Evidence reaches 75.44\%. Moreover, 57.6\% of multi-company questions fail to retrieve all relevant entities, leaving key company evidence absent. This forces the model to conduct comparative reasoning over severely asymmetric or biased contexts.
LT-QA shows a distinct error profile. The 7.64\% Sorting Failure and 4.96\% Distractor Evidence reflect frequent temporal misalignment, with correct-year gold chunks pushed down by similar paragraphs from other years or entangled with multi-year narratives in the top-5 results.

\subsection{Representative Failure Mode Analysis}
\label{sec:case_studies}

To illustrate the error patterns quantified above, we examine three representative failures from different models and task settings. Each case shows the model’s output, the ground‑truth answer, and the corresponding error types from our taxonomy.

\subsubsection{Case 1: Numeric hallucination with reversed comparison.}
\begin{figure}[htbp]
    \centering
    \includegraphics[width=\linewidth]{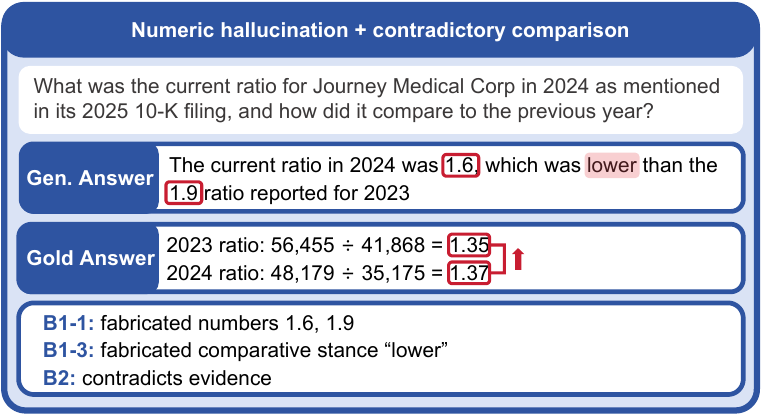}
\caption{Fabricated ratios reverse the true comparison.}
    \label{fig:case1}
\end{figure}

Figure~\ref{fig:case1} illustrates a ratio-reasoning failure where the model fails to extract current assets and current liabilities from the balance sheet to compute the current ratio. It fabricates unsupported numerical values, which in turn leads to the opposite year-over-year trend and wrong comparative conclusion.

\subsubsection{Case 2: Cross-entity evidence blending and attribution failure.}
\begin{figure}[htbp]
    \centering
    \includegraphics[width=\linewidth]{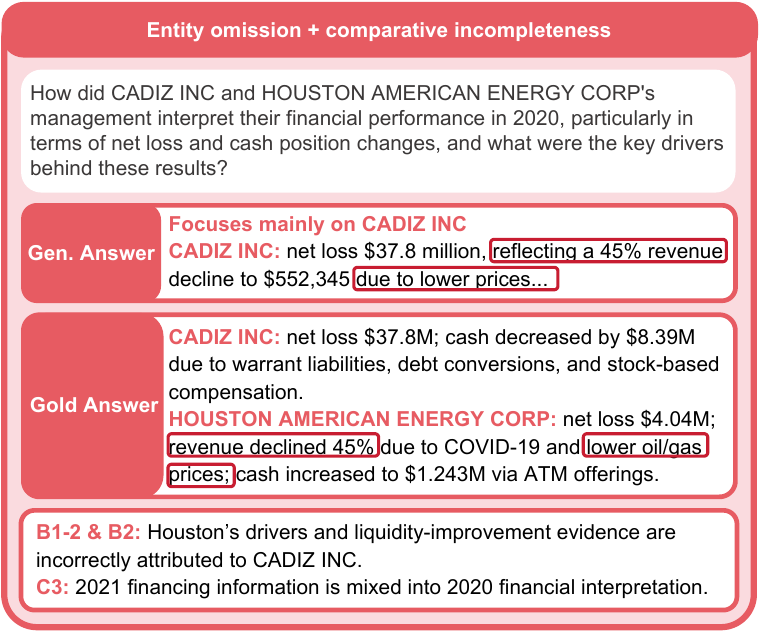}
    \caption{Cross-entity conflation and attribution failure.}
    \label{fig:case2}
\end{figure}

Figure~\ref{fig:case2} illustrates a cross-entity comparison failure where the model conflates the two companies. Houston's liquidity improvement and financing drivers are attributed to CADIZ, and 2021 data is mixed into the 2020 analysis. Consequently, CADIZ's actual cash deterioration is replaced by an incorrect narrative.

\subsubsection{Case 3: Temporal misassignment with comparative stance hallucination.}
\begin{figure}[htbp]
    \centering
    \includegraphics[width=\linewidth]{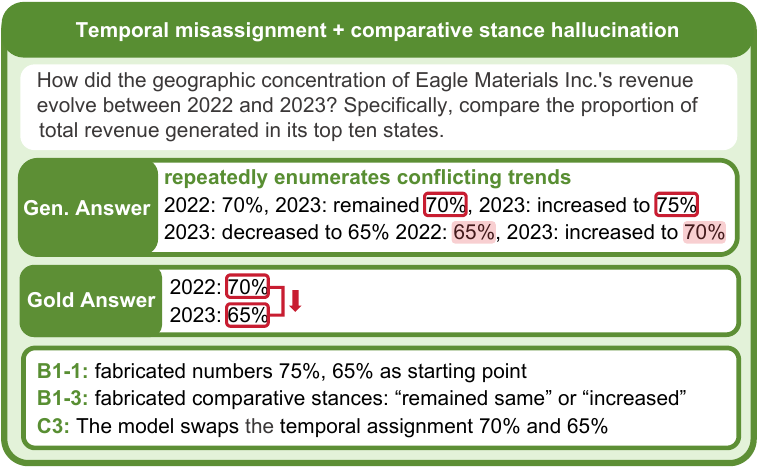}
    \caption{Temporal misassignment and comparative stance hallucination in geographic revenue tracking.}
    \label{fig:case3}
\end{figure}

Figure~\ref{fig:case3} illustrates a longitudinal tracking failure where the model enumerates multiple contradictory trends instead of extracting the correct year-over-year change. It correctly recalls the numerical values (70\% and 65\%) but swaps their temporal assignment, assigning 65\% to 2022 and 70\% to 2023.

\section{Conclusion}
In this work, we introduce Fin-RATE, a benchmark constructed from real-world SEC filings to evaluate LLMs’ financial reasoning across detailed interpretation, enterprise comparison, and longitudinal tracking. Fin-RATE highlights persistent challenges in aligning time-sensitive evidence, synthesizing multi-entity disclosures, and performing longitudinal financial reasoning across fiscal years. While LLMs perform well when reasoning over isolated passages, performance deteriorates sharply in tasks requiring interlinked evidence, such as cross-entity comparison or multi-year trend analysis. In EC-QA and LT-QA, we observe notable drops in clarity and factual consistency, driven by hallucinated comparisons, trend misalignment, and intent misunderstanding. Among different model families, finance-tuned models emerge as the most sensitive to task complexity. While they demonstrate strong command of domain-specific terminology and numerical accuracy, they are prone to hallucinations during complex multi-entity reasoning. This fragility underscores the importance of integrating long-form contexts and deeper logical dependencies into the fine-tuning process.

\begin{acks}
This research was supported by the Roberts Innovation Award 2025 and the Amazon Research Award 2024, and funded by Goldman Sachs AI Research.
\end{acks}

\bibliographystyle{ACM-Reference-Format}
\bibliography{ref}

\appendix
\section{Detailed Corpus and Task Statistics}
\label{sec:dataset_detail}
Our dataset comprises 43 publicly traded companies spanning seven major industry sectors, providing diverse representation across the U.S. economy. Table~\ref{tab:sector-summary} provides a high-level summary of sector distribution. The Energy sector as the most represented with 8 companies. The Basic Materials sector follows with 7 companies, while Technology and Services sectors each contribute 4-5 companies to the dataset. This distribution ensures adequate representation of both capital-intensive industries (Energy, Basic Materials) and knowledge-intensive sectors (Technology, Healthcare), enabling comprehensive evaluation of financial analysis capabilities across diverse business environments.

\section{Detailed Corpus and Task Statistics}
\label{sec:dataset_detail}
Our dataset comprises 43 publicly traded companies spanning seven major industry sectors, providing diverse representation across the U.S. economy. Table~\ref{tab:sector-summary} provides a high-level summary of sector distribution. The Energy sector is the most represented with 10 companies. The Basic Materials sector follows with 7 companies, while Technology and Services sectors each contribute 5-6 companies to the dataset. This distribution ensures adequate representation of both capital-intensive industries (Energy, Basic Materials) and knowledge-intensive sectors (Technology, Healthcare), enabling comprehensive evaluation of financial analysis capabilities across diverse business environments.

\begin{figure}[htbp]
    \centering
    \includegraphics[width=0.75\linewidth]{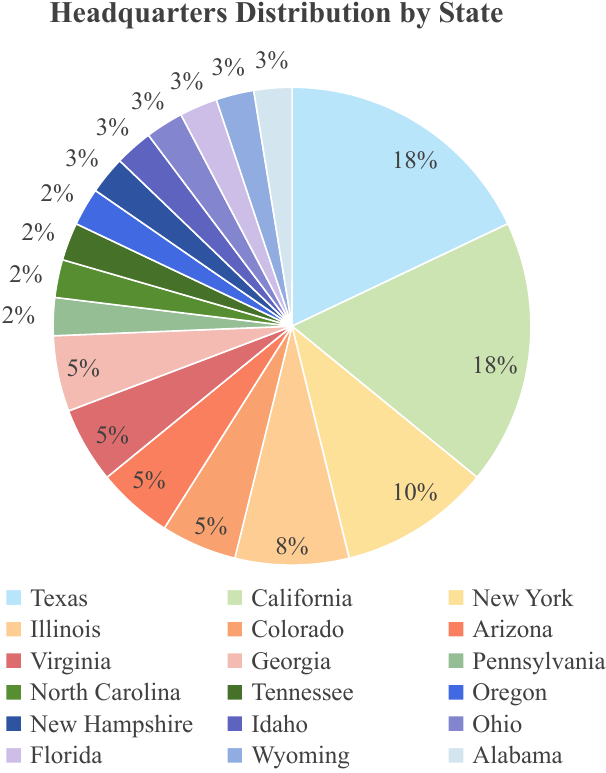}
    \caption{Pie chart illustrating the state-wise distribution of incorporation for 39 U.S.-incorporated companies.}
    \label{fig:company-location-distribution}
\end{figure}

\begin{figure}[htbp]
    \centering
    \includegraphics[width=0.7\linewidth]{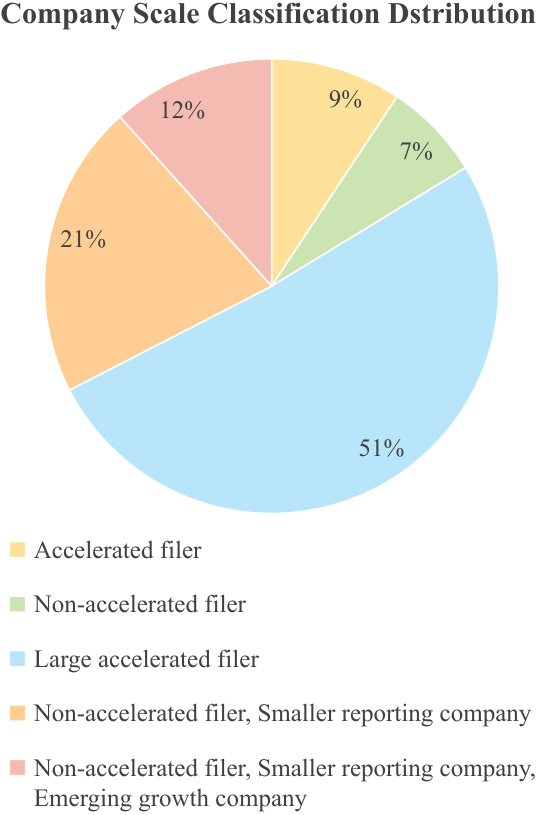}
    \caption{Pie chart illustrating the distribution of the 43 companies by their SEC filing status.}
    \label{fig:company-scale-distribution}
\end{figure}

The geographic location of a company's headquarters provides a crucial dimension for understanding its operational environment, regulatory framework, and potential regional economic linkages. To analyze the geographical distribution of companies in our dataset, we extract headquarters information from official SEC records. Among the 43 companies in Fin-RATE, 39 have identifiable U.S. state-level headquarters, while four are non-U.S. registrants and are excluded from the state-level distribution. Figure~\ref{fig:company-location-distribution} shows the distribution of these 39 companies across U.S. states. Figure~\ref{fig:company-scale-distribution} shows the distribution of the 43 companies by SEC filing status, providing an overview of the company-scale composition in our dataset.

\subsection{Document Length and Type Distribution}
Our dataset comprises SEC filings from 2020 to 2025, totaling 15,311 documents across 11 primary document types. Table~\ref{tab:sec-document-stats} presents the annual distribution, revealing significant variations in filing patterns over time. 
\begin{table*}[t]
\centering
\caption{LLM Focus Guidance for QA Generation by SEC Form Type}
\label{tab:intra_chunk_guidance_keywords}
\renewcommand{\arraystretch}{1.3}
\begin{tabular}{l|p{4.5cm}|p{10.5cm}}
\toprule
\textbf{Form Type} & \textbf{Focus Area} & \textbf{Guidance Keywords} \\
\midrule
\multirow{2}{*}{10-K} & Financial Ratio Analysis & financial ratios, performance metrics, year-over-year comparisons \\
\cline{2-3}
& Performance Trend Analysis & business performance trends, revenue growth, operational efficiency \\
\hline
\multirow{2}{*}{10-Q} & Quarterly Analysis & quarterly financial performance, seasonal trends, period-over-period changes \\
\cline{2-3}
& Cash Flow Analysis & cash flow patterns, working capital management, liquidity position \\
\hline
\multirow{2}{*}{8-K} & Materiality Impact Analysis & material events and their quantitative business impacts \\
\cline{2-3}
& Strategic Significance Analysis & strategic importance and long-term implications of disclosed events \\
\hline
\multirow{2}{*}{DEF 14A} & Executive Compensation Details & executive compensation details, shareholder proposals \\
\cline{2-3}
& Governance Structure & board structure, corporate governance changes \\
\bottomrule
\end{tabular}
\end{table*}
\begin{table*}[t]
\centering
\caption{Individual SEC Filing Items for Focused Analysis (10-K)}
\label{tab:individual_item_EC-QA}
\renewcommand{\arraystretch}{1.3}
\begin{tabular}{p{2.4cm}p{3.8cm}p{5.9cm}} 
\toprule
\textbf{Section} & \textbf{Full Title} & \textbf{Key Analysis Dimensions} \\
\midrule
\textbf{Item 1} & Business & Business operations; competition \& regulation \\
\hline
\textbf{Item 1A} & Risk Factors & Risk identification; industry threats. \\
\hline
\textbf{Item 7A} & Market Risk Disclosures & Risk measurement; sensitivity analysis. \\
\hline
\textbf{Item 7} & MD\&A & Financial interpretation; business trends. \\
\hline
\textbf{Item 8} & Financial Statements & Financial quantification; profitability. \\
\hline
\textbf{Item 3} & Legal Proceedings & Litigation exposure; compliance. \\
\hline
\textbf{Item 10} & Governance & Leadership; board composition. \\
\bottomrule
\end{tabular}
\end{table*}

\begin{table*}[t]
\centering
\caption{Strategic Item Combinations for Cross-Company Comparative Analysis}
\label{tab:multi_item_EC-QA}
\renewcommand{\arraystretch}{1.3}
\begin{tabular}{p{4cm}p{4cm}p{8cm}}
\toprule
\textbf{Combination Type} & \textbf{Items Combined} & \textbf{Comparative Analytical Dimension} \\
\midrule
\textbf{Business Model \& Financial Performance} & 
Item 8 + Item 1 & 
Linking operational business models to resulting financial outcomes across companies. \\
\hline
\textbf{Strategy \& Integrated Risk Profile} & 
Item 1 + Item 1A + Item 7A & 
Evaluating business strategy against a complete (qualitative + quantitative) risk landscape in comparative analysis. \\
\hline
\textbf{Comprehensive Risk \& Legal Assessment} & 
Item 1A + Item 7A + Item 3 & 
Holistic risk evaluation combining market, operational and legal exposures for cross-firm comparison. \\
\hline
\textbf{Governance \& Performance Linkage} & 
Item 8 + Item 10 & 
Examining the relationship between corporate governance structure and financial results across different organizations. \\
\bottomrule
\end{tabular}
\end{table*}
\begin{table*}[t]
\centering
\caption{Focus Areas for Tracking QA Analysis by SEC Filing Item}
\label{tab:tracking_qa_focus}
\renewcommand{\arraystretch}{1.3}
\begin{tabular}{p{1.5cm}p{4cm}p{10.5cm}}
\toprule
\textbf{Section} & \textbf{Primary Focus} & \textbf{Specific Tracking \& Comparative Dimensions} \\
\midrule
\textbf{DEF 14A} & Corporate Governance \& Executive Compensation & Annual changes in board composition, executive pay structure, shareholder proposals, audit details, related-party transactions, director independence, equity plans, risk oversight, and organizational structure. \\
\hline
\textbf{Item 1} & Business Overview \& Strategy & Year-over-year evolution across six core dimensions: Strategy \& Business Model, Financial Performance \& Revenue Structure, Product \& Technology Portfolio, Operations \& Human Capital, Market, Competition \& Customers, and Growth Drivers \& Investment Focus. \\
\hline
\textbf{Item 1A} & Risk Factors & Comparative analysis of risk profiles: identification of new, removed, or modified risks; changes in risk description, severity, or prioritization; adjustments to mitigation strategies; and emerging risk trends. \\
\hline
\textbf{Item 3} & Legal Proceedings & Tracking changes in legal exposure: status of ongoing cases, new filings, resolved matters, case nature, financial implications, and trends in litigation activity. \\
\bottomrule
\end{tabular}
\end{table*}
\begin{table}[h]
\centering
\caption{Sector Summary Statistics}
\label{tab:sector-summary}
\setlength{\tabcolsep}{1.5pt}
\small
\begin{tabular}{ccccc}
\toprule
GICS & Name & Num.& Industry Focus \\
\midrule
10 & Energy & 10 & Oil, Gas, Utilities, Mining \\
15 & Basic Materials & 7 & Building Materials, Apparel, Food \\
25 & Consumer Cyclicals & 4 & Apparel, Retail, Food Services \\
35 & Healthcare & 5 & Pharmaceuticals, Biotechnology \\
40 & Financials & 6 & Banking, Insurance, Securities \\
45 & Technology & 5 & Software, IT Services, Telecom \\
60 & Services & 6 & Construction, Real Estate, Hospitality \\
\bottomrule
\end{tabular}
\end{table}

\begin{table*}[htbp]
\centering
\caption{SEC Filer Status Definitions Based on Public Float and Annual Revenues}
\label{tab:filer-definitions}
\begin{tabularx}{0.95\textwidth}{p{0.25\textwidth}X}
\toprule
\textbf{Filer Status} & \textbf{Qualification Criteria} \\
\midrule
\textbf{Large Accelerated Filer} & An issuer with a public float of \$700 million or more as of the last business day of its most recently completed second fiscal quarter. \\
\midrule
\textbf{Accelerated Filer} & An issuer with a public float of \$75 million or more but less than \$700 million, and with annual revenues of \$100 million or more. \\
\midrule
\textbf{Non-Accelerated Filer} & An issuer that either: (1) has a public float of less than \$75 million; or (2) has a public float between \$75 million and \$700 million with annual revenues below \$100 million. \\
\midrule
\textbf{Smaller Reporting Company} & An issuer that qualifies if either: (1) its public float is less than \$250 million; or (2) its annual revenues are less than \$100 million. This status may overlap with other filer categories. \\
\midrule
\textbf{Emerging Growth Company} & A separate designation for issuers within five years of their initial public offering. Emerging Growth Companies may simultaneously qualify as Smaller Reporting Companies or Non-Accelerated Filers and are eligible for scaled disclosure requirements. \\
\bottomrule
\end{tabularx}
\end{table*}
\begin{table*}[h]
\centering
\caption{Annual Statistics of Chunks by SEC Document Type (2020–2025)}
\label{tab:sec-document-stats}
\begin{tabular}{lcccccccccccc}
\toprule
\multirow{2}{*}{Year} & \multicolumn{11}{c}{Document Type Counts} \\
\cmidrule(lr){2-12}
& 10-K & 10-Q & 424B2 & 424B3 & 424B4 & 6-K & 8-K & DEF-14A & NT & S-8 & SC 13D/13G \\
\midrule
2020 & 321 & 597 & 384 & 92 & 5 & 93 & 559 & 236 & 28 & 46 & 133 \\
2021 & 765 & 828 & 197 & 61 & 66 & 105 & 586 & 256 & 24 & 44 & 129 \\
2022 & 1049 & 546 & 241 & 97 & 55 & 113 & 594 & 289 & 13 & 52 & 249 \\
2023 & 865 & 502 & 183 & 111 & 0 & 113 & 594 & 149 & 29 & 26 & 151 \\
2024 & 642 & 332 & 171 & 83 & 0 & 110 & 489 & 70 & 14 & 39 & 186 \\
2025 & 753 & 99 & 141 & 42 & 0 & 68 & 345 & 107 & 16 & 28 & 0 \\
\midrule
\textbf{Total} & 4395 & 2904 & 1317 & 486 & 126 & 602 & 3167 & 1107 & 124 & 235 & 848 \\
\bottomrule
\end{tabular}
\end{table*}

The SEC document types in our dataset represent the diverse regulatory filings required for publicly traded companies. Each document type serves specific regulatory purposes and contains distinct types of financial information:

\begin{itemize}
\item \textbf{10-K (Annual Report):} Comprehensive annual filing that provides a complete overview of a company's financial performance, business operations, risk factors, and management discussion. It includes audited financial statements, detailed notes to financial statements, and disclosures about corporate governance, making it the most authoritative source for annual financial analysis.

\item \textbf{10-Q (Quarterly Report):} Interim financial report filed quarterly (excluding the fourth quarter, covered by 10-K) containing unaudited financial statements and management discussion. 10-Q filings provide timely updates on company performance between annual reports and include important disclosures about material changes in financial condition.

\item \textbf{424B2/B3/B4 (Prospectus Amendments):} Registration statement amendments filed in connection with securities offerings. These documents contain detailed information about security pricing, underwriting arrangements, and risk disclosures for debt or equity offerings, providing insights into corporate financing activities. \textit{Note: No 424B4 filings were available from 2023 to 2025 in our selected companies.}

\item \textbf{6-K (Foreign Issuer Report):} Periodic report filed by foreign private issuers to disclose material information required in their home countries. These filings often include earnings releases, interim financial information, and other corporate developments that may not be immediately available through other SEC filings.

\item \textbf{8-K (Current Report):} "Current events" report filed to disclose significant corporate events that shareholders should know about, including mergers and acquisitions, executive changes, bankruptcy filings, and material agreements. 8-Ks provide real-time insight into corporate developments between periodic reports.

\item \textbf{DEF-14A (Definitive Proxy Statement):} Comprehensive disclosure document distributed to shareholders before annual meetings, containing detailed information about board of director elections, executive compensation, shareholder proposals, and corporate governance matters.

\item \textbf{NT (Notification of Late Filing):} Formal notification filed when a company cannot submit a required report by the SEC deadline. These filings often include explanations for the delay and estimated filing dates, providing signals about potential corporate distress or operational issues.

\item \textbf{S-8 (Employee Benefit Plan Registration):} Registration statement for securities to be offered to employees under employee benefit plans. While primarily procedural, these filings can provide information about stock-based compensation programs and employee ownership structures.

\item \textbf{SC 13D/13G (Beneficial Ownership Reports):} Mandatory disclosures filed when an individual or entity acquires beneficial ownership of more than 5\% of a class of securities, revealing significant ownership positions, activist investor involvement, and potential changes in corporate control. \textit{Note: No SC 13D/13G filings were available for 2025 in our selected companies.}
\end{itemize}

\subsection{Document Type Distribution by Question Category}

Table~\ref{tab:doc-type-distribution} presents the distribution of SEC document types referenced across all 7,500 questions in our dataset. Annual reports (10-K) constitute the dominant source, referenced in 72.92\% of questions, reflecting their comprehensive nature for financial analysis. Proxy statements (DEF-14A) and quarterly reports (10-Q) are the next most frequently referenced at 8.65\% and 7.79\% respectively, while current reports (8-K) account for 6.17\% of references.

\begin{table}[h]
\centering
\caption{Document Type Distribution Across Questions}
\label{tab:doc-type-distribution}
\begin{tabular}{lcc}
\toprule
\textbf{Document Type} & \textbf{Questions} & \textbf{Percentage} \\
\midrule
10-K & 5,469 & 72.92\% \\
10-Q & 584 & 7.79\% \\
424B2 & 90 & 1.20\% \\
424B3 & 57 & 0.76\% \\
424B4 & 36 & 0.48\% \\
6-K & 41 & 0.55\% \\
8-K & 463 & 6.17\% \\
DEF-14A & 649 & 8.65\% \\
NT & 4 & 0.05\% \\
S-8 & 13 & 0.17\% \\
SC 13D/13G & 94 & 1.25\% \\
\midrule
\textbf{Total} & 7,500 & 100.00\% \\
\bottomrule
\end{tabular}
\end{table}
\subsection{Question Type and Context Analysis}

Our evaluation framework categorizes questions into three types based on their information complexity. Table~\ref{tab:dataset-stats} presents the dataset composition and context requirements for each question type.
\begin{table}[h]
\centering
\caption{Dataset Statistics by Question Type}
\label{tab:dataset-stats}
\begin{tabular}{lccc}
\toprule
\textbf{QA Type} & \textbf{Avg. Chunks/Q} & \textbf{Avg. Words/Q} \\
\midrule
Detail \& Reasoning & 1.00 & 802.39 \\
Enterprise Comparison & 5.61 & 43,674.20 \\
Longitudinal Tracking & 3.05 & 35,697.75 \\
\bottomrule
\end{tabular}
\end{table}

\section{LLM Focus Guidance for Detail \& Reasoning QA}
\label{LLM_focus_guidance_DR}
We guide the generation direction by embedding Targeted Focus Guidance in the prompt as shown in Table \ref{tab:intra_chunk_guidance_keywords}, instructing the LLM to concentrate on annual financial ratios, quarterly trend analysis, or the quantitative impact of material events, among others.

\section{LLM Focus Guidance for Enterprise Comparison QA}
\label{LLM_focus_guidance_EC}
Table \ref{tab:individual_item_EC-QA} and Table \ref{tab:multi_item_EC-QA} detail the SEC filing items and their strategic combinations specifically designed for Enterprise Comparison QA (EC-QA) generation. In practice, Item 1A (Risk Factors) and Item 7A (Market Risk Disclosures) are intrinsically linked and are used together to form a complete qualitative and quantitative risk profile. Item 7 provides narrative, expert-driven analysis of financial results, whereas Item 8 contains standardized numerical data with minimal interpretation, focusing on precise financial figures.

\section{LLM Focus Guidance for Longitudinal Tracking QA}
\label{LLM_focus_guidance_LT}
Table \ref{tab:tracking_qa_focus} summarizes the primary analytical dimensions and tracking objectives associated with key SEC filing items, specifically tailored for generating time-series-based Tracking QA (T-QA) across multiple reporting periods.

\section{Human Verification Criteria}
\label{sec:human_verify}

All QA pairs that pass automated filtering are manually reviewed to ensure strict adherence to the original disclosures and to eliminate unsupported interpretation.

\subsection{Factual Alignment}

This criterion ensures that every answer is strictly grounded in the cited disclosure content and does not introduce information beyond what is explicitly stated.

\begin{itemize}
  \item All statements in the answer must be directly supported by the referenced disclosure content.
  \item No external knowledge, implicit assumptions, or unstated background information is allowed.
  \item If the source text reports numerical outcomes or descriptive facts without evaluative conclusions, the answer must not introduce additional judgments.
\end{itemize}

\subsection{Contribution of Referenced Chunks}

This criterion evaluates whether each referenced chunk plays a necessary and substantive role in the reasoning process required to answer the question.

\begin{itemize}
  \item For QAs involving multiple chunks or cross-document contexts, each referenced chunk must contribute essential information to the final answer.
  \item A chunk is considered contributive only if it provides information such as a financial metric, a temporal reference, or a company-specific fact explicitly used in the answer.
  \item Chunks that only provide general background information or repeat content already covered by other chunks are treated as non-contributive.
  \item If any referenced chunk does not support the reasoning process, the QA is revised or discarded.
\end{itemize}

\subsection{Numerical Verification}

This criterion ensures that all numerical values and calculations in the answer are accurate, and consistent with the cited disclosure.

\begin{itemize}
  \item All numerical values must be manually recalculated and verified against the referenced disclosure content.
  \item Arithmetic operations, percentage calculations, and the consistency of time periods are explicitly checked.
  \item Units and scales must exactly match those used in the original disclosure, such as millions versus billions.
  \item If a numerical result cannot be reproduced using the cited content, the QA is removed from the dataset.
\end{itemize}

\section{Consistency Verification Between LLM and Human Evaluation}
\label{consistency_verification}
To evaluate the reliability of the LLM-as-a-Judge framework and its alignment with human judgment, we conducted a consistency verification experiment. A total of 4,500 representative QA instances are  randomly sampled from the Fin-RATE dataset, covering three major task categories: DR-QA, EC-QA, and LT-QA, with 1,500 instances per category. Each question is paired with a gold-standard answer, key answering points, the corresponding ground truth context, and responses generated by multiple target LLMs.

We use three large language models with distinct architectures and training paradigms, GPT-5, DeepSeek-V3.2, and Qwen3-235B, as automated Judges. For each Judge model, human-alignment evaluation is conducted on the same set of QA instances. All judgments are based on the same questions and the same target-LLM responses to ensure controlled evaluation conditions. To assess inter-judge agreement and evaluate each model’s alignment with human ratings, we compute agreement metrics across model-human pairs. Furthermore, to derive a robust final judgment that integrates divergent evaluations, we perform a grid search (with a step size of 0.02) over possible weight combinations to optimize agreement with human annotations. The resulting optimal weights for combining the three Judges are as follows: for DR-QA, DeepSeek-V3.2 = 0.70, Qwen3-235B = 0.10, GPT-5 = 0.20; for EC-QA, DeepSeek-V3.2 = 0.140, Qwen3-235B = 0.120, GPT-5 = 0.740; and for LT-QA, DeepSeek-V3.2 = 0.140, Qwen3-235B = 0.120, GPT-5 = 0.740.

\begin{table}[htbp]
\centering
\caption{Agreement scores between LLM judges and human evaluations across QA task types. For each QA type (DR-QA, EC-QA, LT-QA), we report the agreement between human scores and those from each LLM Judge, as well as the agreement from the final fused judgment (Multi-Judge), where weights are tuned via grid search with step size 0.02. \textbf{Bold} denote the best performance.}
\label{tab:agreement_results}
\begin{tabular}{@{}l *{3}{c}@{}}
\toprule
\textbf{Judge Model} & \textbf{DR-QA} & \textbf{EC-QA} & \textbf{LT-QA} \\
\midrule
DeepSeek-V3.2  & 84.29 & 82.99 & 81.35 \\
Qwen3-235B     & 84.10 & 84.34 & 84.67 \\
GPT-5          & 79.57 & 90.22 & 90.07 \\
Multi-Judge (Weighted) & \textbf{88.82} & \textbf{90.41} & \textbf{90.46} \\
\bottomrule
\end{tabular}
\end{table}

\section{Fusion Strategy for Multi-Judge Evaluation}
\label{judge_fusion_strategy}

This section details the fusion strategies applied to the evaluation outputs of multiple LLM Judges, including correctness labels, Likert scores, and error type attributions. All fusion methods are based on the optimal task-specific weights derived from the human alignment experiment described in Appendix~\ref{consistency_verification}.

For each QA instance, we follow the same procedure as in the grid search setting of the alignment experiment. Discrete correctness labels are first mapped to ordinal numeric scores, where Failure is mapped to 0, Incorrect to 1, Partial to 3, and Correct to 5. Based on the task-specific weights optimized in Appendix~\ref{consistency_verification}, we then compute a weighted average across the three Judges’ scores to obtain a continuous fused score. This score is mapped back to a final discrete label using the following rule: a score of 0 yields Failure, a score of 1 yields Incorrect, scores of 2–3 yield Partial, and scores of 4–5 yield Correct.

For the five fine-grained dimensions, each LLM Judge provides a Likert score ranging from 1 to 5. The final score for each dimension is computed by applying the same task-specific weights used for correctness labels to the Judges’ Likert scores.

For responses that are not labeled as Correct, each LLM Judge provide one or more error type attributions. During alignment verification, we observed that when an answer was also judged as non-Correct by human annotators, the errors identified by the Judges were generally valid. Therefore, we adopt a union-based fusion strategy: any error type identified by at least one Judge is retained in the final error set for diagnostic analysis. If the fused correctness label is Correct, all associated error types are discarded.

\section{Error Type Taxonomy}
\label{error_cases}
To enable fine-grained failure analysis and facilitate structured heatmap visualizations, we define a compact yet expressive taxonomy of error types. Each INCORRECT judgment is annotated with primary error types. The taxonomy spans four major dimensions:

\subsection{Retrieval-Related Errors}
These errors arise from failures in the retrieval stage of retrieval-augmented generation, where the necessary evidence is missing, misleading, or improperly ranked before generation.

\begin{itemize}
  \item \textbf{Missing Evidence}: 
  The key passage, table, or numerical evidence required to answer the question is never retrieved into the candidate pool, and thus does not appear in the final context. The model must answer without access to the gold evidence.

  \item \textbf{Sorting Failure}: 
  The gold evidence is retrieved but assigned a much lower rank than many less relevant chunks. 
  It often appears near the tail of the top-$K$ list, making it hard for the model to attend to and prone to being partially or fully truncated.

  \item \textbf{Distractor Evidence}: 
  The gold evidence is included in the top-$K$ context, but the context is dominated by noisy, partially relevant, 
  or misleading chunks that are ranked ahead of it or occupy most of the context window. 
  The model focuses on these distractors and reaches an incorrect conclusion even though the gold chunk is technically available.
\end{itemize}
\subsection{Generation-Related Errors}
These errors stem from the generation behavior of the model itself, including factual fabrication or failure to synthesize information.

\begin{itemize}
  \item \textbf{Hallucination}: The generated answer introduces information that directly contradicts the gold answer or key points.
  \begin {itemize}
  \item \textbf{Numeric or Categorical Hallucination}: The generated answer introduces fabricated numbers, percentages, ratings, years, categories, or other hard values that contradict or are absent from evidence.
  \item \textbf{Entity Attribute Hallucination}: The generated answer introduces fabricated attributes, states, policies, or strategies of a single entity that contradict or are absent from evidence.
  \item \textbf{Comparative Stance Hallucination}: The generated answer introduces fabricated comparative statements (A is more/less than B) without sufficient evidence.
  \item \textbf{Trend or Trajectory Hallucination}: The generated answer introduces fabricated trends or trajectories over multiple periods without sufficient evidence
  \end{itemize}
  \item \textbf{Contradicts Evidence}: The model's answer contains internal logical inconsistencies, such as conflicting statements or reasoning that contradicts its own earlier claims.
  \item \textbf{Excessive Inference}: The model draws overgeneralized or speculative conclusions not warranted by facts.
  \item \textbf{Evidence Fusion Failure}: The model fails to correctly combine multiple evidence pieces, e.g., ignoring conflicting values or mishandling complementary data.
\end{itemize}

\subsection{Finance-Specific Numerical \& Semantic Errors}
These errors reflect domain-specific misunderstandings in financial computations, temporal references, or numerical precision.

\begin{itemize}
  \item \textbf{Numerical Precision}: Minor but relevant rounding, tolerance, or percentage-point errors.
  \item \textbf{Units and Scales}: Misuse of financial units, ratios vs absolute values, or incorrect currency references.
  \item \textbf{Time Mismatch}: The model references an incorrect time period, or misidentifies the relevant fiscal year/quarter.
  \item \textbf{Computation Logic}: The model uses correct base values but applies the wrong formula or arithmetic.
\end{itemize}
\subsection{Query and Contextual Understanding Errors}
These errors arise from failures in understanding the question or in handling the long/complex retrieved context.

\begin{itemize}
  \item \textbf{Query Misunderstanding}: The model misinterprets the user’s intent, misidentifies the target entity, or answers the wrong financial metric.
  \begin{itemize}
      \item \textbf{Intent Misunderstanding}: The generated answer misunderstands the true intent of the question.
      \item \textbf{Entity Misidentification}: The generated answer incorrectly identifies key entities (company names, person names, metric names, etc.).
      \item \textbf{Metric Misidentification}: The generated answer incorrectly identifies the asked metric or measurement.
  \end{itemize}
  \item \textbf{Context Window Abuse}: The model fails to extract or prioritize the most relevant parts of a long context, or loses critical infomation due to truncation.
\end{itemize}


\section{Supplementary Evaluation Results}
\subsection{Full-Label Evaluation under Retrieved Context}
\begin{table*}[htbp]
\centering
\small
\caption{
Each model with the top-5 retrieved chunks is evaluated on three task types (DR-QA, EC-QA, LT-QA). Abbreviations consistent with Table~\ref{tab:R}.
}

\begin{tabular}{@{}l|l|ccccc|ccccc|ccccc@{}}
\toprule
\textbf{Model} & \textbf{Retrieve} 
& \multicolumn{5}{c|}{\textbf{DR-QA}} 
& \multicolumn{5}{c|}{\textbf{EC-QA}} 
& \multicolumn{5}{c}{\textbf{LT-QA}} \\
\cmidrule(lr){3-7} \cmidrule(lr){8-12} \cmidrule(lr){13-17}
& & Corr & Part & InCorr & Fail & Acc. (\%) 
  & Corr & Part & InCorr & Fail & Acc. (\%) 
  & Corr & Part & InCorr & Fail & Acc. (\%) \\
\midrule
\multirow{5}{*}{GPT-5}     & BM25+R                                        &  108    &   152   &  113      &  127    &    21.60           & 30     &   92   &     368   &    10  &      6.00        &   62   &  115    &    302    &   21   &     {\ul 12.40}         \\
                             & Hybrid+R                                      &  41    &   52   &    51    &  356    &    8.20           &  47    &   174   &    272    &   7   &    {\ul 9.40}          & 94     &   187   &    211    &   8   &  \textbf{18.80}            \\
                             & VF                                            &   125   &   164   &     210   &  1    &      25.00         &   45   &   190   &   261     &    4  &      9.00        &  58    &   183   &    253    &  6    &   11.60           \\
                             & VF+R                                          &  131    &  171    &   198     &  0    &      \textbf{26.20}         &   54   &  198    &   247     &     1 &    \textbf{10.80}          &  70    &  179    &    247    &    4  &         14.00     \\
                             & VM+R                                          &   111   &   187   &    199    & 3     &     22.20          &   42   &   194   &    261    &    3  &    8.40          &   57   &  189    &   247     &   7   &   11.40           \\

\midrule
\multirow{5}{*}{DeepSeek-V3} & BM25+R                                        & 16   & 50   & 338    & 96   & 3.20           & 0    & 4    & 8      & 488  & 0.00            & 2    & 7    & 205    & 286  & 0.40          \\
                             & Hybrid+R                                      & 7    & 35   & 439    & 19   & 1.40           & 0    & 3    & 192    & 305  & 0.00            & 0    & 4    & 223    & 273  & 0.00            \\
                             & VF                                            & 38   & 128  & 301    & 33   & 7.60           & 5    & 70   & 205    & 220  & 1.00            & 6    & 83   & 151    & 260  & 1.20          \\
                             & VF+R                                          & 41   & 142  & 271    & 46   & 8.20           & 4    & 71   & 177    & 248  & 0.80          & 4    & 88   & 133    & 275  & 0.80          \\
                             & VM+R                                          & 41   & 148  & 286    & 25   & 8.20           & 3    & 73   & 184    & 240  & 0.60          & 8    & 95   & 136    & 261  & 1.60          \\

\midrule
\multirow{5}{*}{Qwen3-235B}  & BM25+R                                        & 72   & 72   & 102    & 254  & 14.40          & 2    & 23   & 425    & 50   & 0.40          & 16   & 53   & 93     & 338  & 3.20          \\
                             & Hybrid+R                                      & 55   & 84   & 265    & 96   & 11.00            & 8    & 126  & 348    & 18   & 1.60          & 26   & 189  & 275    & 10   & 5.20    \\
                             & VF                                            & 90   & 121  & 289    & 0    & 18.00            & 14   & 122  & 345    & 19   & 2.80          & 25   & 148  & 302    & 25   & 5.00            \\
                             & VF+R                                          & 119  & 136  & 245    & 0    & {\ul 23.80} & 18   & 137  & 342    & 3    & 3.60 & 28   & 167  & 288    & 17   & 5.60 \\
                             & VM+R                                          & 108  & 141  & 251    & 0    & 21.60    & 17   & 154  & 314    & 15   & 3.40    & 22   & 186  & 277    & 15   & 4.40          \\

\midrule
\multirow{5}{*}{Fin-R1}      & BM25+R                                        & 23   & 222  & 240    & 15   & 4.60           & 6    & 113  & 205    & 176  & 1.20          & 15   & 172  & 211    & 102  & 3.00            \\
                             & Hybrid+R                                      & 11   & 162  & 307    & 20   & 2.20           & 4    & 116  & 183    & 197  & 0.80          & 12   & 174  & 193    & 121  & 2.40          \\
                             & VF                                            & 27   & 150  & 316    & 7    & 5.40           & 6    & 87   & 297    & 110  & 1.20          & 9    & 159  & 248    & 84   & 1.80          \\
                             & VF+R                                          & 37   & 173  & 277    & 13   & 7.40           & 5    & 116  & 252    & 127  & 1.00            & 11   & 181  & 224    & 84   & 2.20          \\
                             & VM+R                                          & 54   & 173  & 262    & 11   & 10.80          & 4    & 125  & 261    & 110  & 0.80          & 6    & 166  & 252    & 76   & 1.20\\
\bottomrule
\end{tabular}
\end{table*}

\subsection{Likert-Scale Evaluation across Five Dimensions}
\label{sec:likert-scale}
\begin{table}[]
\small
\setlength{\tabcolsep}{2pt}
\caption{
Likert-Scale Evaluation under Gold context Settings across Five Dimensions on DR-QA. 
Abbreviation key: \textbf{InfoCov} = Information Coverage, \textbf{Reason} = Reasoning Chain, \textbf{Fact} = Factual Consistency, \textbf{Clarity} = Clarity of Expression, \textbf{Depth} = Analytical Depth. 
\textbf{Bold} indicates the highest score across all models; \underline{Underlined} indicates the second-highest overall. Colored values are the highest score within each category.
}
\begin{tabular}{@{}lccccc@{}}
\toprule
\multicolumn{1}{c}{Model} & \cellcolor[HTML]{FFFFFF}{\color[HTML]{0F1115} InfoCov} & Reason                               & FactCon                             & Clarity                              & Depth                             \\ \midrule
\multicolumn{6}{c}{\cellcolor[HTML]{FFF3CA}Closed-Source}                                                                                                                                                                                       \\\midrule
GPT-4.1                   & {\color[HTML]{CD9934} 3.99}                            & {\color[HTML]{CD9934} {\ul 4.29}}    & {\color[HTML]{CD9934} \textbf{4.45}} & {\color[HTML]{CD9934} \textbf{4.83}} & {\color[HTML]{CD9934} 3.85}          \\
GPT-4.1-websearch         & 3.97                                                   & 4.26                                 & {\ul 4.44}                           & {\ul 4.82}                           & 3.81                                 \\
GPT-5-websearch           & 3.90                                                   & 4.07                                 & 4.42                                 & 4.75                                 & 3.71                                 \\\midrule
\multicolumn{6}{c}{\cellcolor[HTML]{D9E1F4}Open-Source General}                                                                                                                                                                                \\\midrule
MIMO-V2-Flash             & 3.36                                                   & 3.53                                 & 4.18                                 & 4.51                                 & 3.14                                 \\
DeepSeek-V3               & 3.30                                                   & 3.44                                 & 4.37                                 & 4.59                                 & 3.04                                 \\
DeepSeek-V3.2             & 3.36                                                   & 3.58                                 & 4.37                                 & 4.62                                 & 3.10                                 \\
DeepSeek-R1               & 3.33                                                   & 3.49                                 & 4.32                                 & 4.53                                 & 3.04                                 \\
GPT-OSS-20B               & {\color[HTML]{3166ff} {\ul 4.03}}                      & 4.20                                 & 4.28                                 & {\color[HTML]{3166ff} {\ul 4.82}}    & {\color[HTML]{3166ff} {\ul 3.93}}    \\
Llama-3.3-70B-Instruct    & 3.82                                                   & 3.74                                 & 4.18                                 & 4.03                                 & 3.46                                 \\
Qwen3-8B                  & 2.43                                                   & 2.30                                 & 2.87                                 & 2.64                                 & 2.10                                 \\
Qwen3-14B                 & 3.13                                                   & 3.11                                 & 3.56                                 & 3.44                                 & 2.88                                 \\
Qwen3-30B                 & 3.63                                                   & 3.78                                 & 4.13                                 & 4.59                                 & 3.47                                 \\
Qwen3-235B                & 4.00                                                   & {\color[HTML]{3166ff} 4.21}          & {\color[HTML]{3166ff} 4.39}          & {\color[HTML]{3166ff} {\ul 4.82}}    & 3.92                                 \\\midrule
\multicolumn{6}{c}{\cellcolor[HTML]{E3F2D9}Open-Source Finance}                                                                                                                                                                                \\\midrule
Fin-R1                    & {\color[HTML]{036400} \textbf{4.35}}                   & {\color[HTML]{036400} \textbf{4.48}} & {\color[HTML]{036400} 4.33}          & {\color[HTML]{036400} 4.76}          & {\color[HTML]{036400} \textbf{4.34}} \\
Fino1-14B                 & 3.52                                                   & 3.73                                 & 3.91                                 & 4.09                                 & 3.47                                 \\
FinanceConnect-13B        & 2.46                                                   & 2.57                                 & 3.25                                 & 3.80                                 & 2.30                                 \\
TouchstoneGPT-7B-Instruct & 1.61                                                   & 1.52                                 & 2.40                                 & 2.05                                 & 1.47                                 \\ \bottomrule
\end{tabular}
\end{table}

\begin{table}[htbp]
\centering
\small
\setlength{\tabcolsep}{2pt}
\caption{
Likert-Scale Evaluation under Gold context Settings across Five Dimensions on EC-QA. 
Abbreviation key: \textbf{InfoCov} = Information Coverage, \textbf{Reason} = Reasoning Chain, \textbf{Fact} = Factual Consistency, \textbf{Clarity} = Clarity of Expression, \textbf{Depth} = Analytical Depth. 
\textbf{Bold} indicates the highest score across all models; \underline{Underlined} indicates the second-highest overall. Colored values are the highest score within each category.
}
\begin{tabular}{@{}lccccc@{}}
\toprule
\multicolumn{1}{c}{Model} & \cellcolor[HTML]{FFFFFF}{\color[HTML]{0F1115} InfoCov} & Reason                               & FactCon                             & Clarity                              & Depth                             \\ \midrule
\multicolumn{6}{c}{\cellcolor[HTML]{FFF3CA}Closed-Source}                                                                                                                                                                                       \\\midrule
GPT-4.1                   & {\ul 3.61}                                             & {\ul 3.70}                           & {\ul 3.80}                           & {\ul 4.23}                           & {\ul 3.51}                           \\
GPT-4.1-websearch         & 3.57                                                   & 3.68                                 & 3.79                                 & 4.22                                 & 3.49                                 \\
GPT-5-websearch           & {\color[HTML]{CD9934} \textbf{3.99}}                   & {\color[HTML]{CD9934} \textbf{4.09}} & {\color[HTML]{CD9934} \textbf{4.04}} & {\color[HTML]{CD9934} \textbf{4.53}} & {\color[HTML]{CD9934} \textbf{3.99}} \\\midrule
\multicolumn{6}{c}{\cellcolor[HTML]{D9E1F4}Open-Source General}                                                                                                                                                                                \\\midrule
MIMO-V2-Flash             & {\color[HTML]{3166ff} 2.91}                            & {\color[HTML]{3166ff} 3.05}          & {\color[HTML]{3166ff} 3.43}          & {\color[HTML]{3166ff} 3.88}          & {\color[HTML]{3166ff} 2.87}          \\
DeepSeek-V3               & 1.66                                                   & 1.62                                 & 2.19                                 & 2.10                                 & 1.60                                 \\
DeepSeek-V3.2             & 2.69                                                   & 2.77                                 & 3.49                                 & 3.66                                 & 2.61                                 \\
DeepSeek-R1               & 2.36                                                   & 2.42                                 & 2.98                                 & 3.12                                 & 2.27                                 \\
GPT-OSS-20B               & 1.40                                                   & 1.42                                 & 2.63                                 & 2.53                                 & 1.42                                 \\
Llama-3.3-70B-Instruct    & 2.16                                                   & 2.10                                 & 2.92                                 & 3.05                                 & 2.04                                 \\
Qwen3-8B                  & 1.20                                                   & 1.18                                 & 2.07                                 & 1.51                                 & 1.14                                 \\
Qwen3-14B                 & 1.43                                                   & 1.42                                 & 2.30                                 & 2.08                                 & 1.40                                 \\
Qwen3-30B                 & 2.67                                                   & 2.80                                 & 3.02                                 & 3.70                                 & 2.65                                 \\
Qwen3-235B                & 2.67                                                   & 2.59                                 & 3.10                                 & 3.33                                 & 2.53                                 \\\midrule
\multicolumn{6}{c}{\cellcolor[HTML]{E3F2D9}Open-Source Finance}                                                                                                                                                                                \\\midrule
Fin-R1                    & {\color[HTML]{036400} 1.85}                            & {\color[HTML]{036400} 1.95}          & {\color[HTML]{34696D} 2.51}          & {\color[HTML]{34696D} 2.64}          & {\color[HTML]{34696D} 1.87}          \\
Fino1-14B                 & 1.72                                                   & 1.71                                 & 2.73                                 & 2.34                                 & 1.68                                 \\
FinanceConnect-13B        & 1.13                                                   & 1.13                                 & 2.47                                 & 1.88                                 & 1.12                                 \\
TouchstoneGPT-7B-Instruct & 1.47                                                   & 1.46                                 & 2.20                                 & 1.54                                 & 1.45                                 \\ \bottomrule
\end{tabular}
\end{table}

\begin{table}[htbp]
\centering
\small
\setlength{\tabcolsep}{2pt}
\caption{
Likert-Scale Evaluation under Gold context Settings across Five Dimensions on LT-QA. 
Abbreviation key: \textbf{InfoCov} = Information Coverage, \textbf{Reason} = Reasoning Chain, \textbf{Fact} = Factual Consistency, \textbf{Clarity} = Clarity of Expression, \textbf{Depth} = Analytical Depth. 
\textbf{Bold} indicates the highest score across all models; \underline{Underlined} indicates the second-highest overall. Colored values are the highest score within each category.
}
\begin{tabular}{@{}lccccc@{}}
\toprule
\multicolumn{1}{c}{Model} & \cellcolor[HTML]{FFFFFF}{\color[HTML]{0F1115} InfoCov} & Reason                               & FactCon                             & Clarity                              & Depth                             \\ \midrule
\multicolumn{6}{c}{\cellcolor[HTML]{FFF3CA}Closed-Source}                                                                                                                                                                                       \\\midrule
GPT-4.1                   & {\ul 3.74}                                             & {\ul 3.80}                           & {\color[HTML]{CD9934} \textbf{4.18}} & {\ul 4.40}                           & {\ul 3.52}                           \\
GPT-4.1-websearch         & 3.68                                                   & 3.73                                 & 4.12                                 & 4.29                                 & 3.41                                 \\
GPT-5-websearch           & {\color[HTML]{CD9934} \textbf{3.91}}                   & {\color[HTML]{CD9934} \textbf{3.97}} & {\ul 4.15}                           & {\color[HTML]{CD9934} \textbf{4.45}} & {\color[HTML]{CD9934} \textbf{3.68}} \\\midrule
\multicolumn{6}{c}{\cellcolor[HTML]{D9E1F4}Open-Source General}                                                                                                                                                                                \\\midrule
MIMO-V2-Flash             & {\color[HTML]{3166ff} 3.11}                            & {\color[HTML]{3166ff} 3.24}          & 3.69                                 & {\color[HTML]{3166ff} 4.09}          & {\color[HTML]{3166ff} 2.98}          \\
DeepSeek-V3               & 2.22                                                   & 2.12                                 & 3.06                                 & 2.84                                 & 2.08                                 \\
DeepSeek-V3.2             & 2.95                                                   & 3.04                                 & {\color[HTML]{3166ff} 3.79}          & 3.79                                 & 2.74                                 \\
DeepSeek-R1               & 2.80                                                   & 2.83                                 & 3.62                                 & 3.53                                 & 2.58                                 \\
GPT-OSS-20B               & 1.49                                                   & 1.52                                 & 1.66                                 & 3.21                                 & 1.52                                 \\
Llama-3.3-70B-Instruct    & 2.67                                                   & 2.47                                 & 3.48                                 & 3.14                                 & 2.41                                 \\
Qwen3-8B                  & 1.74                                                   & 1.70                                 & 3.00                                 & 2.18                                 & 1.58                                 \\
Qwen3-14B                 & 1.85                                                   & 1.75                                 & 2.80                                 & 2.46                                 & 1.71                                 \\
Qwen3-30B                 & 2.90                                                   & 2.73                                 & 3.48                                 & 3.72                                 & 2.69                                 \\
Qwen3-235B                & 2.95                                                   & 2.72                                 & 3.41                                 & 3.47                                 & 2.67                                 \\\midrule
\multicolumn{6}{c}{\cellcolor[HTML]{E3F2D9}Open-Source Finance}                                                                                                                                                                                \\\midrule
Fin-R1                    & {\color[HTML]{036400} 2.49}                            & {\color[HTML]{036400} 2.67}          & 2.83                                 & {\color[HTML]{036400} 3.42}          & {\color[HTML]{036400} 2.53}          \\
Fino1-14B                 & 2.34                                                   & 2.17                                 & {\color[HTML]{036400} 3.23}          & 2.69                                 & 2.14                                 \\
FinanceConnect-13B        & 1.31                                                   & 1.28                                 & 2.65                                 & 2.15                                 & 1.27                                 \\
TouchstoneGPT-7B-Instruct & 1.05                                                   & 1.05                                 & 1.13                                 & 1.10                                 & 1.04                                 \\ \bottomrule
\end{tabular}
\end{table}

\begin{table}[]
\small
\setlength{\tabcolsep}{1.8pt}
\caption{
Likert-Scale Evaluation under RAG Settings across Five Dimensions on DR-QA. 
Abbreviation key: \textbf{InfoCov} = Information Coverage, \textbf{Reason} = Reasoning Chain, \textbf{Fact} = Factual Consistency, \textbf{Clarity} = Clarity of Expression, \textbf{Depth} = Analytical Depth. 
\textbf{Bold} indicates the highest score across all models; \underline{Underlined} indicates the second-highest overall. Colored values are the highest score within each category.
}
\begin{tabular}{@{}c|l|ccccc@{}}
\toprule
Model                         & Retrieve & \cellcolor[HTML]{FFFFFF}{\color[HTML]{0F1115} InfoCov} & Reason        & FactCons      & Clarity       & AnaDepth      \\ \midrule
                              & BM25+R   & 2.07                                                   & 2.09          & 2.16          & 2.53          & 2.06          \\
                              & Hybrid+R & {\ul 2.98}                                             & {\ul 3.08}    & {\ul 3.21}    & \textbf{3.97} & {\ul 3.03}    \\
                              & VF       & 2.93                                                   & 3.03          & 3.18          & 3.94          & 2.97          \\
                              & VF+R     & \textbf{3.02}                                          & \textbf{3.10} & \textbf{3.29} & 3.97          & \textbf{3.05} \\
\multirow{-5}{*}{GPT-4.1}     & VM+R     & 2.93                                                   & 3.02          & 3.14          & 3.91          & 2.98          \\\midrule
                              & BM25+R   & 1.01                                                   & 1.01          & 1.15          & 1.04          & 1.01          \\
                              & Hybrid+R & 1.09                                                   & 1.09          & 1.40          & 1.35          & 1.09          \\
                              & VF       & 1.47                                                   & 1.50          & 1.75          & 2.16          & 1.48          \\
                              & VF+R     & 1.48                                                   & 1.48          & 1.75          & 2.08          & 1.46          \\
\multirow{-5}{*}{Deepseek-V3} & VM+R     & 1.41                                                   & 1.46          & 1.80          & 2.13          & 1.44          \\\midrule
                              & BM25+R   & 1.16                                                   & 1.20          & 1.20          & 1.41          & 1.19          \\
                              & Hybrid+R & 2.05                                                   & 2.26          & 2.16          & 3.29          & 2.25          \\
                              & VF       & 2.03                                                   & 2.20          & 2.17          & 3.23          & 2.19          \\
                              & VF+R     & 2.21                                                   & 2.37          & 2.47          & 3.44          & 2.35          \\
\multirow{-5}{*}{Qwen3-235B}  & VM+R     & 2.09                                                   & 2.30          & 2.51          & 3.40          & 2.26          \\\midrule
                              & BM25+R   & 1.19                                                   & 1.17          & 2.49          & 1.81          & 1.16          \\
                              & Hybrid+R & 1.21                                                   & 1.17          & 2.54          & 1.82          & 1.16          \\
                              & VF       & 1.39                                                   & 1.40          & 2.16          & 2.17          & 1.38          \\
                              & VF+R     & 1.38                                                   & 1.38          & 2.57          & 2.24          & 1.36          \\
\multirow{-5}{*}{Fin-R1}      & VM+R     & 1.33                                                   & 1.35          & 2.32          & 2.13          & 1.33          \\ \midrule
\end{tabular}
\end{table}

\begin{table}[]
\small
\setlength{\tabcolsep}{1.8pt}
\caption{
Likert-Scale Evaluation under RAG Settings across Five Dimensions on EC-QA. 
Abbreviation key: \textbf{InfoCov} = Information Coverage, \textbf{Reason} = Reasoning Chain, \textbf{Fact} = Factual Consistency, \textbf{Clarity} = Clarity of Expression, \textbf{Depth} = Analytical Depth. 
\textbf{Bold} indicates the highest score across all models; \underline{Underlined} indicates the second-highest overall. Colored values are the highest score within each category.
}
\begin{tabular}{@{}c|l|ccccc@{}}
\toprule
Model                         & Retrieve & \cellcolor[HTML]{FFFFFF}{\color[HTML]{0F1115} InfoCov} & Reason              & FactCons      & Clarity       & AnaDepth      \\ \midrule
                              & BM25+R   & 2.07                                                   & 2.09                & 2.16          & 2.53          & 2.06          \\
                              & Hybrid+R & {\ul 2.98}                                             & {\ul 3.08}          & {\ul 3.21}    & \textbf{3.97} & {\ul 3.03}    \\
                              & VF       & \textbf{2.93}                                          & \textbf{3.03}       & 3.18          & 3.94          & \textbf{2.97} \\
                              & VF+R     & {\ul \textbf{3.02}}                                    & {\ul \textbf{3.10}} & \textbf{3.29} & 3.97          & \textbf{3.05} \\
\multirow{-5}{*}{GPT-4.1}     & VM+R     & 2.93                                                   & 3.02                & {\ul 3.14}    & 3.91          & 2.98          \\\midrule
                              & BM25+R   & 1.01                                                   & 1.01                & 1.15          & 1.04          & 1.01          \\
                              & Hybrid+R & 1.09                                                   & 1.09                & 1.40          & 1.35          & 1.09          \\
                              & VF       & 1.47                                                   & 1.50                & 1.75          & 2.16          & 1.48          \\
                              & VF+R     & 1.48                                                   & 1.48                & 1.75          & 2.08          & 1.46          \\
\multirow{-5}{*}{Deepseek-V3} & VM+R     & 1.41                                                   & 1.46                & 1.80          & 2.13          & 1.44          \\\midrule
                              & BM25+R   & 1.16                                                   & 1.20                & 1.20          & 1.41          & 1.19          \\
                              & Hybrid+R & 2.05                                                   & 2.26                & 2.16          & 3.29          & 2.25          \\
                              & VF       & 2.03                                                   & 2.20                & 2.17          & 3.23          & 2.19          \\
                              & VF+R     & 2.21                                                   & 2.37                & 2.47          & 3.44          & {\ul 2.35}    \\
\multirow{-5}{*}{Qwen3-235B}  & VM+R     & 2.09                                                   & 2.30                & 2.51          & \textbf{3.40} & 2.26          \\\midrule
                              & BM25+R   & 1.19                                                   & 1.17                & 2.49          & 1.81          & 1.16          \\
                              & Hybrid+R & 1.21                                                   & 1.17                & 2.54          & 1.82          & 1.16          \\
                              & VF       & 1.39                                                   & 1.40                & 2.16          & 2.17          & 1.38          \\
                              & VF+R     & 1.38                                                   & 1.38                & 2.57          & 2.24          & 1.36          \\
\multirow{-5}{*}{Fin-R1}      & VM+R     & 1.33                                                   & 1.35                & 2.32          & 2.13          & 1.33          \\ \midrule 
\end{tabular}
\end{table}

\begin{table}[]
\small
\setlength{\tabcolsep}{1.8pt}
\caption{
Likert-Scale Evaluation under RAG Settings across Five Dimensions on LT-QA. 
Abbreviation key: \textbf{InfoCov} = Information Coverage, \textbf{Reason} = Reasoning Chain, \textbf{Fact} = Factual Consistency, \textbf{Clarity} = Clarity of Expression, \textbf{Depth} = Analytical Depth. 
\textbf{Bold} indicates the highest score across all models; \underline{Underlined} indicates the second-highest overall. Colored values are the highest score within each category.
}
\begin{tabular}{@{}c|l|ccccc@{}}
\toprule
\textbf{Model}                & \textbf{Retrieve} & \cellcolor[HTML]{FFFFFF}{\color[HTML]{0F1115} InfoCov} & Reason        & FactCons      & Clarity       & AnaDepth      \\ \midrule
                              & BM25+R            & 2.39                                                   & 2.41          & 2.63          & 2.85          & 2.34          \\
                              & Hybrid+R          & \textbf{3.16}                                          & \textbf{3.17} & \textbf{3.57} & \textbf{3.98} & \textbf{3.07} \\
                              & VF                & {\ul 2.83}                                             & {\ul 2.96}    & {\ul 3.28}    & {\ul 3.82}    & {\ul 2.86}    \\
                              & VF+R              & 2.82                                                   & 2.96          & 3.26          & 3.82          & 2.83          \\
\multirow{-5}{*}{GPT-4.1}     & VM+R              & 2.72                                                   & 2.89          & 3.24          & 3.77          & 2.79          \\\midrule
                              & BM25+R            & 1.06                                                   & 1.05          & 1.12          & 1.13          & 1.05          \\
                              & Hybrid+R          & 1.09                                                   & 1.11          & 1.46          & 1.41          & 1.08          \\
                              & VF                & 1.40                                                   & 1.40          & 1.89          & 2.05          & 1.39          \\
                              & VF+R              & 1.41                                                   & 1.43          & 1.88          & 2.04          & 1.41          \\
\multirow{-5}{*}{Deepseek-V3} & VM+R              & 1.45                                                   & 1.46          & 1.93          & 2.06          & 1.43          \\\midrule
                              & BM25+R            & 1.51                                                   & 1.53          & 1.63          & 1.80          & 1.52          \\
                              & Hybrid+R          & 2.35                                                   & 2.44          & 2.73          & 3.40          & 2.40          \\
                              & VF                & 2.06                                                   & 2.19          & 2.45          & 3.22          & 2.13          \\
                              & VF+R              & 2.20                                                   & 2.34          & 2.74          & 3.41          & 2.27          \\
\multirow{-5}{*}{Qwen3-235B}  & VM+R              & 2.23                                                   & 2.45          & 2.78          & 3.48          & 2.33          \\\midrule
                              & BM25+R            & 1.41                                                   & 1.34          & 2.73          & 2.03          & 1.31          \\
                              & Hybrid+R          & 1.39                                                   & 1.30          & 2.66          & 1.98          & 1.29          \\
                              & VF                & 1.45                                                   & 1.44          & 2.41          & 2.24          & 1.41          \\
                              & VF+R              & 1.38                                                   & 1.36          & 2.74          & 2.19          & 1.33          \\
\multirow{-5}{*}{Fin-R1}      & VM+R              & 1.43                                                   & 1.42          & 2.51          & 2.16          & 1.41          \\ \midrule 
\end{tabular}
\end{table}

\subsection{Error Pattern Analysis across Model Families}
We further analyze how different model families exhibit distinct failure patterns by grouping models into three categories: closed-source, open-source finance-tuned, and open-source general models. 
For each category, we report the distribution of Level 1 and Level 2 error types based on our taxonomy.  
Level 1 captures high-level reasoning failures—including hallucination (B1), contradictions (B2), excessive inference (B3), evidence fusion failures (B4), numerical issues (C1–C2), temporal mismatches (C3), computation-logic errors (C4), and comprehension failures (D1–D2). 
Level 2 refines the two most prevalent categories, B1 and D1, into eight subtypes: numeric or categorical hallucination (B1-1), entity-attribute hallucination (B1-2), comparative-stance hallucination (B1-3), trend or trajectory hallucination (B1-4), intent misunderstanding (D1-1), entity misidentification (D1-2), metric misidentification (D1-3), and context-window abuse (D1-4). 
\begin{figure*}[t]
    \centering
    \includegraphics[width=\textwidth]{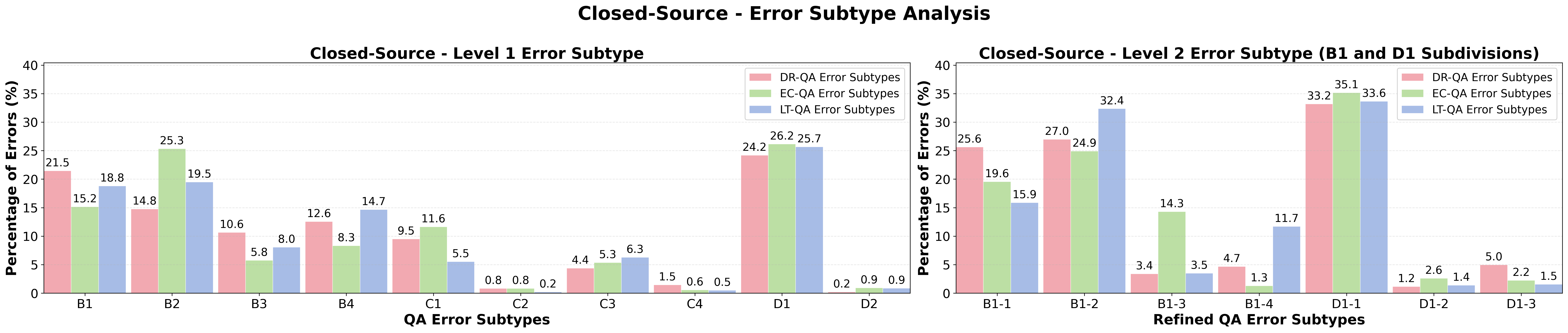}
    \caption{Level 1 and Level 2 error distributions for Closed-Source models}
    \label{fig:error_closedsource}
\end{figure*}

\begin{figure*}[t]
    \centering
    \includegraphics[width=\textwidth]{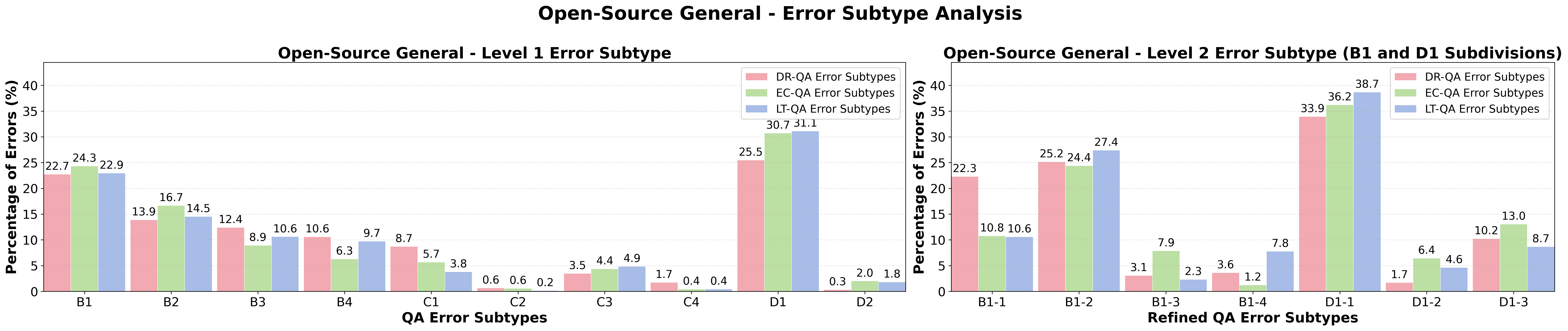}
    \caption{Level 1 and Level 2 error distributions for Open-Source General models}
    \label{fig:error_opensource}
\end{figure*}

\begin{figure*}[t]
    \centering
    \includegraphics[width=\textwidth]{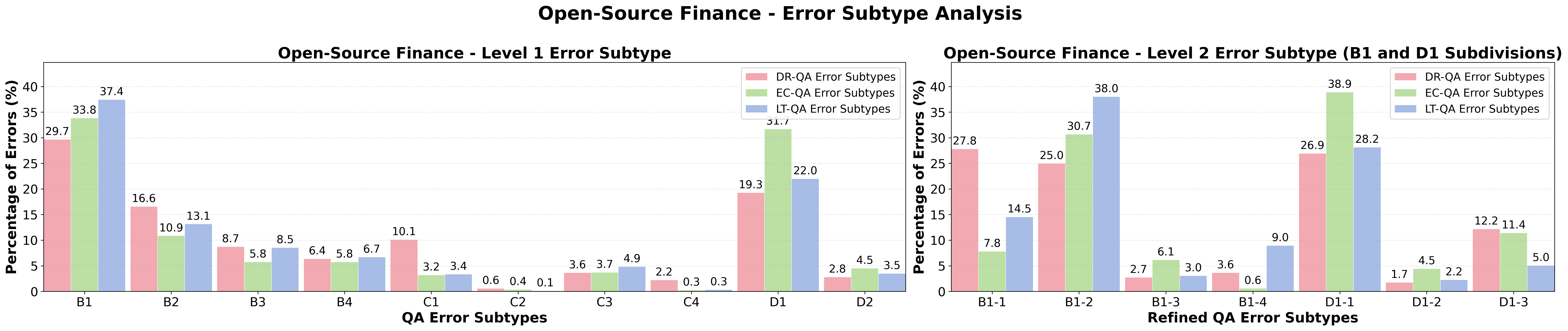}
    \caption{Level 1 and Level 2 error distributions for Open-Source Finance models}
    \label{fig:error_financetuned}
\end{figure*}

\section{Illustrative Example Questions}

\subsection{Detail \& Reasoning QA}
\begin{figure*}
    \centering
    \includegraphics[width=0.69\textwidth]{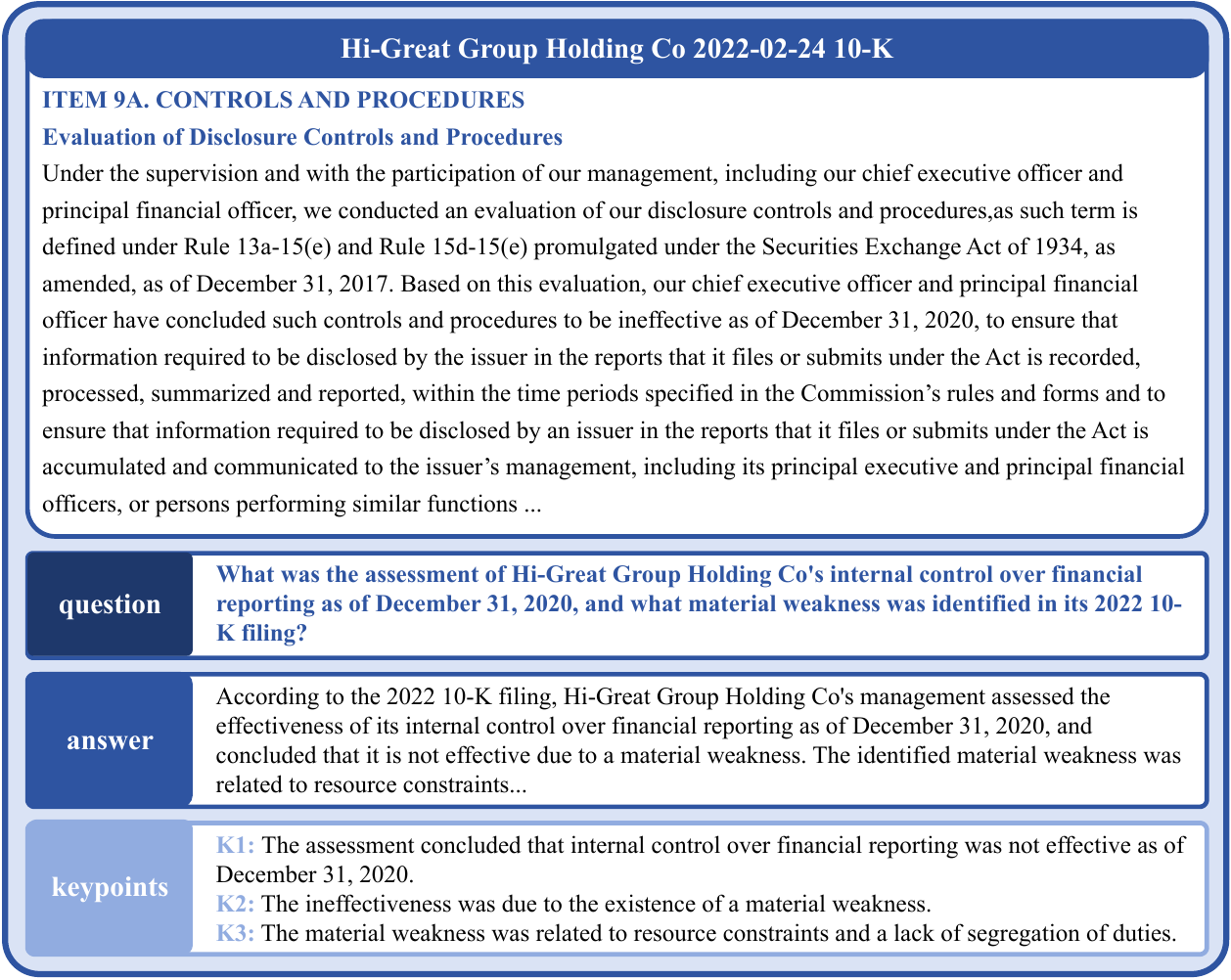}
    \caption{Example for Detail \& Reasoning QA (DR) question}
    \label{fig:DR-example}
\end{figure*}

Detail \& Reasoning questions focus on precise information extraction
and localized reasoning within a single document segment. As shown in
Figure~\ref{fig:DR-example}, the question requires identifying specific factual
statements and reasoning over closely related sentences, without relying on
cross-document or cross-temporal aggregation.

\subsection{Enterprise Comparison QA}
\begin{figure*}
    \centering
    \includegraphics[width=0.69\textwidth]{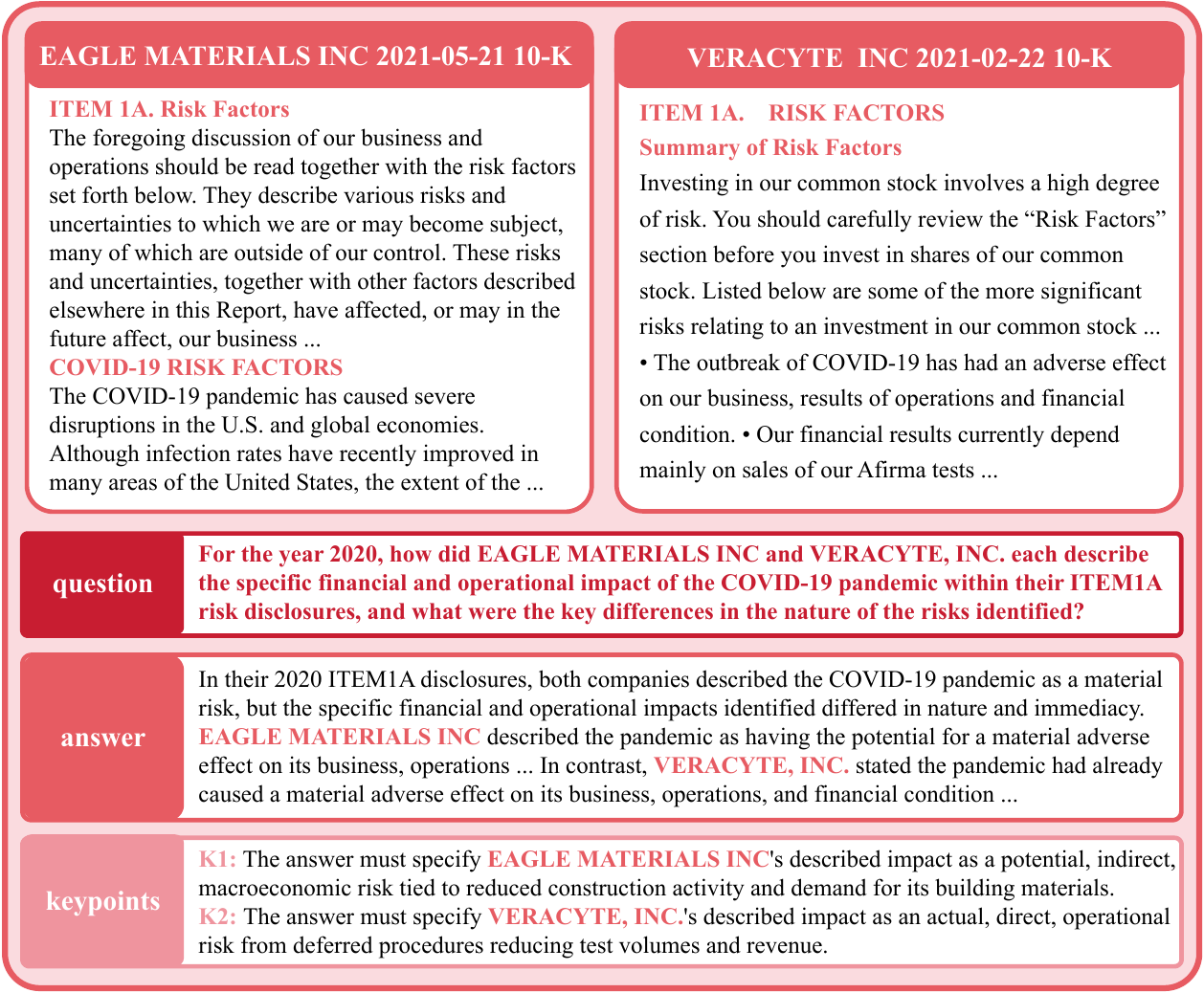}
    \caption{Example for Enterprise Comparison (EC) question}
    \label{fig:EC-example}
\end{figure*}

Enterprise Comparison questions require models to compare financial or
operational attributes across multiple companies. As illustrated in
Figure~\ref{fig:EC-example}, the model must align heterogeneous disclosures from
different firms and reason over relative differences rather than isolated facts.

\subsection{Longitudinal Tracking QA}
\begin{figure*}
    \centering
    \includegraphics[width=0.69\textwidth]{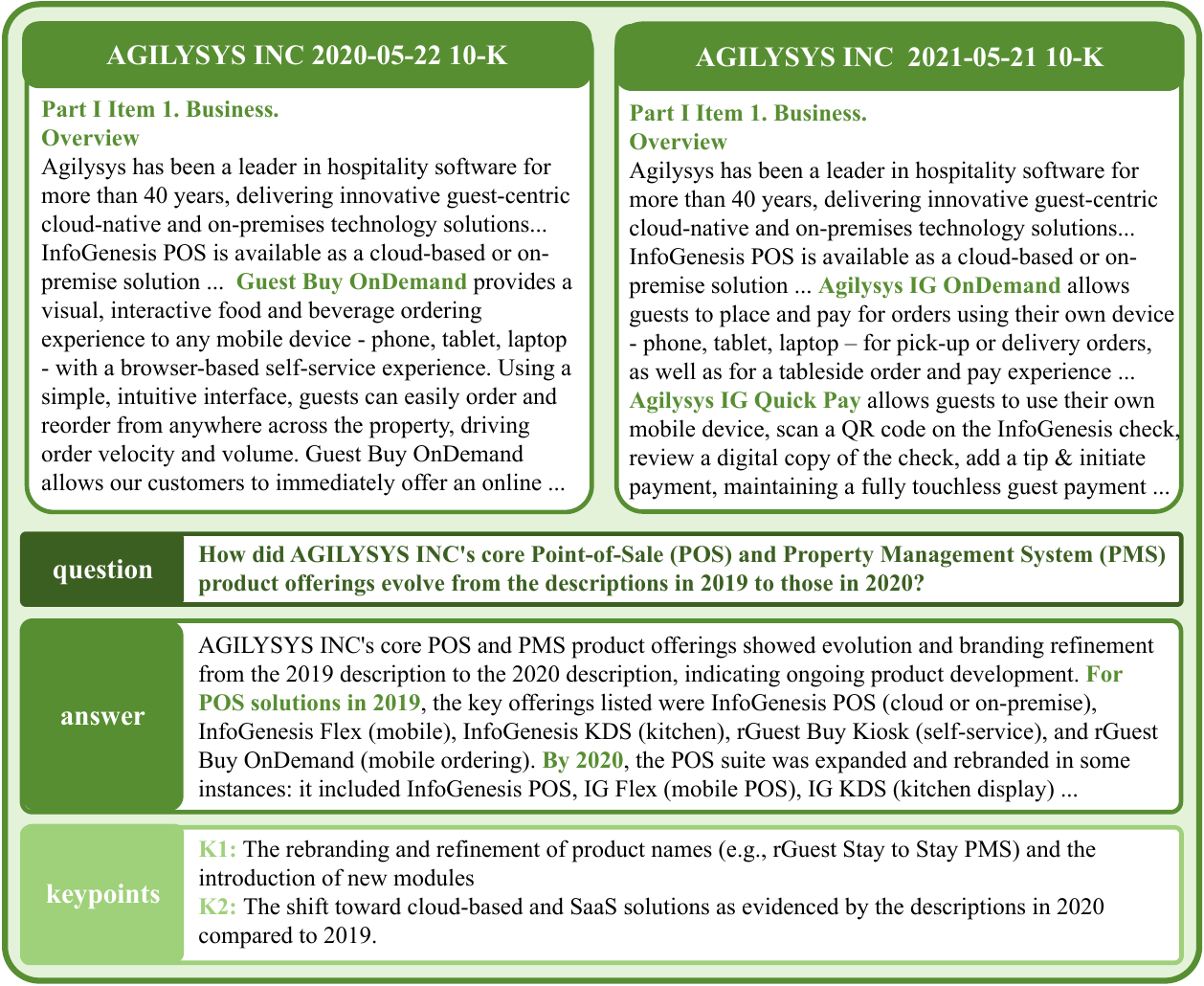}
    \caption{Example for Longitudinal Tracking (LT) question}
    \label{fig:LT-example}
\end{figure*}

Longitudinal Tracking questions examine a model’s ability to track changes
in financial conditions, policies, or disclosures over time. As shown in
Figure~\ref{fig:LT-example}, answering such questions requires integrating
information across multiple reporting periods and reasoning about temporal
trends or evolutions.

\section{Experiment Details}
\subsection{LLM Details}
We evaluate a diverse set of large language models (LLMs) covering different
training paradigms, parameter scales, and domain specializations, as summarized
in Table~\ref{tab:models_scale}. Open-source models are evaluated either via local inference on NVIDIA A100/A6000 GPUs (\(\leq 70\)B) or through APIs for larger models like Qwen3-235B, and all models use a unified decoding temperature of 0.7.
\begin{table}[h]
\centering
\caption{Model Type and Scale Comparison}
\label{tab:models_scale}
\begin{tabular*}{\linewidth}{@{\extracolsep{\fill}} p{0.35\linewidth} p{0.32\linewidth} p{0.12\linewidth} @{}}
\toprule
\textbf{Model Name} & \textbf{Type} & \textbf{Scale} \\
\midrule
GPT-5-websearch & Closed-Source & N/A \\ 
GPT-4.1 & Closed-Source & N/A \\ 
GPT-4.1-websearch & Closed-Source & N/A \\ 
MIMO-V2-Fash & Open-Source General & 309B \\ 
DeepSeek-V3 & Open-Source General & 685B \\ 
DeepSeek-V3.2 & Open-Source General & 685B \\ 
DeepSeek-R1 & Open-Source General & 685B \\ 
GPT-OSS-20B & Open-Source General & 20B \\ 
Llama-3.3-70B-Instruct & Open-Source General & 70B \\
Qwen3-8B & Open-Source General & 8B \\ 
Qwen3-14B & Open-Source General & 14B \\ 
Qwen3-30B-A3B-Instruct-2507 & Open-Source General & 30B \\ 
Qwen3-235B & Open-Source General & 235B \\ 
Fin-R1 & Open-Source Finance & 7B \\ 
Fino1-14B & Open-Source Finance & 14B \\ 
FinanceConnect-13B & Open-Source Finance & 13B \\
TouchstoneGPT-7B-Instruct & Open-Source Finance & 7B \\
\bottomrule
\end{tabular*}
\end{table}

\subsection{Retrieval Methods}

To simulate realistic document access in financial analysis scenarios, we evaluate
models under a retrieval-augmented generation (RAG) setting. We adopt three
representative retrieval strategies that are widely used in both academic research
and industrial systems.

\textbf{Lexical Retrieval (BM25).}
BM25 is used as a sparse retrieval baseline to capture exact term matching and
keyword-based relevance, which remains a strong and interpretable method for
financial documents containing formal terminology.

\textbf{Dense Vector Retrieval.}
We employ dense semantic retrieval using two embedding models:
\textit{all-MiniLM-L6-v2}, a general-purpose sentence embedding model, and
\textit{finance-embeddings-investopedia}, a finance-oriented embedding model
trained on domain-specific corpora. This setup allows us to examine the effect
of domain-adapted representations in dense retrieval.

\textbf{Hybrid Retrieval.}
We further evaluate a hybrid retrieval strategy implemented via Elasticsearch,
which combines BM25 with dense vector similarity. In this hybrid setting, the
dense embeddings are generated using all-MiniLM-L6-v2, and the final retrieval
scores are computed by integrating lexical and semantic signals.

\section{Prompts for QA Generation}
\label{appendix:D}
\subsection{Detail \& Reasoning QA}
The prompts used in the QA generation system are designed with the following principles:

\begin{enumerate}
    \item \textbf{Dynamic Variable Insertion}: Prompts incorporate context-specific variables such as report year, and document type to ensure each generated question is tailored.
    \item \textbf{Structured Output Control}: Prompts enforce strict formatting rules (e.g., "Q1:", "A1:") to facilitate automated parsing and validation.
    \item \textbf{Content Grounding}: Each prompt includes explicit instructions to base responses solely on provided document content, avoiding external knowledge.
    \item \textbf{Diversity Enhancement}: Randomization elements are added to prompts to encourage varied question types and analytical angles, reducing repetition.
\end{enumerate}

The QA generation process follows a two-stage approach: question generation followed by answer generation with validation.

\subsubsection{Stage 1: Question Generation Prompt}
\label{sec:question_generation_prompt}

This prompt is used to generate analytical questions based on a chunk, which includes:
\begin{itemize}
    \item Contextual variables (company, year, filing type)
    \item Instructions for generating a fixed number of questions
    \item Requirements for question uniqueness and document relevance
    \item Formatting rules for easy parsing
\end{itemize}

\begin{tcolorbox}[
    colback=ldr-light-primary,
    colframe=ldr-dark-secondary,
    title=Prompt Template: Question Generation,
    breakable
]
You are an expert financial analyst specializing in SEC filings analysis. 

\textbf{TASK}: Generate exactly \{target\_qa\_count\} analytical questions based on the provided SEC filing document.

You are analyzing a \{filing\_type\} filing for \{company\_name\} for the year \{year\}. This is \{company\_name\}'s \{year\} annual financial report. Generate exactly \{target\_qa\_count\} sophisticated analytical questions based ONLY on the content provided.\\
\{filing\_guidance\} \\
\{text\_analysis\_guidance\} \\

\textbf{CRITICAL REQUIREMENTS}:\\
- Generate EXACTLY \{target\_qa\_count\} questions
- Each question MUST explicitly mention \{company\_name\} and \{year\}\\
- Questions should be analytical and require detailed answers\\
- Focus on financial analysis, business operations, risk factors, and strategic implications\\
- Questions should be answerable based on the provided document\\
- Each question must be UNIQUE and reflect specific content from the document\\

\textbf{MANDATORY FORMAT REQUIREMENTS}:\\
- Start each question with \texttt{"Q"} followed by the number and a colon (e.g., Q1:, Q2:, Q3:)\\
- NO other formats are acceptable (no Q1., Q1), Q1-, etc.)\\
- Each question MUST contain the exact text \{company\_name\} and \{year\}\\
- Do not use any markdown formatting, asterisks, or special characters around Q labels\\

\textbf{EXAMPLE FORMAT (follow this exactly)}:\\
Q1: [Question - MUST include \{company\_name\} and \{year\} and be specific to the document content]\\
Q2: [Question - MUST include \{company\_name\} and \{year\} and be specific to the document content]\\
Q3: [Question - MUST include \{company\_name\} and \{year\} and be specific to the document content]\\

\textbf{YOU MUST Return in the following format}:\\
Q1: [Question - MUST include \{company\_name\} and \{year\}]\\
Q2: [Question - MUST include \{company\_name\} and \{year\}]\\
Q3: [Question - MUST include \{company\_name\} and \{year\}]\\

\textbf{Document Content}: \{text\}
\end{tcolorbox}

\textbf{Filing-Specific Guidance}

The following filing-specific instructions are dynamically inserted into the \{text\_analysis\_guidance\} field of the Question Generation prompt based on document type:

\begin{tcolorbox}[
    colback=ldr-light-primary,
    colframe=black!15!ldr-dark-border,
    title=10-K Filing Guidance,
    breakable
]

\begin{itemize}
    \item Generate analytical questions about financial ratios, performance metrics, and year-over-year comparisons
    \item Create questions analyzing business performance trends, revenue growth, and operational efficiency
\end{itemize}

\end{tcolorbox}

\begin{tcolorbox}[
    colback=ldr-light-primary,
    colframe=black!15!ldr-dark-border,
    title=10-Q Filing Guidance,
    breakable
]

\begin{itemize}
    \item Generate questions about quarterly financial performance, seasonal trends, and period-over-period changes
    \item Create questions analyzing cash flow patterns, working capital management, and liquidity position
\end{itemize}

\end{tcolorbox}

\begin{tcolorbox}[
    colback=ldr-light-primary,
    colframe=black!15!ldr-dark-border,
    title=8-K Filing Guidance,
    breakable
]

\begin{itemize}
    \item Generate questions about material events and their quantitative business impacts
    \item Create questions evaluating strategic importance and long-term implications of disclosed events
\end{itemize}

\end{tcolorbox}

\begin{tcolorbox}[
    colback=ldr-light-primary,
    colframe=black!15!ldr-dark-border,
    title=6-K Filing Guidance,
    breakable
]

\begin{itemize}
    \item Generate questions about foreign operations, currency impacts, and international reporting requirements
    \item Create questions about regulatory compliance matters and cross-border considerations
\end{itemize}

\end{tcolorbox}

\begin{tcolorbox}[
    colback=ldr-light-primary,
    colframe=black!15!ldr-dark-border,
    title=SC 13G Filing Guidance,
    breakable
]

\begin{itemize}
    \item Generate questions about beneficial ownership patterns, stake sizes, and ownership changes
    \item Create questions analyzing investment strategies, positions, and portfolio implications
\end{itemize}

\end{tcolorbox}

\textbf{Content Analysis Guidance}
To ensure generated questions are specific and grounded in the actual document content rather than generic templates, the system includes detailed content analysis guidance. These prompts direct the LLM to focus on concrete details, unique document features, and diverse analytical perspectives.
\begin{tcolorbox}[
    colback=ldr-light-primary,
    colframe=ldr-dark-secondary!60!black,
    title=Part 1: Content-Based Question Generation,
    breakable
]

\begin{itemize}
    \item Analyze the specific content provided in the document below
    \item Generate questions that are UNIQUE to this specific document section
    \item Focus on specific details, numbers, events, or unique aspects mentioned in the text
    \item Avoid generic financial analysis questions unless specifically relevant to this content
    \item Each question should reflect something SPECIFIC found in the provided text
    \item Questions should demonstrate deep understanding of the document's unique content
\end{itemize}

\end{tcolorbox}

\begin{tcolorbox}[
    colback=ldr-light-primary,
    colframe=ldr-dark-secondary!60!black,
    title=Part 2: Content Analysis Requirements,
    breakable
]

\begin{itemize}
    \item Look for specific financial figures, percentages, or ratios mentioned
    \item Identify unique business operations, strategies, or events described
    \item Find specific risk factors, challenges, or opportunities discussed
    \item Notice any industry-specific details or regulatory mentions
    \item Extract specific time periods, locations, or business segments mentioned
    \item Focus on concrete data points rather than general concepts
\end{itemize}

\end{tcolorbox}

\begin{tcolorbox}[
    colback=ldr-light-primary,
    colframe=ldr-dark-secondary!60!black,
    title=Part 3: Question Diversity Requirements,
    breakable
]

\begin{itemize}
    \item Each question should explore a DIFFERENT aspect of the document
    \item Avoid repetitive question patterns
    \item Mix analytical, quantitative, and qualitative question types
    \item Include questions about specific business operations, not just financial metrics
    \item Ask about specific events, decisions, or strategies mentioned in the text
\end{itemize}

\end{tcolorbox}

\textbf{Randomization Instructions}
\label{sec:randomization_prompts}

To prevent repetitive question patterns and enhance the diversity of generated content, the system incorporates a randomization mechanism. This approach injects variability into each generation cycle by dynamically selecting different analytical angles and focus areas from predefined pools. The randomization occurs at two levels: the analytical perspective and the content emphasis.

\begin{tcolorbox}[
    colback=ldr-light-primary,
    colframe=blue!50!black,
    title=Randomization Instruction Template,
    breakable
]
\begin{itemize}
    \item Focus on \{selected\_angle\} approach
    \item Emphasize \{selected\_focus\} in your questions
    \item Ensure each question explores different aspects of the document
    \item Avoid repeating similar question structures from previous generations
\end{itemize}

\end{tcolorbox}

\begin{tcolorbox}[
    colback=ldr-light-primary,
    colframe=blue!50!black,
    title=Analytical Angle Pool,
    breakable
]
\begin{itemize}
    \item quantitative financial analysis
    \item qualitative business strategy analysis
    \item operational efficiency assessment
    \item risk and opportunity evaluation
    \item competitive positioning analysis
    \item regulatory compliance review
    \item market performance evaluation
    \item strategic decision impact analysis
\end{itemize}

\end{tcolorbox}

\begin{tcolorbox}[
    colback=ldr-light-primary,
    colframe=blue!50!black,
    title=Focus Area Pool,
    breakable
]
\begin{itemize}
    \item specific financial metrics and ratios
    \item unique business operations and processes
    \item distinctive market strategies and initiatives
    \item particular risk factors and mitigation strategies
    \item specific regulatory challenges and responses
    \item unique competitive advantages and disadvantages
    \item particular operational challenges and solutions
    \item specific investment decisions and their rationale
\end{itemize}

\end{tcolorbox}

\subsubsection{Stage 2: Answer Generation Prompt}
\label{sec:answer_generation_prompt}

This prompt generates comprehensive answers for individual questions while requesting log probability data for validation.

\begin{tcolorbox}[
    colback=ldr-light-primary,
    colframe=ldr-dark-border!50!black,
    title=QA Generation Prompt Template,
    breakable
]
You are an expert financial analyst specializing in SEC filings analysis. 

TASK: Answer the following question based on the provided SEC filing document.

You are analyzing a \{filing\_type\} filing for \{company\_name\} for the year \{year\}.

QUESTION: \{question\}

CRITICAL REQUIREMENTS:
- Provide a comprehensive answer based ONLY on the information provided in the document
- Include specific data, calculations, and analysis where applicable
- Show step-by-step work for any calculations
- Be accurate and professional
- Use ONLY the information provided in the document

Answer:
\end{tcolorbox}

\subsection{Enterprise Comparison QA}
\subsubsection{Stage 1: Summarization Generation Prompt}
This prompt is used to generate comprehensive summaries of individual SEC filing sections. It includes:
\begin{itemize}
    \item Contextual variables including company name, section heading, focus area
    \item Requirements for detailed, integrated analysis across chunks
    \item Instructions for structured data extraction and organization
    \item Specific output format with JSON schema for parsing
    \item Constraints ensuring factual accuracy and data completeness
\end{itemize}
\begin{tcolorbox}[
    colback=ec-light-primary,
    colframe=ec-dark-border,
    title=\textbf{Item Summary Generation Prompt Template},
    fonttitle=\bfseries,
    breakable,
    before skip=10pt,
    after skip=15pt
]
\texttt{<|begin\_of\_text|><|start\_header\_id|>system\\<|end\_header\_id|>} \\
You are a senior financial analyst specializing in SEC filing analysis. Your task is to analyze the following document chunks from \{company\_name\}'s \{section\_heading\} section and provide a comprehensive, structured, and detailed summary.

\textbf{Company:} \{company\_name\} \\
\textbf{Section:} \{section\_heading\} \\
\textbf{Focus Area:} \{self.item\_summary\_guidance.get(item, "General analysis of the section content")\}

\textbf{Analysis Requirements:}
1. \textbf{Comprehensive Summary}: Provide a detailed, integrated summary that synthesizes information from all chunks\\
2. \textbf{Key Points Extraction}: Extract 5-10 most important points with specific details\\
3. \textbf{Data Extraction}: Identify and extract specific numerical data, metrics, percentages, and quantitative information\\
4. \textbf{Structured Organization}: Organize information logically by themes or categories\\
5. \textbf{Specificity}: Use exact figures, dates, and facts from the source material\\
6. \textbf{Context Awareness}: Note any important context, trends, or changes mentioned

\textbf{INTEGRATED DOCUMENT CONTENT TO ANALYZE:} \{context\_text\}

\textbf{Required Output Format (JSON):}
\lstset{
    breaklines=true,    
    breakatwhitespace=true, 
    postbreak=\raisebox{0ex}[0ex][0ex]{\ensuremath{\hookrightarrow\space}},
    basicstyle=\ttfamily\small,
    columns=fullflexible
}

\begin{lstlisting}
{
    "summary": "A comprehensive, detailed summary integrating all chunks. Must be specific, factual, and include quantitative data where available. Length: 300-500 words.",
    "key_points": [
        "Key point 1 with specific details",
        "Key point 2 with numbers/dates/facts",
        "... (5-10 key points)"
    ],
    "quantitative_data": {
        "metrics": ["metric 1", "metric 2", "..."],
        "numbers": ["specific number with context"],
        "percentages": ["percentage with context"],
        "dates": ["important date with context"]
    },
    "focus_area": "{focus_area}",
    "themes": ["theme 1", "theme 2", "..."],
    "strengths": ["identified strength 1", "..."],
    "risks": ["identified risk 1", "..."],
    "strategic_points": ["strategic point 1", "..."]
}
\end{lstlisting}

\textbf{IMPORTANT:}
- Use ONLY information from the provided chunks above\\
- Do NOT fabricate, assume, or infer data not explicitly stated\\
- Extract ALL quantitative information (numbers, percentages, dates, metrics)\\
- Be specific and detailed - avoid vague or generic statements\\
- The summary should be comprehensive enough to enable detailed comparison analysis later

Return ONLY the JSON object, no additional text or explanations. \\
\texttt{<|eot\_id|><|start\_header\_id|>assistant\\<|end\_header\_id|>}
\end{tcolorbox}

\textbf{Item-Specific Summary Guidance}
This guidance maps each SEC filing item to its specific analytical focus for targeted summarization:

\begin{tcolorbox}[
    colback=ec-light-primary,
    colframe=ec-dark-border!70!black,
    title=\textbf{Item-Specific Summary Guidance},
    fonttitle=\bfseries,
    breakable
]
\begin{itemize}
    \item \textbf{Item 1 - Business:} Focus on understanding the company's core business model, operations, and market positioning
    \item \textbf{Item 1A - Risk Factors:} Focus on identifying key risks, challenges, and potential threats disclosed by the company
    \item \textbf{Item 8 - Financial Statements:} Focus on analyzing financial performance, trends, and key metrics
    \item \textbf{Item 7 - MD\&A:} Focus on extracting management's perspective on financial condition and operational results
    \item \textbf{Item 3 - Legal Proceedings:} Focus on assessing legal, regulatory, and compliance matters
    \item \textbf{Item 10 - Governance:} Focus on analyzing corporate governance structure and executive leadership
\end{itemize}

Each focus area guides the summarization process to extract relevant information for subsequent comparative analysis.
\end{tcolorbox}

\subsubsection{Stage 2: QA Generation Prompt}
This prompt is used to generate enterprise comparison QAs based on summarizations generated in stage 1. It includes:

\begin{itemize}
    \item Task Definition: Clear instruction to generate comparative Q\&A pairs
    \item Focus Constraints: Section-based limitations and question type guidance
    \item Information Availability Requirements: Critical constraint ensuring questions are only generated when both companies have relevant data
    \item Quality Standards: Detailed requirements for both questions and answers
    \item Output Format: JSON structure with \texttt{"qa\_pairs"} array
\end{itemize}

\begin{tcolorbox}[
    colback=ec-light-primary,
    colframe=ec-dark-border,
    title=QA Generation Prompt Template,
    fonttitle=\bfseries,
    breakable
]
\texttt{<|begin\_of\_text|><|start\_header\_id|>system\\<|end\_header\_id|>} \\
You are a financial analyst expert at generating insightful comparative questions and comprehensive answers based on company annual reports. You always output valid JSON objects. \\
\texttt{<|eot\_id|><|start\_header\_id|>user<|end\_header\_id|>} \\
Based on the following context information about \{company1\} and \{company2\} for the year \{year\}, generate \{num\_pairs\} complete question-answer pairs for comparative analysis.

\textbf{FOCUS AREA:} \texttt{\{sections\_str\}}

\{question\_type\_instruction\}

\textbf{FOCUS GUIDANCE:}
\{focus\_guidance\}

\textbf{CONTEXT INFORMATION:}
\{context\_text\}

\textbf{CRITICAL REQUIREMENT - INFORMATION AVAILABILITY:}\\
- You MUST ONLY generate questions about topics where BOTH \{company1\} AND \{company2\} have relevant information in the context
- Before generating each question, verify that the context contains comparable information for BOTH companies
- DO NOT generate questions that would require information from only one company or where one company's information is missing
- If information is only available for one company, DO NOT ask about that topic

\textbf{REQUIREMENTS FOR EACH QUESTION:}\\
1. MUST explicitly mention both company names: \{company1\} and \{company2\} \\
2. MUST explicitly mention the year: \{year\} \\
3. MUST focus on comparative analysis between the two companies \\
4. MUST be specific and directly related to the provided context information \\
5. MUST strictly follow the QUESTION TYPE instruction above \(single item depth vs. multi-item integration\) \\
6. MUST align with the focus guidance provided above \\
7. MUST be suitable for financial or business analysis \\
8. MUST be written in clear, professional English \\
9. MUST address different aspects to ensure variety across the \{num\_pairs\} questions \\
10. For multi-item combinations: Each question MUST reference multiple sections and show their interrelationship \\
11. CRITICAL: MUST only ask about topics where BOTH companies have information - verify this before generating the question \\

\textbf{REQUIREMENTS FOR EACH ANSWER:}\\
1. MUST be comprehensive and detailed \\
2. MUST be based on the provided context information \\
3. MUST compare aspects between \{company1\} and \{company2\} \\
4. MUST be written in clear, professional English \\
5. MUST provide specific insights and comparisons \\
6. For multi-item combinations: Answers MUST integrate information from multiple sections \\
7. CRITICAL: MUST provide information for BOTH companies - DO NOT use phrases like \texttt{"not mentioned"}, \texttt{"not available"}, \texttt{"not disclosed"}, \texttt{"information is not provided"}, or similar statements indicating missing information \\
8. If you cannot provide information for both companies, DO NOT generate this question-answer pair - skip it and generate a different one \\

\textbf{IMPORTANT:} Ensure questions are DISTINCT and cover different aspects. For single items, vary the specific details being explored. For combinations, vary the types of relationships being analyzed.

Generate exactly \{num\_pairs\} distinct question-answer pairs. Format your response as a JSON object with a key \texttt{"qa\_pairs"} containing an array of objects, where each object has \texttt{"question"} and \texttt{"answer"} keys. \\
\texttt{<|eot\_id|><|start\_header\_id|>assistant\\<|end\_header\_id|>}
\end{tcolorbox}

\textbf{Single Item Guidence: Deep Dive Analysis}

This guidance configures the prompt for \textbf{granular, in-depth analysis} of individual sections. 
\begin{tcolorbox}[
    colback=ec-light-primary,
    colframe=ec-dark-border!70!black,
    title=QUESTION TYPE: SINGLE ITEM DEEP DIVE,
    fonttitle=\bfseries,
    breakable
]
- Generate questions that explore DEPTH within this single section
- Focus on specific details, examples, methodologies, or granular aspects
- Avoid broad or general questions - instead ask about concrete specifics
- Each question should drill into particular elements within this section
- Questions should be narrow, detailed, and section-specific
\end{tcolorbox}

\textbf{Multi-item Guidance: Integrated Cross-sectional Analysis}

This guidance configures the prompt for \textbf{cross-sectional, integrated analysis} that connects multiple report sections. The approach emphasizes:
\begin{tcolorbox}[
    colback=ec-light-primary,
    colframe=ec-dark-border!70!black,
    title=QUESTION TYPE: MULTI-ITEM INTEGRATED ANALYSIS,
    fonttitle=\bfseries,
    breakable
]
\texttt{<|begin\_of\_text|><|start\_header\_id|>system\\<|end\_header\_id|>} \\
\textbf{QUESTION TYPE: MULTI-ITEM INTEGRATED ANALYSIS} \\
- Generate questions that EXPLICITLY connect information from MULTIPLE sections, do not only use information from only one section \\
- Each question MUST require information from at least 2 different sections to answer \\
- Focus on relationships, interactions, and cross-sectional insights \\
- Ask "how" and "why" questions that show how sections relate to each other \\
- Avoid questions that could be answered using only one section \\
- Questions should demonstrate integrated analysis across sections \\
\texttt{<|eot\_id|><|start\_header\_id|>assistant\\<|end\_header\_id|>}
\end{tcolorbox}

\textbf{Single Item Focus}

For each designated section, questions are generated that drill deeply into specific details, concrete examples, and nuanced aspects, emphasizing \textbf{depth over breadth}, producing narrow, section-specific questions that explore granular differences in disclosure practices, methodologies, and prioritization strategies within each standalone section.
\begin{tcolorbox}[
    colback=ec-light-primary,
    colframe=ec-dark-border!50!black,
    title=Item 1A. Risk Factors - Individual Item Focus,
    fonttitle=\bfseries,
    breakable
]
\texttt{<|begin\_of\_text|><|start\_header\_id|>system\\<|end\_header\_id|>} \\
\textbf{INDIVIDUAL ITEM FOCUS - Generate questions that:} \\
1. Drill deep into specific risk categories (operational, financial, regulatory, market, etc.) \\
2. Compare detailed risk descriptions, quantification methods, and disclosure specificity \\
3. Analyze granular differences in risk prioritization and presentation styles \\
4. Focus on concrete examples and specific risk factors mentioned in each company's disclosure \\
5. Avoid general comparisons - instead ask about specific risk types, mitigation strategies, or risk assessment methodologies \\
6. Questions should be narrow and detailed, exploring depth within this single section \\
\texttt{<|eot\_id|><|start\_header\_id|>assistant\\<|end\_header\_id|>}
\end{tcolorbox}

\begin{tcolorbox}[
    colback=ec-light-primary,
    colframe=ec-dark-border!50!black,
    title=Item 7. Management's Discussion and Analysis of Financial Condition and Results of Operations - Individual Item Focus,
    fonttitle=\bfseries,
    breakable
]
\texttt{<|begin\_of\_text|><|start\_header\_id|>system\\<|end\_header\_id|>} \\
\textbf{INDIVIDUAL ITEM FOCUS - Generate questions that:} \\
1. Focus on specific financial metrics, trends, or performance indicators discussed in MD\&A \\
2. Compare detailed explanations of revenue drivers, cost factors, or operational changes \\
3. Analyze specific forward-looking statements, guidance, or management outlook \\
4. Explore granular differences in how each company explains specific financial results or operational events \\
5. Ask about concrete examples, specific periods, or particular business segments discussed \\
6. Questions should be narrow and detailed, exploring depth within this single section \\
\texttt{<|eot\_id|><|start\_header\_id|>assistant\\<|end\_header\_id|>}
\end{tcolorbox}

\begin{tcolorbox}[
    colback=ec-light-primary,
    colframe=ec-dark-border!50!black,
    title=Item 3. Legal Proceedings - Individual Item Focus,
    fonttitle=\bfseries,
    breakable
]
\texttt{<|begin\_of\_text|><|start\_header\_id|>system\\<|end\_header\_id|>} \\
\textbf{INDIVIDUAL ITEM FOCUS - Generate questions that:} \\
1. Focus on specific legal cases, claims, or proceedings mentioned by each company \\
2. Compare detailed information about case types, potential outcomes, or settlement amounts \\
3. Analyze specific litigation strategies, legal reserves, or disclosure practices \\
4. Explore granular differences in how each company categorizes or prioritizes legal matters \\
5. Ask about concrete examples of specific legal proceedings or their potential impacts \\
6. Questions should be narrow and detailed, exploring depth within this single section \\
\texttt{<|eot\_id|><|start\_header\_id|>assistant\\<|end\_header\_id|>}
\end{tcolorbox}

\begin{tcolorbox}[
    colback=ec-light-primary,
    colframe=ec-dark-border!50!black,
    title={Item 10. Directors, Executive Officers, and Corporate Governance - Individual Item Focus},
    fonttitle=\bfseries,
    breakable
]
\texttt{<|begin\_of\_text|><|start\_header\_id|>system\\<|end\_header\_id|>} \\
\textbf{INDIVIDUAL ITEM FOCUS - Generate questions that:} \\
1. Focus on specific governance structures, board committees, or executive roles \\
2. Compare detailed information about board composition, director qualifications, or executive compensation \\
3. Analyze specific governance policies, codes of conduct, or corporate governance practices \\
4. Explore granular differences in how each company structures its governance framework \\
5. Ask about concrete examples of governance mechanisms or specific governance-related disclosures \\
6. Questions should be narrow and detailed, exploring depth within this single section \\
\texttt{<|eot\_id|><|start\_header\_id|>assistant\\<|end\_header\_id|>}
\end{tcolorbox}

\textbf{Multi-item Focus}

The multi-item focus configuration enables \textbf{integrated, relational analysis} across multiple annual report sections by explicitly connecting information from different areas to generates questions that can explore \textbf{interrelationships and systemic connections}. Questions require synthesis of information from at least two distinct sections, focusing on how different business components interact, influence each other, and collectively shape the company's overall profile and performance.

\begin{tcolorbox}[
    colback=ec-light-primary,
    colframe=ec-dark-border!50!black,
    title=Financial + Business Analysis - Core business model and financial performance relationship,
    fonttitle=\bfseries,
    breakable
]
\texttt{<|begin\_of\_text|><|start\_header\_id|>system\\<|end\_header\_id|>} \\
\textbf{MULTI-ITEM COMBINATION FOCUS - Generate questions that:} \\
1. EXPLICITLY connect information from BOTH Item 1 (Business) AND Item 8 (Financial Statements) \\
2. Ask about HOW business operations (from Item 1) translate into specific financial metrics (from Item 8) \\
3. Compare the relationship between business model characteristics and financial performance indicators \\
4. Analyze how operational strategies or business segments correlate with revenue, costs, or profitability \\
5. Explore cross-sectional insights: how business descriptions explain financial results, or how financial data reflects business operations \\
6. Questions MUST reference both sections and show their interrelationship - avoid questions that could be answered by only one section \\
\texttt{<|eot\_id|><|start\_header\_id|>assistant\\<|end\_header\_id|>}
\end{tcolorbox}

\begin{tcolorbox}[
    colback=ec-light-primary,
    colframe=ec-dark-border!50!black,
    title=Business + Risk Analysis - Business strategy and risk exposure relationship,
    fonttitle=\bfseries,
    breakable
]
\texttt{<|begin\_of\_text|><|start\_header\_id|>system\\<|end\_header\_id|>} \\
\textbf{MULTI-ITEM COMBINATION FOCUS - Generate questions that:} \\
1. EXPLICITLY connect business strategy from Item 1 (Business) with risk factors from Item 1A (Risk Factors) \\
2. Ask about HOW specific business operations or strategies create or mitigate particular risks \\
3. Compare how each company's business model characteristics relate to their identified risk exposures \\
4. Analyze the alignment between strategic initiatives and risk management approaches \\
5. Explore how business decisions or market positions influence risk profiles \\
6. Questions MUST reference both sections and show their interrelationship - avoid questions that could be answered by only one section \\
\texttt{<|eot\_id|><|start\_header\_id|>assistant\\<|end\_header\_id|>}
\end{tcolorbox}

\begin{tcolorbox}[
    colback=ec-light-primary,
    colframe=ec-dark-border!50!black,
    title=Comprehensive Risk Analysis - Complete risk assessment framework,
    fonttitle=\bfseries,
    breakable
]
\texttt{<|begin\_of\_text|><|start\_header\_id|>system\\<|end\_header\_id|>} \\
\textbf{MULTI-ITEM COMBINATION FOCUS - Generate questions that:} \\
1. EXPLICITLY integrate information from Item 1A (Risk Factors), Item 3 (Legal Proceedings), AND Item 7A (Market Risk) \\
2. Ask about HOW different types of risks (operational, legal, market) interact or relate to each other \\
3. Compare the comprehensiveness and integration of each company's overall risk management framework \\
4. Analyze how different risk categories are prioritized, disclosed, or managed together \\
5. Explore cross-risk insights: how legal risks relate to operational risks, or how market risks connect to other risk factors \\
6. Questions MUST reference multiple risk sections and show their interrelationship - avoid questions that could be answered by only one section \\
\texttt{<|eot\_id|><|start\_header\_id|>assistant\\<|end\_header\_id|>}
\end{tcolorbox}

\begin{tcolorbox}[
    colback=ec-light-primary,
    colframe=ec-dark-border!50!black,
    title=Governance and Performance - Corporate governance structure and financial performance relationship,
    fonttitle=\bfseries,
    breakable
]
\texttt{<|begin\_of\_text|><|start\_header\_id|>system\\<|end\_header\_id|>} \\
\textbf{MULTI-ITEM COMBINATION FOCUS - Generate questions that:} \\
1. EXPLICITLY connect governance structure from Item 10 (Governance) with financial performance from Item 8 (Financial Statements) \\
2. Ask about HOW governance practices, board composition, or executive leadership relate to specific financial outcomes \\
3. Compare how governance characteristics correlate with financial performance metrics \\
4. Analyze the relationship between governance mechanisms and financial results or strategic decisions \\
5. Explore how governance quality or structure may influence financial performance or operational efficiency \\
6. Questions MUST reference both sections and show their interrelationship - avoid questions that could be answered by only one section \\
\texttt{<|eot\_id|><|start\_header\_id|>assistant\\<|end\_header\_id|>}
\end{tcolorbox}

\subsection{Longitudinal Tracking QA}
\subsubsection{Stage 1: Summarization Generation Prompt}
\label{sec:item_summarization_prompts}

These prompts are used to generate structured summaries for specific SEC filing items before conducting cross-year comparison analysis.

\textbf{Item 1: Business Overview Summary Prompt}
\label{sec:item1_summary_prompt}

\begin{tcolorbox}[
    colback=lt-light-primary!80,
    colframe=lt-dark-border,
    title=Item 1 Business Overview Summary Prompt,
    breakable,
    fontupper=\small,
    before upper={\sloppy\parindent0pt\parskip5pt,
    breakable}
]
Please act as a rigorous financial analyst. Summarize the following business overview text from \texttt{\{company\_name\}} for year \texttt{\{year\}} from the following 6 dimensions:

1. \textbf{Strategy \& Business Model Evolution} - Summarize strategic direction, business model shifts, strategic initiatives, and major strategic decisions.\\
2. \textbf{Financial Performance \& Revenue Structure} - Summarize revenue sources, financial metrics, revenue composition, profitability trends, and key financial indicators.\\
3. \textbf{Product \& Technology Portfolio} - Summarize product offerings, technology platforms, product development, innovation initiatives, and portfolio composition.\\
4. \textbf{Operations \& Human Capital} - Summarize operational structure, facilities, employee counts, organizational changes, and operational efficiency.\\
5. \textbf{Market, Competition \& Customers} - Summarize market position, competitive landscape, customer base, market segments served, and market dynamics.\\
6. \textbf{Growth Drivers \& Investment Focus} - Summarize growth strategies, investment priorities, capital allocation, expansion plans, and growth initiatives.\\

\textbf{Text to Summarize:}
\texttt{\{text\}}

Please provide a comprehensive summary covering all 6 dimensions. For each dimension, provide key facts, numbers, and important information. Use exact numbers and data from the text. Format your response as a structured text covering each dimension clearly.

Return only the summary text, no additional explanation.
\end{tcolorbox}

\textbf{Item 1A: Risk Factors Summary Prompt}
\label{sec:item1a_summary_prompt}

\begin{tcolorbox}[
    colback=lt-light-primary!80,
    colframe=lt-dark-border,
    title=Item 1A Risk Factors Summary Prompt,
    breakable,
    fontupper=\small,
    before upper={\sloppy\parindent0pt\parskip5pt,
    breakable}
]
Please act as a rigorous financial analyst. Summarize the following risk factors text from \texttt{\{company\_name\}} for year \texttt{\{year\}}.

\textbf{Your Task:}
Extract and summarize all risks mentioned in the text. For each risk, provide a concise summary based on how it is actually described in the original text. Do not force a fixed structure - instead, summarize what the text actually says about each risk.

\textbf{Focus on:}
- \textbf{All risks mentioned} in the text (identify each distinct risk)\\
- \textbf{Risk names/titles} (if explicitly named)\\
- \textbf{Risk categorization} (if any explicit categories are mentioned)\\
- \textbf{Key information from the original text} about each risk - summarize the actual description as it appears, including:\\
  * What the text says about the nature of the risk\\
  * Any specific details, examples, or context mentioned\\

\textbf{Important Guidelines:}
- Follow the structure and emphasis of the original text\\
- if the text focuses on certain aspects, your summary should reflect that\\
- Do not create a fixed template\\
- instead, summarize what the text actually describes about the risk itself\\
- Keep summaries concise and focused on the risk's nature and key characteristics\\
- Preserve the actual wording and phrasing when describing the risk, especially for technical terms or specific risk characteristics\\

\textbf{Text to Summarize:}
\texttt{\{text\}}

Please provide a structured summary that lists all risks mentioned, with each risk summarized based on its actual description in the original text, focusing on the risk's nature and characteristics.

Return only the summary text, no additional explanation.
\end{tcolorbox}

\subsubsection{Stage 2: QA Generation Prompt}
\label{sec:cross_year_tracking_prompts}

These prompts generate comparative tracking questions and answers by analyzing specific SEC filing items across multiple years.

\textbf{DEF 14A Tracking Analysis Prompt}
\label{sec:def14a_tracking_prompt}

This prompt is designed to generate tracking questions and answers by comparing DEF 14A (Definitive Proxy Statement) documents across different years. The focus is on identifying changes and trends in corporate governance, executive compensation, and shareholder matters.

\begin{tcolorbox}[
    colback=lt-light-primary!80,
    colframe=lt-dark-accent!70!black,
    title=DEF 14A Tracking Analysis Prompt,
    breakable,
    fontupper=\small,
    before upper={\sloppy\parindent0pt\parskip5pt
}
]
You are a professional financial analyst specializing in SEC filings analysis. Your task is to generate tracking questions and answers by comparing DEF 14A (Definitive Proxy Statement) documents from \texttt{\{company\_name\}} across different years (\texttt{\{years\_desc\}}).\\
DEF 14A documents contain critical information about:\\
- Board of Directors composition, qualifications, and elections\\
- Executive compensation (salaries, bonuses, stock options, benefits)\\
- Corporate governance practices and policies\\
- Shareholder proposals and voting matters\\
- Audit committee reports and auditor information\\
- Related party transactions\\
- Director independence and conflicts of interest\\
- Equity compensation plans\\
- Risk oversight and management\\
- Company structure and employee composition\\

Focus on identifying CHANGES and TRENDS across the years, such as:

1. Changes in board composition (directors joining, leaving, age changes)\\
2. Changes in executive compensation structure and amounts\\
3. Changes in governance policies\\
4. Changes in company structure\\
5. Changes in employee composition\\
6. Trends in financial metrics\\
7. Changes in audit fees and auditor information\\
8. Changes in equity compensation plans\\

IMPORTANT REQUIREMENTS:

- Each question MUST explicitly mention the company name \texttt{"\{company\_name\}"} and the specific year(s) being compared\\
- Generate 3-4 high-quality question-answer pairs that highlight meaningful changes or comparisons between the years\\
- Each question should be specific and focused on observable changes\\
- Each answer should reference specific data points from the documents\\
- Each answer should be clear, factual, and based on the provided content\\
- Questions should be useful for tracking company evolution over time

Format your response as a JSON array, where each item is a dictionary with \texttt{"question"} and \texttt{"answer"} keys.

\texttt{\{comparison\_text\}}

Please generate the tracking Q\&A pairs in JSON format:
\end{tcolorbox}

\textbf{Item 3: Legal Proceedings Tracking Prompt}
\label{sec:item3_tracking_prompt}

\begin{tcolorbox}[
    colback=lt-light-primary!80,
    colframe=lt-dark-accent!70!black,
    title=Item 3 Legal Proceedings Tracking Prompt,
    breakable,
    fontupper=\small,
    before upper={\sloppy\parindent0pt\parskip5pt,
    breakable}
]
Please act as a rigorous financial analyst. Based on the text passages I provide from different years, generate a set of "tracking analysis Q\&A pairs."

\textbf{Your Task and Objectives:}
1. Generate high-quality tracking analysis questions and answers. \textbf{For Item 1, generate 12-18 Q\&A pairs (1-2 per dimension). For other items, generate 3 Q\&A pairs.}\\
2. Questions must \textbf{span and compare multiple time points (years) mentioned in the texts}, focusing on the evolution of the same entity, indicator, or theme.\\
3. \textbf{Answers must comprehensively utilize information from all provided text passages}, none should be omitted.\\

\textbf{Special Focus for Item 3 - Legal Proceedings:}
- Focus specifically on \textbf{legal proceedings} and litigation matters.\\
- Track changes in ongoing legal cases, new cases filed, resolved cases, and any significant developments.\\
- Compare the status, nature, and impact of legal proceedings between the two years.\\
- Analyze any trends in legal exposure, settlement amounts, or case outcomes.\\

\textbf{Specific Requirements for Question Design (all must be satisfied):}
- \textbf{Form and Depth}: Questions must be complete, meaningful interrogative sentences, not overly simple. They should simulate analytical questions that financial analysts would ask in their daily work when writing reports or making decisions.\\
- \textbf{Content Focus}: Questions must explicitly mention the company name (e.g., \texttt{"\{company\_name\}"}) and the involved years (e.g., \texttt{"\{year1\} and \{year2\}"}).\\
- \textbf{Data Precision}: All cited numbers, dates, percentages, etc., must strictly use the exact wording from the original text. \textbf{Any form of rounding or estimation is prohibited}.\\
- \textbf{Analysis Requirements}: Questions should guide answers that include description, comparison, cause analysis, and inference.\\

\textbf{Specific Specifications for Answer Generation (all must be satisfied):}\\
- \textbf{Format}: Answers must be coherent, fluent paragraph-style text. \textbf{The use of bullet points, numbered lists, or point-by-point statements is prohibited}.\\
- \textbf{CRITICAL: Number Accuracy and Traceability} - This is of utmost importance:\\
  - \textbf{Every number, percentage, date, and numerical value MUST be taken EXACTLY from the provided text passages.} No rounding, approximation, or modification is allowed.\\
  - \textbf{All numbers must have a clear source.} Every numerical value in the answer must be traceable to a specific location in the provided text passages.\\
  - \textbf{When Calculations are Involved}: If analysis requires calculations:\\
    * Use ONLY exact numbers from the original text as inputs\\
    * Show calculation steps clearly and step by step\\
    * Verify all calculations are mathematically correct\\
    * Use the exact calculation results (no rounding unless the original text rounds)\\
    * Ensure all intermediate and final numerical results are accurate\\
- \textbf{Content Composition}:\\
  a. \textbf{Factual Description and Comparison}: Based on texts from different years, clearly describe key facts, changes, and continuity of the theme along the timeline.\\
  b. \textbf{Analysis and Inference}: Provide reasonable analysis of the causes, impacts, or trends of changes. Logical inferences based on provided information are allowed, but \textbf{excessive speculation without textual basis or introduction of external knowledge is strictly prohibited}.\\
  c. \textbf{Comprehensive Integration}: Answers must organically integrate all provided text information, demonstrating connections between information.\\

\textbf{Text Materials for Analysis:}

--- Text from \texttt{\{year1\}} ---
\texttt{\{text1\}}

--- Text from \texttt{\{year2\}} ---
\texttt{\{text2\}}

Now, please generate tracking analysis Q\&A pairs that meet all the above requirements. \textbf{For Item 1, generate 12-18 Q\&A pairs (1-2 per dimension). For other items, generate 3 Q\&A pairs.} Format your response as a JSON array with the following structure:\\
\begin{verbatim}
[
  {
    "question": "...",
    "answer": "..."
  },
  ...
]
\end{verbatim}

Only return the JSON array, no additional text. Do not include a "years" field in your response.
\end{tcolorbox}

\textbf{Item 1: Business Overview Tracking Prompt}
\label{sec:item1_tracking_prompt}

\begin{tcolorbox}[
    colback=lt-light-primary!80,
    colframe=lt-dark-accent!70!black,
    title=Item 1 Business Overview Tracking Prompt,
    breakable,
    fontupper=\small,
    before upper={\sloppy\parindent0pt\parskip5pt,
    breakable}
]
Please act as a rigorous financial analyst. Based on the text passages I provide from different years, generate a set of "tracking analysis Q\&A pairs."

\textbf{Your Task and Objectives:}
1. Generate high-quality tracking analysis questions and answers. \textbf{For Item 1, generate 12-18 Q\&A pairs (1-2 per dimension). For other items, generate 3 Q\&A pairs.}\\
2. Questions must \textbf{span and compare multiple time points (years) mentioned in the texts}, focusing on the evolution of the same entity, indicator, or theme.\\
3. \textbf{Answers must comprehensively utilize information from all provided text passages}, none should be omitted.\\

\textbf{Special Focus for Item 1 - Business Overview:}
Generate questions that focus on tracking and comparing changes across the following key dimensions. \textbf{IMPORTANT: Generate 1-2 Q\&A pairs for EACH of the 6 dimensions below (total 12-18 Q\&A pairs):}

1. \textbf{Strategy \& Business Model Evolution} - Track changes in strategic direction, business model shifts, strategic initiatives, and major strategic decisions. Generate 1-2 Q\&A pairs for this dimension.\\
2. \textbf{Financial Performance \& Revenue Structure} - Track changes in revenue sources, financial metrics, revenue composition, profitability trends, and key financial indicators. Generate 1-2 Q\&A pairs for this dimension.\\
3. \textbf{Product \& Technology Portfolio} - Track changes in product offerings, technology platforms, product development, innovation initiatives, and portfolio composition. Generate 1-2 Q\&A pairs for this dimension.\\
4. \textbf{Operations \& Human Capital} - Track changes in operational structure, facilities, employee counts, organizational changes, and operational efficiency. Generate 1-2 Q\&A pairs for this dimension.\\
5. \textbf{Market, Competition \& Customers} - Track changes in market position, competitive landscape, customer base, market segments served, and market dynamics. Generate 1-2 Q\&A pairs for this dimension.\\
6. \textbf{Growth Drivers \& Investment Focus} - Track changes in growth strategies, investment priorities, capital allocation, expansion plans, and growth initiatives. Generate 1-2 Q\&A pairs for this dimension.

Ensure questions span multiple years and focus on the evolution and changes in these critical business areas. You must generate a total of 6-12 Q\&A pairs (1-2 per dimension).

\textbf{Specific Requirements for Question Design (all must be satisfied):}\\
- \textbf{Form and Depth}: Questions must be complete, meaningful interrogative sentences, not overly simple. They should simulate analytical questions that financial analysts would ask in their daily work when writing reports or making decisions.\\
- \textbf{Content Focus}: Questions must explicitly mention the company name (e.g., \texttt{"\{company\_name\}"}) and the involved years (e.g., \texttt{"\{year1\} and \{year2\}"}).\\
- \textbf{Data Precision}: All cited numbers, dates, percentages, etc., must strictly use the exact wording from the original text. \textbf{Any form of rounding or estimation is prohibited}.\\
- \textbf{Analysis Requirements}: Questions should guide answers that include description, comparison, cause analysis, and inference.\\

\textbf{Specific Specifications for Answer Generation (all must be satisfied):}\\
- \textbf{Format}: Answers must be coherent, fluent paragraph-style text. \textbf{The use of bullet points, numbered lists, or point-by-point statements is prohibited}.\\
- \textbf{CRITICAL: Number Accuracy and Traceability} - This is of utmost importance:\\
  - \textbf{Every number, percentage, date, and numerical value MUST be taken EXACTLY from the provided text passages.} No rounding, approximation, or modification is allowed.\\
  - \textbf{All numbers must have a clear source.} Every numerical value in the answer must be traceable to a specific location in the provided text passages.\\
  - \textbf{When Calculations are Involved}: If analysis requires calculations:\\
    * Use ONLY exact numbers from the original text as inputs\\
    * Show calculation steps clearly and step by step\\
    * Verify all calculations are mathematically correct\\
    * Use the exact calculation results (no rounding unless the original text rounds)\\
    * Ensure all intermediate and final numerical results are accurate\\
- \textbf{Content Composition}:\\
  a. \textbf{Factual Description and Comparison}: Based on texts from different years, clearly describe key facts, changes, and continuity of the theme along the timeline.\\
  b. \textbf{Analysis and Inference}: Provide reasonable analysis of the causes, impacts, or trends of changes. Logical inferences based on provided information are allowed, but \textbf{excessive speculation without textual basis or introduction of external knowledge is strictly prohibited}.\\
  c. \textbf{Comprehensive Integration}: Answers must organically integrate all provided text information, demonstrating connections between information.\\

\textbf{Text Materials for Analysis:}

--- Text from \texttt{\{year1\}} ---
\texttt{\{text1\}}

--- Text from \texttt{\{year2\}} ---
\texttt{\{text2\}}

Now, please generate tracking analysis Q\&A pairs that meet all the above requirements. \textbf{For Item 1, generate 12-18 Q\&A pairs (1-2 per dimension). For other items, generate 3 Q\&A pairs.} Format your response as a JSON array with the following structure:\\
\begin{verbatim}
[
  {
    "question": "...",
    "answer": "..."
  },
  ...
]
\end{verbatim}

Only return the JSON array, no additional text. Do not include a "years" field in your response.
\end{tcolorbox}

\textbf{Item 1A: Risk Factors Tracking Prompt}
\label{sec:item1a_tracking_prompt}

\begin{tcolorbox}[
    colback=lt-light-primary!80,
    colframe=lt-dark-accent!70!black,
    title=Item 1A Risk Factors Tracking Prompt,
    breakable,
    fontupper=\small,
    before upper={\sloppy\parindent0pt\parskip5pt}
]
Please act as a rigorous financial analyst. Based on the text passages I provide from different years, generate a set of "tracking analysis Q\&A pairs."

\textbf{Your Task and Objectives:}\\
1. Generate high-quality tracking analysis questions and answers. \textbf{For Item 1, generate 12-18 Q\&A pairs (1-2 per dimension). For other items, generate 3 Q\&A pairs.}\\
2. Questions must \textbf{span and compare multiple time points (years) mentioned in the texts}, focusing on the evolution of the same entity, indicator, or theme.\\
3. \textbf{Answers must comprehensively utilize information from all provided text passages}, none should be omitted.

\textbf{Special Focus for Item 1A - Risk Factors:}\\
- Focus specifically on \textbf{comparing risk factors} mentioned in both years.\\
- Identify which risks have changed (new risks added, risks removed, risks modified).\\
- Compare how the \textbf{same risk} is described differently between the two years (if applicable).\\
- Track changes in risk severity, risk descriptions, risk prioritization, and risk mitigation strategies.\\
- Analyze whether risks have increased, decreased, or remained stable, and what factors might explain these changes.\\
- Note any emerging risks in the later year and any risks that are no longer mentioned.

\textbf{Specific Requirements for Question Design (all must be satisfied):}
- \textbf{Form and Depth}: Questions must be complete, meaningful interrogative sentences, not overly simple. They should simulate analytical questions that financial analysts would ask in their daily work when writing reports or making decisions.\\
- \textbf{Content Focus}: Questions must explicitly mention the company name (e.g., \texttt{"\{company\_name\}"}) and the involved years (e.g., \texttt{"\{year1\} and \{year2\}"}).\\
- \textbf{Data Precision}: All cited numbers, dates, percentages, etc., must strictly use the exact wording from the original text. \textbf{Any form of rounding or estimation is prohibited}.\\
- \textbf{Analysis Requirements}: Questions should guide answers that include description, comparison, cause analysis, and inference.

\textbf{Specific Specifications for Answer Generation (all must be satisfied):}\\
- \textbf{Format}: Answers must be coherent, fluent paragraph-style text. \textbf{The use of bullet points, numbered lists, or point-by-point statements is prohibited}.\\
- \textbf{CRITICAL: Number Accuracy and Traceability} - This is of utmost importance:\\
  - \textbf{Every number, percentage, date, and numerical value MUST be taken EXACTLY from the provided text passages.} No rounding, approximation, or modification is allowed.\\
  - \textbf{All numbers must have a clear source.} Every numerical value in the answer must be traceable to a specific location in the provided text passages.\\
  - \textbf{When Calculations are Involved}: If analysis requires calculations:\\
    * Use ONLY exact numbers from the original text as inputs\\
    * Show calculation steps clearly and step by step\\
    * Verify all calculations are mathematically correct\\
    * Use the exact calculation results (no rounding unless the original text rounds)\\
    * Ensure all intermediate and final numerical results are accurate\\
- \textbf{Content Composition}:\\
  a. \textbf{Factual Description and Comparison}: Based on texts from different years, clearly describe key facts, changes, and continuity of the theme along the timeline.\\
  b. \textbf{Analysis and Inference}: Provide reasonable analysis of the causes, impacts, or trends of changes. Logical inferences based on provided information are allowed, but \textbf{excessive speculation without textual basis or introduction of external knowledge is strictly prohibited}.\\
  c. \textbf{Comprehensive Integration}: Answers must organically integrate all provided text information, demonstrating connections between information.\\

\textbf{Text Materials for Analysis:}

--- Text from \texttt{\{year1\}} ---
\texttt{\{text1\}}

--- Text from \texttt{\{year2\}} ---
\texttt{\{text2\}}

Now, please generate tracking analysis Q\&A pairs that meet all the above requirements. \textbf{For Item 1, generate 12-18 Q\&A pairs (1-2 per dimension). For other items, generate 3 Q\&A pairs.} Format your response as a JSON array with the following structure:
\begin{verbatim}
[
  {
    "question": "...",
    "answer": "..."
  },
  ...
]
\end{verbatim}

Only return the JSON array, no additional text. Do not include a "years" field in your response.
\end{tcolorbox}

\subsection{Generate Scoring Points for Gold Answers}
To establish objective evaluation criteria for question-answer pairs, the system includes a specialized prompt that extracts essential scoring points from reference (``gold'') answers. This process identifies the critical elements that must be present for an answer to be considered complete and correct. The generated scoring points serve as a benchmark for subsequent automated evaluation of LLM-generated answers.

\begin{tcolorbox}[
    colback=score-light,
    colframe=score-dark,
    title=\textbf{Scoring Points Generation},
    breakable,
    fonttitle=\bfseries
]
You are an expert evaluator. Analyze the following question-answer pair and identify the CRITICAL key points that MUST be correctly addressed and mentioned for the answer to be considered correct.

Question: \texttt{\{question\}}

Answer: \texttt{\{answer\}}

Your task: Generate 3-5 key points that represent the ESSENTIAL elements that MUST be:
1. CORRECTLY answered - Each key point must be factually accurate and correctly addressed
2. EXPLICITLY mentioned - Each key point must be clearly stated or implied in the answer
3. COMPLETE - Missing any of these key points would make the answer incomplete or incorrect

Each key point should:
- Represent a MANDATORY requirement for a correct answer to this specific question
- Be specific enough that you can verify whether the answer correctly addresses it
- Cover the core information that the question is asking for
- Be essential - if the answer fails on this point, the answer is wrong or incomplete
- Be based ONLY on what the question requires and what the answer provides

Focus on identifying what MUST be correct and what MUST be mentioned. These are 
non-negotiable requirements for a correct answer.

Please output the result in the following JSON format:
\texttt{\{}
  \texttt{"key\_points": }[
    \texttt{"Key point 1 description"},
    \texttt{"Key point 2 description"},
    \texttt{"Key point 3 description"}
  ]
\texttt{\}}

Make sure to return only valid JSON without any additional text or explanation.
\end{tcolorbox}

\section{Prompts for QA Quality Assurance}
\label{sec:quality_assure}
The accuracy evaluation prompt serves as a self-consistency check that validates whether generated answers are factually correct and contextually grounded.
\subsection{Detail \& Reasoning QA}
\begin{tcolorbox}[
    colback=validate-light,
    colframe=validate-dark,
    title=QA Quality Evaluation Prompt for DR QA,
    fonttitle=\bfseries,
    breakable
]
\texttt{<|begin\_of\_text|><|start\_header\_id|>system\\<|end\_header\_id|>} \\
You are an expert financial analyst evaluating Q\&A pairs for accuracy and quality. \\

\textbf{EVALUATION CRITERIA:} \\
1. \textbf{CORRECT ANSWER}: All facts, numbers, and calculations are accurate based on the provided context \\
2. \textbf{INCORRECT ANSWER}: Contains factual errors, wrong numbers, or fabricated information \\
3. \textbf{FAILURE TO ANSWER}: Refuses to answer or says information unavailable \\
4. \textbf{PARTIAL ANSWER}: Contains some correct information but has minor issues or incomplete analysis \\
\textbf{CRITICAL RULES:} \\
- Numbers in answers MUST match numbers in context exactly \\
- NO fabricated data (made-up figures, dates, or facts not in context) \\
- NO external knowledge (industry averages, peer data, historical data not provided) \\
- Calculations must use only data present in context \\
- For investment funds with recent inception dates, reject historical trend questions \\
- If context is provided, verify ALL numerical claims against it \\
- If NO context is available, mark as FAILURE since verification is impossible \\
- For PARTIAL answers: If the answer contains reasonable inferences based on the context or other parts of the answer that help understand the answer without going off-topic, consider it CORRECT rather than PARTIAL \\
\textbf{SCORING SYSTEM (1-5 points):} \\
- 5 points: CORRECT - Perfect accuracy, comprehensive analysis, all facts verified \\
- 4 points: CORRECT - Mostly accurate with minor omissions, reasonable inferences \\
- 3 points: PARTIAL - Some correct information but missing key details or minor errors \\
- 2 points: INCORRECT - Significant factual errors or fabricated information \\
- 1 point: FAILURE - Cannot answer or completely wrong \\
\textbf{RESPONSE FORMAT:} \\
ANALYSIS: [Brief analysis of accuracy and quality] \\
VERDICT: [CORRECT/INCORRECT/PARTIAL/FAILURE] \\
SCORE: [1-5] \\
\texttt{<|eot\_id|><|start\_header\_id|>user<|end\_header\_id|>} \\
\textbf{QUESTION:} \texttt{\{question\}} \\
\textbf{ANSWER:} \texttt{\{answer\}} \\
\textbf{CONTEXT:} \texttt{\{context\}} \\
Evaluate this Q\&A pair for accuracy and provide a score: \\
\texttt{<|eot\_id|><|start\_header\_id|>assistant\\<|end\_header\_id|>}
\end{tcolorbox}

\subsection{Enterprise Comparison QA}
\begin{tcolorbox}[
    colback=validate-light,
    colframe=validate-dark,
    title=QA Quality Evaluation Prompt for EC QA,
    fonttitle=\bfseries,
    breakable
]
\texttt{<|begin\_of\_text|><|start\_header\_id|>system\\<|end\_header\_id|>} \\
You are a rigorous financial analyst evaluating enterprise comparison Q\&A pairs for accuracy, completeness, and comparative validity. \\

\textbf{EVALUATION CRITERIA:} \\
1. \textbf{CORRECT ANSWER}: The answer accurately compares \emph{all specified companies} using information fully supported by the provided context \\
2. \textbf{INCORRECT ANSWER}: Contains factual errors, incorrect comparisons, fabricated information, or unsupported claims \\
3. \textbf{FAILURE TO ANSWER}: The question cannot be answered for \emph{both companies} based on the provided context \\
4. \textbf{PARTIAL ANSWER}: Contains some correct comparative information but is incomplete, imbalanced, or insufficiently supported \\

\textbf{CRITICAL RULES:} \\
- The answer MUST include information for \textbf{all companies involved in the comparison} \\
- The answer MUST perform an explicit comparison (not isolated descriptions) \\
- NO phrases indicating missing information are allowed (e.g., ``not mentioned'', ``not disclosed'', ``not available'') \\
- If the context does not provide comparable information for \textbf{both companies}, mark as \textbf{FAILURE TO ANSWER} \\
- ALL factual statements and numerical values MUST be traceable to the provided context \\
- NO external knowledge, assumptions, or industry-wide generalizations \\
- If the comparison is one-sided or omits a key company, mark as PARTIAL or FAILURE depending on severity \\

\textbf{SCORING SYSTEM (1--5 points):} \\
- 5 points: CORRECT — Fully accurate, balanced comparison, all facts traceable, comprehensive analysis \\
- 4 points: CORRECT — Mostly accurate comparison with minor omissions, all claims supported \\
- 3 points: PARTIAL — Some correct comparison but missing key aspects or showing imbalance \\
- 2 points: INCORRECT — Major factual errors, invalid comparisons, or unsupported claims \\
- 1 point: FAILURE — Cannot perform a valid comparison based on the context \\

\textbf{RESPONSE FORMAT:} \\
ANALYSIS: [Brief analysis of comparative accuracy, completeness, and traceability] \\
VERDICT: [CORRECT / INCORRECT / PARTIAL / FAILURE] \\
SCORE: [1--5] \\

\texttt{<|eot\_id|><|start\_header\_id|>user<|end\_header\_id|>} \\
\textbf{QUESTION:} \texttt{\{question\}} \\
\textbf{ANSWER:} \texttt{\{answer\}} \\
\textbf{CONTEXT:} \texttt{\{context\}} \\

Evaluate this enterprise comparison Q\&A pair for accuracy and provide a score: \\
\texttt{<|eot\_id|><|start\_header\_id|>assistant\\<|end\_header\_id|>}
\end{tcolorbox}

\subsection{Longitudinal Tracking QA}
\begin{tcolorbox}[
    colback=validate-light,
    colframe=validate-dark,
    title=QA Quality Evaluation Prompt for LT QA,
    fonttitle=\bfseries,
    breakable
]
\texttt{<|begin\_of\_text|><|start\_header\_id|>system\\<|end\_header\_id|>} \\
You are a rigorous financial analyst evaluating Q\&A pairs for accuracy and quality. \\

\textbf{EVALUATION CRITERIA:} \\
1. \textbf{CORRECT ANSWER}: All facts, numbers, dates, and calculations are accurate and traceable to the provided context \\
2. \textbf{INCORRECT ANSWER}: Contains factual errors, wrong numbers, fabricated information, or untraceable numerical claims \\
3. \textbf{FAILURE TO ANSWER}: Question cannot be answered based on the provided context \\
4. \textbf{PARTIAL ANSWER}: Contains some correct information but has minor issues, incomplete analysis, or limited traceability \\

\textbf{CRITICAL RULES:} \\
- Numbers in answers MUST match numbers in context exactly (no rounding, approximation, or modification unless explicitly allowed) \\
- EVERY number, percentage, date, and numerical value in the answer MUST have a clear source in the context \\
- NO fabricated data (made-up figures, dates, or facts not in context) \\
- NO external knowledge (industry averages, peer data, historical data not provided) \\
- Calculations must use only data present in context and be mathematically correct \\
- Verify ALL numerical claims against the context \\
- If context does NOT contain sufficient information to answer the question, mark as FAILURE TO ANSWER \\
- For calculations: verify input numbers, calculation steps, logic, and final results \\
- If the answer contains any number without a clear source in the context, mark as INCORRECT \\

\textbf{SCORING SYSTEM (1-5 points):} \\
- 5 points: CORRECT - Perfect accuracy, all numbers traceable, comprehensive analysis, all facts verified \\
- 4 points: CORRECT - Mostly accurate with minor omissions, all numbers traceable, reasonable inferences \\
- 3 points: PARTIAL - Some correct information but missing key details, minor errors, or limited traceability \\
- 2 points: INCORRECT - Significant factual errors, untraceable numbers, or fabricated information \\
- 1 point: FAILURE - Cannot answer based on context or completely wrong \\

\textbf{RESPONSE FORMAT:} \\
ANALYSIS: [Brief analysis of accuracy, traceability, and quality] \\
VERDICT: [CORRECT/INCORRECT/PARTIAL/FAILURE] \\
SCORE: [1-5] \\

\texttt{<|eot\_id|><|start\_header\_id|>user<|end\_header\_id|>} \\
\textbf{QUESTION:} \texttt{\{question\}} \\
\textbf{ANSWER:} \texttt{\{answer\}} \\
\textbf{CONTEXT:} \texttt{\{context\}} \\
Evaluate this Q\&A pair for accuracy and provide a score: \\
\texttt{<|eot\_id|><|start\_header\_id|>assistant\\<|end\_header\_id|>}
\end{tcolorbox}

\section{Prompts for Evaluation}
\label{evaluation_prompt}
\subsection{Prompts for Answer Generation at Test Time}
\begin{tcolorbox}[
    colback=eval-light,
    colframe=eval-dark,
    title=Prompt for Generating LLM Answer in EC QA Test,
    fonttitle=\bfseries,
    breakable
]
\texttt{<|begin\_of\_text|>\\<|start\_header\_id|>system<|end\_header\_id|>} \\
You are a SEC filing financial analysis expert. \\
- Answer in English only. \\
- Output only the final answer. Do not include chain-of-thought or \texttt{<think>} sections. \\
\texttt{<|eot\_id|>\\<|start\_header\_id|>user<|end\_header\_id|>} \\
\textbf{Question:} \{question\} \\
\textbf{Context:} \{content\} \\
Based on the provided context, please answer the question. Remember to output only the final answer. \\
\texttt{<|eot\_id|>\\<|start\_header\_id|>assistant<|end\_header\_id|>}
\end{tcolorbox}

\subsection{Prompts for Automated Evaluation}
\begin{tcolorbox}[
    colback=eval-light,
    colframe=eval-dark,
    title=LLM as Judge Evaluation Prompt,
    fonttitle=\bfseries,
    breakable
]
\texttt{<|begin\_of\_text|><|start\_header\_id|>system\\<|end\_header\_id|>} \\
You are an expert evaluator for financial Q\&A tasks with retrieved evidence. \\

\textbf{CORE DEFINITIONS:} \\
- Use ONLY the provided GOLD ANSWER / KEY POINTS as reference. \\
- The generated answer may contain additional correct information beyond what's in the gold answer/key points. \\
- \textbf{CRITICAL CONSTRAINT: Without access to the full context/retrieved evidence, you cannot determine if additional information is unsubstantiated.} \\
- \textbf{Therefore: Only penalize for DIRECT CONTRADICTIONS with gold answer/key points.} \\
- \textbf{DO NOT penalize for information merely absent from gold answer/key points.} \\

\textbf{VERDICT LABELS (use exactly one):} \\
1. \textbf{CORRECT:} \\
   - All key points are correctly covered (explicitly or through reasonable paraphrase) \\
   - No factual errors or contradictions with gold answer/key points \\
   - Answer is relevant and complete \\
   - \textbf{May include additional correct information not in gold answer/key points} \\
2. \textbf{PARTIAL:} \\
   - Some key points are correctly covered, but important ones are missing \\
   - Missing key points are essential for answering the question \\
   - No factual errors or contradictions with gold answer/key points \\
   - Answer is relevant but incomplete \\
   - \textbf{May include additional correct information not in gold answer/key points} \\
3. \textbf{INCORRECT:} \\
   - \textbf{Contains DIRECT CONTRADICTIONS with gold answer/key points} \\
   - \textbf{Has clear factual errors that conflict with provided reference} \\
   - Answer is relevant but wrong \\
   - \textbf{DO NOT mark as incorrect for missing gold answer/key points info if answer has alternative correct info} \\
4. \textbf{FAILURE:} \\
   - Answer is completely irrelevant to the question, OR \\
   - Answer fails to address the question’s topic, intent, or target entities, OR \\
   - Answer consists of off-topic, generic, or hallucinated content, OR \\
   - Answer is empty, null, or contains no substantive response \\
   
\textbf{KEY POINTS EVALUATION RULES:} \\
- Check each key point against the generated answer. \\
- For each key point, label as one of: \\
  \texttt{PRESENT} (correctly mentioned), \texttt{PARTIAL} (partially addressed), \\
  \texttt{MISSING} (not addressed), \texttt{INCORRECT} (addressed but factually wrong). \\
- Consider numeric/date/percent equivalence and allow reasonable paraphrases. \\

\textbf{ERROR TYPE TAXONOMY (choose NONE if VERDICT=CORRECT):} \\
B) \textit{Generation-related} \\
  B1. \texttt{Hallucination}: answer not entailed by retrieved evidence \\
    - \textbf{Hallucination = Information that CONTRADICTS gold answer/key points} \\
    - \textbf{NOT hallucination = Information absent from but not contradicting gold answer/key points} \\
    - B1 Error Subtypes (select the applicable label(s)):\\
        - B1-1: Numeric or Categorical Hallucination - Fabricated
        numbers, percentages, ratings, years, categories, or other hard values that contradict or are absent from evidence. \\
        Examples: Making up specific percentages, case counts, or numerical comparisons when evidence doesn't provide them.\\
        - B1-2: Entity Attribute Hallucination - Fabricated attributes, states, policies, or strategies of a single entity that contradict or are absent from evidence.\\
        Examples: Claiming a company has an ESG rating system, specific litigation status, or strategic policy when evidence doesn't mention it.\\
        - B1-3: Comparative Stance Hallucination - Fabricated comparative statements (A is more/less than B) without sufficient evidence support.\\
        Examples: Claiming "A has significantly lower risk than B" or "A is more diversified than B" when evidence doesn't support such comparisons.\\
        - B1-4: Trend or Trajectory Hallucination - Fabricated trends or trajectories over multiple periods without sufficient evidence.\\
        Examples: Claiming "continuously increasing", "steadily declining", or "long-term trend" when evidence only covers partial periods or doesn't support such strong trend statements.\\
  B2. \texttt{Contradicts Evidence}: contains internal logical inconsistencies, explicitly conflicts with evidence it mentioned before\\
  B3. \texttt{Excessive Inference}: generalizes beyond a reasonable range based on the evidence \\
  B4. \texttt{Evidence Fusion Failure}: fails to correctly synthesize multiple evidence pieces (complementary or conflicting) \\

C) \textit{Finance-specific numeric \& semantic errors} \\
  C1. \texttt{Numerical Precision}: rounding/tolerance mistakes; \% vs bps confusion \\
  C2. \texttt{Units and scales}: millions vs billions; ratio vs absolute confusion; currency/unit mismatch \\
  C3. \texttt{Time mismatch}: wrong period (e.g., annual vs quarterly, wrong FY/Q) \\
  C4. \texttt{Computation Logic}: uses correct data but computes incorrectly (formula/arithmetic error) \\

D) \textit{Query and context errors} \\
  D1. \texttt{Query misunderstanding}: misidentifies intent, key entity, or asked metric \\
    - D1 Error Types (select the applicable label(s)):\\
        - D1-1: Intent Misunderstanding - The generated answer misunderstands the true intent of the question.\\
        - D1-2: Entity Misidentification - Incorrectly identifies key entities (company names, person names, metric names, etc.).\\
        - D1-3: Metric Misidentification - Incorrectly identifies the asked metric or measurement.\\
  D2. \texttt{Context window abuse}: loses key info due to length limits or fails to prioritize relevant parts \\

\textbf{ERROR TAGGING RULES:} \\
- Output 1 PRIMARY error group (B/C/D) and 1 PRIMARY subtype (B1..D2) when VERDICT != CORRECT. \\
- Optionally output up to 2 SECONDARY subtypes if multiple issues contribute. \\
- Prefer the MOST CAUSAL error: e.g., if evidence is present but model ignores it -> B/C; if question misunderstood -> D. \\

\textbf{RESPONSE FORMAT (strict):} \\
ALL SECTIONS BELOW ARE MANDATORY. Do not omit any section. Use exactly the headings and labels as shown. Do not add extra text outside this format. \\

\textbf{ANALYSIS:} [Concise analysis of answer quality, groundedness, and any numeric/unit/period issues] \\

\textbf{KEY POINTS:} \\
1. [PRESENT|PARTIAL|MISSING|INCORRECT] - brief justification \\
2. [PRESENT|PARTIAL|MISSING|INCORRECT] - brief justification \\
... (one line per key point) \\

\textbf{KEY POINTS SUMMARY:} matched=<int>; partial=<int>; missing=<int>; incorrect=<int> \\
\textbf{DIMENSIONAL SCORES:} \\
1. Information Coverage: [1-5]\\
- Includes all query-critical facts/constraints needed to answer.\\
- Avoids spending space on irrelevant details that don’t support the answer.\\
2. Reasoning Chain: [1-5]\\
- Provides a logical sequence linking evidence → intermediate conclusions → final answer.\\
- Not just paraphrasing; shows why the conclusion follows.\\
3. Factual Consistency: [1-5]\\
- Every stated claim is supported by the given evidence/context.\\
- No contradictions with evidence; no unsupported additions.\\
4. Clarity of Expression: [1-5]\\
- Main answer is easy to find; structure is organized (e.g., bullet points, clear sentences).\\
- Minimal redundancy; no “burying the lead” with unnecessary text.\\
5. Analytical Depth: [1-5]\\
- Selects and prioritizes relevant evidence rather than summarizing everything.\\
- Synthesizes/comparisons/inferences are reasonable and grounded in evidence.\\
- Produces a decisive, query-directed outcome (e.g., classification, comparison, recommendation).

\textbf{ERROR TYPE:} \\
\textbf{PRIMARY\_GROUP: }[GENERATION\_RELATED|\\FINANCE\_NUMERIC\_SEMANTIC|QUERY\_CONTEXT|\\NONE] \\
\textbf{PRIMARY\_SUBTYPE: }[B1|B2|B3|B4|C1|C2|C3|C4|D1|D2|\\NONE] \\
\textbf{SECONDARY\_SUBTYPES: }[<subtype>|<subtype>\\|NONE] \\
\textbf{EVIDENCE\_IDS\_USED: }[comma-separated ids from the provided evidence; or NONE] \\

\textbf{VERDICT:} [CORRECT|INCORRECT|PARTIAL|FAILURE] \\
(All sections are mandatory; the VERDICT line must contain only one of the listed labels.) \\
\texttt{<|eot\_id|><|start\_header\_id|>user<|end\_header\_id|>} \\

\textbf{QUESTION:} \{question\} \\
\textbf{GOLD ANSWER:} \{gold\_answer\} \\
\textbf{GENERATED ANSWER:} \{generated\_answer\} \\

\textbf{KEY POINTS TO CHECK:} \{kp\_block\} \\

\texttt{<|eot\_id|><|start\_header\_id|>assistant\\<|end\_header\_id|>}
\end{tcolorbox}

\end{document}